\documentclass[12pt,notitlepage,letterpaper]{article}
\usepackage{amssymb}
\usepackage{amsmath}
\usepackage{amsfonts}
\usepackage{verbatim}
\usepackage{geometry}
\usepackage[scaled]{helvet}
\usepackage{graphicx}
\usepackage{subfig}
\usepackage{amssymb}
\usepackage{amsmath}
\usepackage{amsfonts}
\usepackage{scalefnt}
\usepackage{pgflibraryarrows}
\usepackage{pgflibrarysnakes}
\usepackage{xcolor}
\usepackage{hyperref}
\usepackage{caption}
\usepackage[onehalfspacing]{setspace}
\usepackage{hyperref}
\usepackage{lipsum}
\usepackage{todonotes}
\usepackage{threeparttable}
\usepackage{longtable}
\usepackage{xr}

\normalfont
\newtheorem{theorem}{Theorem}

\newtheorem{corollary}{Corollary}

\newtheorem{definition}{Definition}

\newtheorem{lemma}{Lemma}

\newtheorem{proposition}{Proposition}
\newtheorem{remark}{Remark}

\newenvironment{proof}[1][Proof]{\noindent\textbf{#1.} }{\ \rule{0.5em}{0.5em}}
\hypersetup{colorlinks=true,linktoc=all,linkcolor=[rgb]{0.00,0.00,0.63}}
\let\OLDthebibliography\thebibliography
\renewcommand\thebibliography[1]{
  \OLDthebibliography{#1}
  \setlength{\parskip}{0pt}
  \setlength{\itemsep}{0pt plus 0.3ex}
}

\input{tcilatex}

\begin{document}

\title{\textbf{Monotone Equilibrium}\\
\textbf{in Matching Markets with Signaling\thanks{%
This paper extends the first part of our paper, \textquotedblleft Designing
a Competitive Monotone Signaling Equilibrium.\textquotedblright\ We are very
grateful to Marzena Rostek (Editor) and two anonymous referees for their
comments and suggestions on the original paper, which allow us to greatly
improve the quality of this paper. For their comments on the original paper,
we would also like to thank Maxim Ivanov, Shuo Liu, Gabor Virag, and seminar
participants at the 2022 Australasia Meeting of the Econometric Society and
the 17th European Meeting on Game Theory. Han and Shin gratefully
acknowledge support from the Social Sciences and Humanities Research Council
of Canada, respectively.}}}
\author{Seungjin Han\thanks{%
Dept. of Economics, McMaster University, Canada. Email: hansj@mcmaster.ca}
\and Alex Sam\thanks{%
Dept. of Economics, McMaster University, Canada. Email: sama1@mcmaster.ca}
\and Youngki Shin\thanks{%
Dept. of Economics, McMaster University, Canada. Email: shiny11@mcmaster.ca}}
\date{January 26, 2024}
\maketitle

\begin{abstract}
We introduce a notion of competitive signaling equilibrium (CSE) in
one-to-one matching markets with a continuum of heterogeneous senders and
receivers. We then study monotone CSE where equilibrium outcomes - sender
actions, receiver reactions, beliefs, and matching - are all monotone in the
stronger set order. We show that if the sender utility is
monotone-supermodular and the receiver's utility is weakly
monotone-supermodular, a CSE is stronger monotone if and only if it passes
Criterion D1 (Cho and Kreps (1987), Banks and Sobel (1987)). Given any
interval of feasible reactions that receivers can take, we fully
characterize a unique stronger monotone CSE and establishes its existence
with quasilinear utility functions.

\medskip


\end{abstract}









\section{Introduction\label{sec_intro}}

If players move sequentially, action choices made by players (senders) on
one side who moves early make the other players (receivers) on the other
side form a belief on their types. Such a belief plays a critical role in
matching between a sender (e.g., a worker, a seller, an entrepreneur) and a
receiver (e.g., a firm, a buyer, a lender) in a matching market. This is
because a sender's action and type both affect a receiver's utility and a
receiver's reaction choice (e.g., a firm's wage offer) depends on reactions
that other receivers are willing to take, which in turn depends on their
beliefs on a sender's type.

Not surprisingly, a belief on a sender's type conditional on her action is
the source of multiple equilibria. To provide a unified and sharper analysis
of signaling equilibrium in a large matching market, we develop a theory of
monotone signaling equilibrium where all equilibrium outcomes - actions,
reactions, beliefs, and matching - are monotone in the \emph{stronger set
order}.\footnote{%
Consider two sets $A$ and $B$ in the power set $P(Y)$ for $Y$, a lattice
with a given relation $\geq $. We say that $A\leq _{c}B$, read
\textquotedblleft $A$ is \emph{completely lower} than $B$\textquotedblright\
in the stronger set order if for every $a\in A$ and $b\in B,$ $a\leq b.$}.

\paragraph{Notion of competitive signaling equilibrium}

For the analysis of stronger monotone signaling equilibrium, we first
propose a notion of competitive signaling equilibrium (CSE) in a large
matching market with a continuum of heterogeneous senders and receivers in
terms of their types. Senders first move by choosing their action. A single
sender has no significant market power in a large market. Therefore, when
she chooses her action, each sender takes as given the market reaction
function $\tau $ that specifies the reaction of the receiver who is matched
with a sender on the market as a function of her action. A sender's action
choice is optimal when not only there is no incentive for her to choose any
action that any other sender chooses but also there is no profitable sender
deviation to an off-path action.

After observing actions chosen by senders, receivers form a belief about the
sender's type conditional on her action. An equilibrium matching outcome is
characterized by receivers' reaction choices and matching such that, given
senders' action choices and a belief about the sender's type conditional on
her action, there is no pair of sender type $z$\ and receiver type $x$\ who
are strictly better off by forming a new match with an alternative reaction
chosen by receiver type $x$. The market reaction function $\tau $ is \emph{%
endogenous} in that in equilibrium the receivers' reaction choices confirm
the market reaction function $\tau $ that senders take as given when they
choose their action.

\paragraph{Stronger monotone CSE}

We show that if the sender's utility satisfies the monotone-supermodular
condition in Liu and Pei (2020), a belief $\mu $ is stronger monotone (i.e., 
$s^{\prime }\leq s$ $\Rightarrow $ supp $\mu (s^{\prime })\leq _{c}$ supp $%
\mu (s)$) if and only if it passes Criterion D1 (Cho and Kreps (1987), Cho
and Sobel (1990), Banks and Sobel (1987)). That is, the stronger
monotonicity of a belief is the full implication of Criterion D1.

It is worth mentioning Cho and Sobel monotonicity of beliefs\footnote{%
Suppose that an action $s$ is chosen by some sender type $z$ on the
equilibrium path. Cho and Sobel monotonicity means that a receiver should
believe that $s^{\prime }>s$ is not chosen by a lower sender type than $z$.}
(Cho and Sobel (1990)). It is only a partial implication of Criterion D1 but
is instrumental for selecting a separating equilibrium as a unique D1
equilibrium when the set of feasible reactions is not bounded. However, if
it is bounded, we may not have a separating equilibrium. In this case, we
need to derive a unique stronger monotone equilibrium (D1 equilibrium) using
the stronger monotonicity of beliefs, which is the full implication of
Criterion D1. The stronger monotonicity of beliefs implies that the support
of a belief on the sender's type conditional on any off-path action $%
s^{\prime }$ is a singleton $\{z^{\prime }\}$ and that no sender types have
an incentive to deviate to $s^{\prime }$ if and only if sender type $%
z^{\prime }$ has no incentive. This property is crucial in deriving a
stronger monotone equilibrium with a bounded set of feasible reactions.

Applying Milgrom and Shannon's Monotone Selection to each receiver's maximal
utility function, we establish the following stronger monotone signaling
theorem (Theorem \ref{theorem_stronger_monotone_eq}): Suppose that the
sender's utility is monotone-supermodular and the receiver's utility is
weakly monotone-supermodular respectively.\footnote{%
The monotonicity of the receiver's utility is imposed only over his type.}
Then, a CSE is stronger monotone if and only if $\mu $ passes Criterion D1.

A market matching function specifies a subset of receiver types $m(s)$ who
are matched with senders with action $s$ in equilibrium. In a stronger
monotone CSE, a market matching function admits separating, pooling, or
semi-pooling.

\paragraph{Characterization and existence of stronger monotone CSE}

As an application of stronger monotone CSE, we consider a model where
players have quasilinear utility functions with respect to the receiver's
reaction and the set of feasible reactions is a closed interval in $%
\mathbb{R}
_{+}$. We show that given any interval of feasible reactions, a stronger
monotone CSE is \emph{unique} and \emph{well-behaved}. A \textquotedblleft
well-behaved\textquotedblright\ equilibrium is characterized by the two
threshold sender types (See Theorem \ref{theorem1} for full
characterization). The lower threshold sender type specifies the lowest
sender type who enters the market, whereas any senders above the upper
threshold sender type pool their actions. Any sender between the two
threshold types separates themselves.

If the two threshold types are the same, it becomes a pooling equilibrium.
If the upper threshold type is the supremum of the sender types and greater
than the lower threshold type, it becomes a separating equilibrium. If the
upper threshold type is less than the supremum of the sender types but
greater than the lower threshold type, separating and pooling coexist in the
well-behaved equilibrium. In the separating part of the equilibrium,
matching is \emph{assortative} in terms of sender action and receiver type
(and hence in terms of sender type and receiver type), whereas in the
pooling part, it is \emph{random}.

The characterization of a well-behaved equilibrium with two threshold sender
types leads to the following results. If the upper bound of feasible
reactions is sufficiently high, a unique stronger monotone CSE is
separating. If the upper bound of feasible reactions is too low to induce a
separating CSE, a unique stronger monotone CSE is \emph{strictly}
well-behaved with both separating and pooling: there is a discontinuity at
the upper threshold sender type. If there is only one feasible reaction, a
unique stronger monotone CSE is pooling.

Given an interval of feasible reactions, the existence of a unique
well-behaved stronger monotone CSE depends on the existence of the two
threshold sender types and the solution for differential equation for the
belief on the sender's action in the separating part. We show that they 
\emph{exist} and are \emph{unique} given technical assumptions (Theorem \ref%
{thm_unique_separating_eq}, Lemmas \ref{lemmaA}, \ref{lemma_pooling}, and %
\ref{lemma_well_behaved}). It is important that the existence is
established, given any closed interval of feasible reactions because the
interval of feasible reactions can be a policy decision made by the
government (e.g., wage floor and wage ceiling in a labour market, interest
rate ceiling in a loan market, etc.).

\paragraph{Differentiable separating part}

We also establish that if a stronger monotone CSE has a separating part,
that part is differentiable given technical assumptions (Theorem \ref%
{thm_differentiable_sep_eq}). This differentiability results contribute to
the literature. In a sequential move game with one informed agent (sender)
and (possibly) multiple uninformed agents, Mailath (1987) takes a
reduced-form approach, setting up the informed agent's utility function
incorporating the optimal actions of uninformed agents who choose after the
informed agent. He identified a set of properties of the informed agent's
reduced-form utility function that ensures the monotonicity and
differentiability of her equilibrium action strategy in a separating
equilibrium. A belief monotonicity is assumed as part of the properties. In
contrast to the reduced-form approach, we directly derive the monotonicity
of the equilibrium functions (matching, action, reaction and belief
functions) and the differentiability of the separating part, carefully
handling the market clearing condition to get those properties for the
belief function.

Focusing on a separating equilibrium, Hopkins (2012) applied the
differentiability results in Mailath (1987) to a two-sided matching model by
imposing the restriction that there is no complementary between receiver
type $x$ and sender action $s$ in the receiver's utility (equivalently the
match surplus when the receiver's utility is quasilinear with respect to the
receiver's reaction). This restriction gets rid of a matching effect on the
marginal productivity of a sender's action. Such a restriction is not needed
for establishing our differentiability result.\textbf{\ }

\subsection{Related literature}

In a two-period signaling game between one sender and one receiver, Liu and
Pei (2020) identify the monotone-supermodularity condition on the sender's
utility function to derive the monotonicity of a sender's equilibrium mixed
strategy (in the strong set order). They show that when the receiver has
more than two actions, the monotone-supermodularity condition is not enough
to guarantee that the sender's equilibrium strategy is monotone. This is
because their model allows the receiver to use mixed strategies. When the
receiver has more than two reactions, not every pair of distributions over
the receiver's reactions can be completely ranked in terms of the first
order stochastic dominance. As a result, Liu and Pei (2020) provided
additional conditions on players' payoff functions to ensure that the
sender's equilibrium mixed strategy is monotone for his reaction choice.

However, in our large matching market, because a receiver's reaction choice
is deterministic, such an issue does not arise and the
monotone-supermodularity condition on the sender's utility function alone is
sufficient to derive the monotonicity of both players' equilibrium
strategies. Furthermore, we show the equivalence between the Criterion D1
and the stronger monotone beliefs and its implications. While the set of
types, actions and reactions are all finite in Liu and Pei, they can be
finite or continuous in our model.

Mensch (2020) considers general dynamic incomplete information games with
multiple players where types are continuous but actions are finite. Without
imposing a monotonicity condition on players' payoff functions, he shows
that increasing difference/supermodular properties alone can ensure that a
player's best reply is increasing in the strong set order \emph{if} all
other players' strategies are monotone. This is the basis of the existence
of a monotone pure-strategy equilibrium in which all players' equilibrium
strategies are monotone in the strong set order.\footnote{%
It does not however exclude the existence of non-monotone equilibria.}

His challenge is how to deal with the discontinuity in players' beliefs
conditional on off-path actions, when applying equilibrium existence results
from static Bayesian games (Athey (2001), McAdams (2002), Van Zandt and
Vives (2007), and Reny (2011)) to dynamic Bayesian games. This is because
the discontinuity in players' beliefs makes their expected payoffs in the
future periods also discontinuous. To overcome this problem, he perturbs a
player's belief about their opponents' strategies to ensure that all actions
are taken with positive probability. This perturbation preserves the
continuity of beliefs in the actual strategies taken and it allows for use
of equilibrium existence results from static Bayesian games. Taking the
limit as the perturbation vanishes yields an equilibrium of the original
game where the beliefs are monotone in the stronger set order. What we show
is that the monotone-supermodular condition excludes the possibility of
non-monotone equilibria. Furthermore, we uncover the equivalence between
Criterion D1 and stronger monotone beliefs. The implication of the
equivalence is crucial to establish the existence of a unique stronger
monotone CSE given any interval of reactions. Furthermore, for the existence
and uniqueness results on a stronger monotone CSE, we do not apply
equilibrium existence results from static Bayesian games, nor did we take
the reduced-form approach from Mailath (1987) as explained earlier.

Our notion of CSE relates with Kurlat and Scheuer (2021), who also introduce
two-sided heterogeneity in competitive signaling. Their model shows how an
action (a costly signal) chosen by a sender can induce multiple reactions
from different receivers. For that, they assume that a receiver type has no
direct effect on his utility but it rather represents his ability to obtain
some direct signal on the type of a sender (e.g., ability to screen a
worker's type through interviews, etc.). Such ability is valuable in their
model because a sender's action is a pure waste in that it does not enhance
her productivity. In our matching model, a receiver's type directly affects
his utility, creating complementarity between receiver type and sender type
(and action). We show that this complementarity is the driving force behind
a sender's action decision in a general model where the sender's
productivity may or may not be affected by her action. While their
refinement cannot exclude multiple equilibria in general, we show that the
stronger monotone beliefs (i.e., the full implication of Criterion D1)
induces a sharp characterization of a unique stronger monotone CSE given any
reaction interval in a matching market with two-sided heterogeneity and
signaling.

One may think of our model as an extension of the pre-match investment
competition through a sender-receiver framework. Pre-match investment
competition studies whether pre-match competition to match with a better
partner can solve the hold-up problem of non-contractible pre-match
investment that prevails when a match is considered in isolation (e.g.\
Grossman and Hart (1986) and Williamson (1986)). Cole, Mailath, and
Postlewaite (1995), Rege (2008), and Hoppe, Moldovanu, and Sela (2009)
consider pre-match investment with incomplete information and
non-transferable utility without monetary transfers (i.e., no reaction
choice by a receiver). Therefore, the sender-receiver framework does not
apply. Pre-match investment with incomplete information in Hopkins (2012)
includes the transferable-utility case but only with no restrictions on
transfers. A separating equilibrium is the focus in Hopkins (2012).

Equilibrium matching with pre-match investment in a large market with
complete information was formulated in Cole, Mailath and Postlewaite (2001)
with two separate reward schedules for agents on two sides of the market
when a match generates monetary surplus and each agent has a separable
utility function. We generalize this framework in a large market with
incomplete information and arbitrary utility functions. The key difference
is that we embed the equilibrium matching property in the market matching
function, given the market-clearing market reaction function that
endogenously emerges as a result of optimal action/reaction decisions made
by market participants.

Signaling games are known to generate multiple equilibria. Different types
of signaling games, such as Spence's (1973) costly signaling and Crawford
and Sobel's (1982) costless cheap-talk, require different notions of
equilibrium refinement. Chen, Kartik, and Sobel (2008) proposed the no
incentive to separate (NITS) condition to select a unique equilibrium in the
Crawford and Sobel (henceforth, CS), and CS related cheap-talk games.
According to them, an equilibrium satisfies the NITS condition if the sender
of the lowest type weakly prefers the equilibrium outcome to revealing her
type. However, the NITS condition generally has no bite in selecting a
unique equilibrium in costly signaling models like ours. This is because one
can easily construct a belief that satisfies the NITS condition for an
equilibrium of costly signaling games.\footnote{%
All equilibria in Section 13.C in Mas-Collel, Whinston, and Green (1995)
satisfy the NITS condition because all those equilibria adopt beliefs that
put all the weights on the low type conditional on no education.} On the
other hand, equilibrium refinement techniques, such as Criterion D1 by Banks
and Sobel (1987) and Cho and Kreps (1987), which are typically used in
costly signaling models, also do not have the power to refine in cheap-talk
games. As Chen, Kartik, and Sobel (2008) argue, communication is costless so
that one can support any equilibrium outcome with an equilibrium in which
all messages are sent on the equilibrium path so arguments that limit the
set of out-of-equilibrium beliefs have no power to refine.

Kartik (2009) studied the classic strategic communicating setting of CS when
the sender bears a cost of lying, producing a similar structure of
equilibrium for a different reason. In his environment, the sender (almost)
always claims to be a higher type. As her type increases, the sender
eventually runs out of types to mimic for separation when the sender's type
set is bounded. This induces an equilibrium with a threshold sender type
such that all sender types below that separate and all sender types above
that pool on the highest messages. Therefore, his equilibrium has a
structure similar to our strictly well-behaved CSE. In Kartik (2009), there
is no restriction on the receiver's feasible reactions but such an
equilibrium is induced by the sender's upward lying, whereas in our paper, a
strictly well-behaved CSE occurs when the upper bound of feasible reactions
is too low to induce separating everywhere.

\section{Competitive signaling equilibrium\label{section2.2}}

There is a continuum of senders and receivers. They can be interpreted as
sellers and buyers, workers and firms, or entrepreneurs and investors.
Receivers and senders are all heterogeneous in terms of types. The sender's
type set is $Z$ and the receiver's type set is $X.$ Assume that the measures
of senders and receivers are one respectively. Let $G$ and $H$ denote
(probability) distributions for sender types and receiver types
respectively. $G$ and $H$ are public information. However, each sender's
type is her own private information and each receiver's type is his own
private information.

The timing of competitive signaling and matching unfolds over three stages:

\begin{enumerate}
\item[S1.] Senders and receivers decide whether to stay out of the market or
not.

\item[S2.] Prior to entering the market, a sender who decides to enter the
market chooses her (observable) action $s\in S$.

\item[S3.] Receivers observe the whole distribution of actions chosen by
senders on the market. Each receiver on the market chooses his reaction $%
t\in T$ as he is matched with a sender on the market.
\end{enumerate}

When a sender of type $z$ chooses action $s$ and matches with a receiver of
type $x$ who takes reaction $t,$ the sender's utility is $u(t,s,z)$ and the
receiver's utility is $g(t,s,z,x)$. In the example with workers and firms,
the utilities for a sender (worker) of type $z$ and a receiver (firm) of
type $x$ are $u(t,s,z)=t-c(s,z)$ and $g(t,s,z,x)=v(x,s,z)-t$, respectively.
Note that $t$ is the monetary transfer from a firm to his worker, $c(s,z)$
is the cost of choosing education $s\in S$ for a worker of type $z$, and $%
v(x,s,z)$ is the monetary value of the output produced by the worker in a
match.

The reservation utility for every agent corresponds to staying out of the
market and it is equal to zero. We assume that a sender takes the null
action $\eta \in S$ to stay out of the market such that $\eta <s$\ for all $%
s\neq \eta $ (e.g., $\eta =0$ if $S=%
\mathbb{R}
_{+}$). Each of $S,T,X,$ and $Z$ is totally ordered.

We formulate a notion of competitive signaling equilibrium, based on the
notion of stable matching (Definition \ref{def_stable_matching}). We
abstract the model from the actual matching process in S3 by imposing the
equilibrium property of\textbf{\ }a matching outcome (reaction choices by
receivers in the market and matching), given senders' action choices and a
belief on the sender's type conditional on her action.

Let $\sigma (z)$ be the optimal action chosen by a sender of type $z$. Given 
$\sigma :Z\rightarrow S$, we denote the image set of $\sigma $ by $\sigma
(Z) $. Let 
\begin{equation*}
S^{\ast }:=\sigma (Z)\backslash \{\eta \}
\end{equation*}%
denote the set of actions chosen by senders who enter the market for
matching.

A market reaction function $\tau :S^{\ast }\rightarrow T$ specifies a
receiver's reaction conditional on a sender's equilibrium action $s$. Note
that the domain of the market reaction function is $S^{\ast }$, not the
entire set of feasible actions $S$. This is sufficient when each sender's
action choice has no incentive to choose an off-path action, i.e., no
profitable sender deviation to an off-path action. We will incorporate this
condition into the sender's optimal action choice.

All receivers share a common belief, denoted by $\mu (s)\in \Delta (Z)$, on
a sender's type conditional on her action $s\in S$. Every sender's belief on
every other sender's type conditional on her action is also the same as $\mu
(s)$. When the receiver of type $x$ is matched with a sender with action $s$
by taking a reaction $\tau (s),$ his (expected) utility is $\mathbb{E}_{\mu
(s)}\left[ g(\tau (s),s,z,x)\right] $. If $\{z|\sigma (z)=s\}$ is a
singleton, then $\mu \left( s\right) $ becomes a degenerate probability
distribution. In a separating equilibrium, $\mu \left( s\right) $ is a
degenerate probability distribution for all $s\in S^{\ast }.$%

A receiver's matching problem can be formulated as follows: 
\begin{equation}
\max_{s\in S^{\ast }}\;\mathbb{E}_{\mu (s)}\left[ g(\tau (s),s,z,x)\right] 
\text{ s.t. }\mathbb{E}_{\mu (s)}\left[ g(\tau (s),s,z,x)\right] \geq 0.
\label{FP}
\end{equation}%
We use the notation $\xi (x)$ as the action of the sender whom the receiver
of type $x$ optimally chooses as his match partner. If (\ref{FP}) has a
solution for $x\in X,$ $\xi (x)$\ is the solution. Otherwise, $\xi (x)=\eta
. $ Note that $X^{\ast }$ be the set of receiver types such that $\xi (x)$\
is a solution for (\ref{FP}).

Consider a sender's action choice problem. Let $\sigma (z)\in S^{\ast }$ be
the optimal action for a sender of type $z$ if

\begin{enumerate}
\item[(i)] it solves the following problem, 
\begin{equation}
\max_{s\in S^{\ast }}\;u(\tau (s),s,z)\text{ s.t. }u(\tau (s),s,z)\geq 0,
\label{WP3}
\end{equation}

\item[(ii)] there is no profitable sender deviation to an off-path action $%
s^{\prime }\notin $ $\sigma (Z)$. (See Definition \ref%
{def_profitable_sender_deviation} below for the definition of a profitable
sender deviation).
\end{enumerate}

Note that $\sigma (z)=\eta $ becomes the optimal action for a sender of type 
$z$ if there is no solution for (\ref{WP3}) and there is no profitable
sender deviation to an off-path action $s^{\prime }\notin $ $\sigma (Z)$.

We formulate the profitable sender deviation in Definition \ref%
{def_profitable_sender_deviation}. Let $X^{\ast }\subset X$ be the set of
receivers who enter the market and $B(X^{\ast })$ is the Borel sigma-algebra
on $X^{\ast }$. For all $s\in S^{\ast },$ let $m(s)\in B(X^{\ast })$ be the
set of receiver types who are matched with a sender with $s$. Therefore, $%
m:S^{\ast }\rightarrow B(X^{\ast })$ is a set-valued matching function. For
all $x\in X^{\ast }$, $m^{-1}(x)\in S^{\ast }$ denotes the action chosen by
a sender with whom a receiver of type $x$ is matched, i.e., $x\in m\left(
m^{-1}(x)\right) $.

\begin{definition}
\label{def_profitable_sender_deviation}Given $\{\sigma ,\mu ,\tau ,m\}$,
there is a profitable sender deviation to an off-path action if there exists 
$z$ for which there are an action $s^{\prime }\notin $ $\sigma (Z)$ and a
reaction $t^{\prime }\in T$ such that, for some $x^{\prime }\in X^{\ast }$, 
\begin{align}
& \text{(a) }\mathbb{E}_{\mu (s^{\prime })}\left[ g(t^{\prime },s^{\prime
},z^{\prime },x^{\prime })\right] >\mathbb{E}_{\mu (m^{-1}(x^{\prime }))}%
\left[ g\left( \tau \left( m^{-1}(x^{\prime })\right) ,m^{-1}(x^{\prime
}),z^{\prime },x^{\prime }\right) \right] \text{ and }
\label{receiver_matching_utility} \\
& \text{(b) }%
\begin{array}{l}
u(t^{\prime },s^{\prime },z)>u(\tau (\sigma (z)),\sigma (z),z)\text{ if }%
\sigma (z)\in S^{\ast }, \\ 
u(t^{\prime },s^{\prime },z)>0,\text{ otherwise.}%
\end{array}
\label{sender_matching_partner}
\end{align}
\end{definition}

Note that $z^{\prime }$ on each side of (\ref{receiver_matching_utility}) is
the random variable governed by $\mu (s^{\prime })$ and $\mu
(m^{-1}(x^{\prime }))$, respectively. If (\ref{receiver_matching_utility})
and (\ref{sender_matching_partner}) are not satisfied, there is no
profitable sender deviation to an off-path action. The reason is that those
are the conditions for sender $z$ to get a higher utility by forming a match
with a receiver on the market. Let $B(S^{\ast })$\ be the Borel
sigma-algebra on $S^{\ast }$.

\begin{definition}
\label{def_stable_matching}Given $(\sigma ,\mu ),$ $\{\tau ,m\}$ is an \emph{%
equilibrium matching outcome} if

\begin{enumerate}
\item[(i)] $m$ is stable, i.e., there is no pair of a sender with action $s$
and a receiver of type $x\notin m\left( s\right) $ such that, for some $%
t^{\prime }\in T$, some $z$ with $\sigma (z)=s\in S^{\ast }$, 
\begin{align}
\mbox{(a) }& \mathbb{E}_{\mu (s)}\left[ g(t^{\prime },s,z^{\prime },x)\right]
>\mathbb{E}_{\mu (m^{-1}(x))}\left[ g\left( \tau \left( m^{-1}(x)\right)
,m^{-1}(x),z^{\prime },x\right) \right] ,  \label{stable_matching1} \\
\mbox{(b) }& u(t^{\prime },s,z)>u(\tau (s),s,z).  \label{stable_matching2}
\end{align}

\item[(ii)] $\tau $\emph{\ }clears the markets, i.e., for all $A\in
B(S^{\ast })$ such that $H\left( \left\{ x|x\in m(\xi (x))\text{, }\xi
(x)\in A\right\} \right) =G\left( \left\{ z|\sigma \left( z\right) \in
A\right\} \right) $,
\end{enumerate}
\end{definition}

Note that $z^{\prime }$ on each side of (\ref{stable_matching1}) is the
random variable governed by $\mu (s)$ and $\mu (m^{-1}(x))$, respectively.
Condition (i) implies that the induced matching function $m$\ characterizes
equilibrium matches such that no two agents can be strictly better off by
forming a new match. A receiver of type $x\in X^{\ast }$ is matched with a
sender whose action is $\xi (x)\in S^{\ast }$ and takes a reaction $\tau
(\xi (x))$ when $\{\tau ,m\}$ is an equilibrium matching outcome given $%
(\sigma ,\mu )$. Condition (ii) implies that the market-clearing reaction
function $\tau $ induces a measure preserving matching function $m$.

It is worthwhile to mention that equilibrium matching in a large market with
complete information was formulated in Cole, Mailath and Postlewaite (2001)
with two separate reward schedules for agents on two sides of the market
when a match generates monetary surplus and each agent has a separable
utility function.\footnote{%
See Definition 2 in Cole, Mailath and Postlewaite (2001) for their
definition of stability.} We generalize it in a large market with incomplete
information and arbitrary utility functions. The difference is that we embed 
\textbf{the }equilibrium matching property in the market matching function $m
$ (Definition \ref{def_stable_matching}.(i)), given the market clearing
market reaction function $\tau $ that endogenously emerged as a result of
optimal decisions $\xi $ and $\sigma $ made by receivers and senders
(Definition \ref{def_stable_matching}.(ii)).

Now we define a competitive signaling equilibrium with incomplete
information. The consistency of $\mu $ follows the definition in Ramey
(1996).

\begin{definition}
\label{definition1}$\{\sigma ,\mu ,\tau ,m\}$ constitutes a \emph{%
competitive signaling equilibrium} (CSE) with incomplete information if

\begin{enumerate}
\item for all $z\in Z$, $\sigma (z)$ is optimal

\item $\mu $ is \emph{consistent}:

\begin{enumerate}
\item if $s\in \sigma (Z)$ satisfies $G(\{z|\sigma (z)=s\})>0,$ then $\mu
(s) $ is determined from $G$ and $\sigma ,$ using Bayes' rule.

\item if $s\in \sigma (Z)$ but $G(\{z|\sigma (z)=s\})=0$, then $\mu (s)$ is
any probability distribution with supp $\mu (s)=$ cl $\left\{ z|\sigma
(z)=s\right\} $

\item if $s\notin \sigma (Z)$, then $\mu (s)$ is unrestricted.
\end{enumerate}

\item given $(\sigma ,\mu ),$ $\{\tau ,m\}$ is \emph{an equilibrium matching
outcome.}
\end{enumerate}
\end{definition}

Note that since a single agent has no significant market power in a large
matching market, agents take the market reaction function $\tau $ as given
when they make their decision and that $\tau $ is fully reinforced by their
optimal decisions through the market clearing condition. Therefore, $\tau $
is endogenous similar to how prices are determined endogenously in the
general equilibrium framework. Our notion indeed follows the tradition of
how to formulate a competitive equilibrium in a large matching market in the
literature (e.g., Mailath, Postlewaite, Samuelson (2013, 2017), Peters
(2001), Cole, Mailath, Postlewaite (2001) among many). Our contribution is
to formulate the new notion of competitive \textquotedblleft
signaling\textquotedblright\ equilibrium, combining signaling and matching
in a large market with two-sided heterogeneity.

All proofs can be found in the appendix.

\section{Stronger monotone CSE\label{section_montone_equilibrium}}

Given the indeterminacy of the off-equilibrium-path beliefs, an equilibrium
refinement called \emph{Criterion D1} was developed by Cho and Kreps (1987)
and Banks and Sobel (1987). It restricts the off-equilibrium-path beliefs.
Following Cho and Kreps (1987), we define Criterion D1 as follows. Given an
equilibrium $\{\sigma ,\mu ,\tau ,m\}$, we define type $z$'s equilibrium
utility $U(z)$ as $U(z):=u(\tau \left( \sigma \left( z\right) \right)
,\sigma \left( z\right) ,z)$ for all $z\in Z$.

\begin{definition}[Criterion D1]
\label{definition_Criterion_D1}Fix any $s\notin $ $\sigma (Z)$ and any $t\in
T$. Suppose that there is a non-empty set $Z^{\prime }\subset Z$ such that
the following is true: for each $z\notin Z^{\prime }$, there exists $%
z^{\prime }$ such that 
\begin{equation}
u(t,s,z)\geq U(z)\Longrightarrow u(t,s,z^{\prime })>U(z^{\prime })\text{.}
\label{Criterion_D1}
\end{equation}%
Then, the equilibrium is said to violate Criterion D1 unless it is the case
that supp $\mu (s)\subset Z^{\prime }$.
\end{definition}

Intuitively, following the observation for an off-equilibrium-path action $%
s, $ zero posterior weight is placed on a type $z$ whenever there is another
type $z^{\prime }$ that has a stronger incentive to deviate from the
equilibrium in the sense that type $z^{\prime }$ would strictly prefer to
deviate for any given $t$ that would give type $z$ a weak incentive to
deviate.

We can equivalently define Criterion D1 by the contrapositive of (\ref%
{Criterion_D1}), that is 
\begin{equation}
u(t,s,z^{\prime })\leq U(z^{\prime })\Longrightarrow u(t,s,z)<U(z).
\label{Criterion_D12}
\end{equation}%
Upon observing an off-equilibrium action $s$, zero posterior weight is
placed on a type $z$ whenever a type $z$ is strictly worse off by deviating
for any $t$ that would make type $z^{\prime }$ weakly worse with the same
deviation.

For monotone equilibrium analysis, we first compare two set orders: the
strong set order (Veinnott (1989)) and the stronger set order (Shannon
(1995)).

\begin{definition}[Set orders]
Consider two sets $A$ and $B$ in the power set $P(Y)$ for $Y$ a lattice with
a given relation $\geq $.

\begin{enumerate}
\item We say that $A\leq _{s}B$, read \textquotedblleft $A$ is smaller than $%
B$\textquotedblright\ in the strong set order\ if $a\in A$ and $b\in B$ $%
\Longrightarrow $ $a\wedge b\in A$ and $a\vee b\in B.$

\item We say that $A\leq _{c}B$, read \textquotedblleft $A$ is completely
lower than $B$\textquotedblright\ in the stronger set order if for every $%
a\in A$ and $b\in B,$ $a\leq b.$
\end{enumerate}
\end{definition}

If $A\leq _{c}B$, then $A\leq _{s}B.$ However, the converse may not be true.
Therefore, the stronger set order is stronger than the strong set order. For
example, consider $Y=%
\mathbb{R}
.$ $A=[0,2]$ and $B=[1,3].$ Then, $A\leq _{s}B$ but $A\nleq _{c}B$ because
when $A\leq _{c}B$, the intersection of $A$ and $B$ is either the empty set
or a singleton.

Most studies on comparative statics or monotone equilibrium analysis employ
the strong set order. We employ the stronger set order.

\begin{definition}[Stronger set order]
Given a partially ordered set $K$ with the given relation $\geq $, a
set-valued function $M:K\rightarrow P(Y)$ is monotone non-decreasing in the
stronger set order if $k^{\prime }\leq k$ implies that $M(k^{\prime })\leq
_{c}M(k)$.
\end{definition}

The monotonicity of $\sigma $ and $\tau $ is defined in terms of the
stronger set order. Note that $\sigma $ and $\tau $ are single-valued
functions. We see them as a special set-valued function whose co-domain is a
set of singletons. 
In this case, the stronger set order is identical to the strong set order.
The difference is evident in a belief function. Consider a belief function $%
\mu :S\rightarrow \Delta (Z)$. The monotonicity of a belief function is
defined by the stronger set order on the supports of the probability
distributions. A belief function is non-decreasing in the stronger set order
if $s^{\prime }\leq s$ implies supp $\mu (s^{\prime })\leq _{c}$ supp $\mu
(s)$.

We also use the stronger set order for the monotonicity of a matching
function $m:S^{\ast }\rightarrow B(X^{\ast }).$ A matching function is
non-decreasing in the stronger set order if $s^{\prime }\leq s$ implies $%
m(s^{\prime })\leq _{c}m(s)$. Note that pooling or semi-pooling is allowed
because the co-domain of a matching function $m$ is $B(X^{\ast })$, the
Borel sigma-algebra on $X^{\ast }$ (See Section \ref{Sec_Eq_w_lower_bound}).
In a pooling CSE, $S^{\ast }=\{s^{\ast }\}$ is a singleton and $m(s^{\ast })$
is the set of receivers who are matched with senders with $s^{\ast }$. In a
semi-pooling CSE, only a positive measure of senders of type $z_{h}$ and
above can choose the same action $s_{h}.$ Then, $m(s_{h})$ is the set of
receivers who are matched with senders with $s_{h}$, and it has the same
positive measure.

Now we define the stronger monotone CSE as follows.

\begin{definition}[Stronger Monotone Equilibrium]
\label{definition_monotone_eq}A CSE $\left\{ \sigma ,\mu ,\tau ,m\right\} $
is stronger monotone if $\sigma ,$ $\mu ,\tau ,$ and $m$ are non-decreasing
in the stronger set order.
\end{definition}

We impose the following assumptions for $u$.

\begin{description}
\item[Assumption A] $u(t,s,z)$ is (i) decreasing in $s$, increasing in $t$
and $z$, and satisfies (ii) the strict single crossing property in $%
((t,s);z). $\footnote{%
Let $A$ be a lattice, $\Theta $ be a partially ordered set and $f:A\times
\Theta \rightarrow 
\mathbb{R}
.$ Then, $f$ satisfies the \emph{single crossing property} in $(a;\theta )$
if for $a^{\prime }>a^{\prime \prime }$ and $\theta ^{\prime }>\theta
^{\prime \prime },$ (i) $f(a^{\prime },\theta ^{\prime \prime })\geq
f(a^{\prime \prime },\theta ^{\prime \prime })$ implies $f(a^{\prime
},\theta ^{\prime })\geq f(a^{\prime \prime },\theta ^{\prime })$ and (ii) $%
f(a^{\prime },\theta ^{\prime \prime })>f(a^{\prime \prime },\theta ^{\prime
\prime })$ implies $f(a^{\prime },\theta ^{\prime })>f(a^{\prime \prime
},\theta ^{\prime })$. If $f(a^{\prime },\theta ^{\prime \prime })\geq
f(a^{\prime \prime },\theta ^{\prime \prime })$ implies $f(a^{\prime
},\theta ^{\prime })>f(a^{\prime \prime },\theta ^{\prime })$ for every $%
\theta ^{\prime }>\theta ^{\prime \prime },$ then $f$ satisfies the \emph{%
strict single crossing property} in $(a;\theta )$.}
\end{description}

Given Assumption A, the stronger monotonicity of $\sigma $ and $\tau $ comes
from Lemma \ref{lemma_monotone_eq_A'} below.

\begin{lemma}
\label{lemma_monotone_eq_A'}Consider an equilibrium $\left\{ \sigma ,\mu
,\tau ,m\right\} $. If Assumptions A is satisfied, the equilibrium satisfies
the following properties: (i) $\sigma $ is non-decreasing in $z,$ (ii) $\mu $
is non-decreasing in the subset of domain, $\sigma (Z),$ with respect to the
stronger set order: for $s,s^{\prime }\in $ $\sigma (Z),$ $s\geq s^{\prime }$
implies supp $\mu (s^{\prime })\leq _{c}$ supp $\mu (s)$, (iii) $\tau $ is
increasing.
\end{lemma}

Lemma \ref{lemma_monotone_eq_A'}.(iii) shows that $\mu $ is stronger
monotone only in the subset of domain, $\sigma (Z)$ with the monotone
supermodular property of the sender's utility. Proposition \ref%
{theorem_monotone_belief} is crucial in establishing the equivalence between
Criterion D1 and the stronger monotonicity of $\mu $ in the entire $S$
(Corollary \ref{corollary_monotone_belief}).

\begin{proposition}
\label{theorem_monotone_belief}Let $\sigma $ and $\mu $ be a sender action
function and a belief function in equilibrium respectively. If Assumption A
is satisfied, the belief $\mu (s)$ conditional on $s\notin $ $\sigma (Z)$
that passes Criterion D1 is unique and it is characterized as follows:

\begin{enumerate}
\item If $s$ belongs to the interval of off-path sender actions induced by
the discontinuity of $\sigma $ at $z$, then supp $\mu (s)=\{z\}.$

\item Let $\overline{z}$ be the least upper bound of $Z$ if it exists. If $%
s>\sigma (\overline{z}),$ then supp $\mu (s)=\{\overline{z}\}$.

\item Let $\underline{z}$ be the greatest lower bound of $Z$ if it exists.
If $s<\sigma (\underline{z}),$ then supp $\mu (s)=\{\underline{z}\}$.
\end{enumerate}
\end{proposition}

Proposition \ref{theorem_monotone_belief} leads to Corollaries \ref%
{corollary_monotone_belief} and \ref{corollary_monotone_belief1} below.

\begin{corollary}
\label{corollary_monotone_belief}Let $\sigma $ and $\mu $ be a sender action
function and a belief function in equilibrium, respectively. If Assumption A
is satisfied, $\mu $ passes Criterion D1 if and only if it is non-decreasing
in the stronger set order.
\end{corollary}

Corollary \ref{corollary_monotone_belief} above shows the equivalence
between Criterion D1 and the stronger monotonicity of $\mu $ in the entire $%
S $. The stronger monotonicity of $\mu $ implies that for any $s$ in the
interval of off-path sender actions induced by the discontinuity of $\sigma $
at an interior sender type $z$, the support of $\mu (s)$ is a singleton and
it is $\{z\}.$ This implication is satisfied if and only if $\mu $ satisfies
Criterion D1 given the monotone-supermodular condition for the sender's
utility. Cho and Sobel monotonicity of $\mu $ does not lead to this
implication although any belief function $\mu $ that passes Criterion D1
satisfies Cho and Sobel monotonicity.\footnote{%
Suppose that an action $s$ is chosen by some sender type $z$ on the
equilibrium path. Cho and Sobel monotonicity means that a receiver should
believe that $s^{\prime }>s$ is not chosen by a lower sender type than $z$.}

\begin{corollary}
\label{corollary_monotone_belief1} According to Lemma \ref%
{theorem_monotone_belief} in the appendix, the support of the belief $\mu (s)$
conditional on $s\notin $ $\sigma (Z)$ is a singleton if it passes Criterion
D1. This implies that if the unique type in the support of the belief $\mu
(s)$ is weakly worse off by deviating to $s\notin $ $\sigma (Z),$ any other
type is strictly worse off with the same deviation.
\end{corollary}

The proof of Corollary \ref{corollary_monotone_belief1} is straightforward,
so it is omitted. Cho and Sobel monotonicity of beliefs (Cho and Sobel
(1990)) is a partial implication of Criterion D1 but it is instrumental for
the selection of a separating equilibrium as a unique D1 equilibrium: Among
those who chose the same action, the highest sender type always has a
profitable upward deviation given Cho and Sobel monotonicity, so a pooled
action cannot be sustained in a D1 equilibrium. However, this argument does
not apply if a receiver cannot reward such an upward deviation with a higher
reaction when the upper bound of feasible reactions $T$ is too low. In this
case, we can derive a unique D1 equilibrium using the stronger monotonicity
of beliefs, which is the full implication of Criterion D1. Lemma \ref%
{corollary_monotone_belief} and Corollary \ref{corollary_monotone_belief1}
are crucial in deriving a stronger monotone equilibrium because of the
possibility of bunching on the top.

To establish the stronger monotonicity of an equilibrium, $\left\{ \sigma
,\mu ,\tau ,m\right\} $, we still need to identify sufficient conditions
under which $m$ is non-decreasing in the stronger set order. We impose
Assumption B for $g$ and apply the Milgrom-Shannon Monotone Selection
Theorem (Milgrom and Shannon (1994)).

\begin{description}
\item[Assumption B] (i) $g(t,s,z,x)$ is supermodular\footnote{%
Given a lattice $A,$ $f:A\rightarrow 
\mathbb{R}
$ is \emph{supermodular} if $f(a\wedge b)+f(a\vee b)\geq $ $f(a)+f(b)$ for
all $a$ and $b$ in $A$. $f:A\rightarrow 
\mathbb{R}
$ is \emph{strictly} \emph{supermodular} if $f(a\wedge b)+f(a\vee b)>$ $%
f(a)+f(b)$ for all unordered $a$ and $b$ in $A$.} in $(t,s,z)$ and satisfies
the single crossing property in $(\left( t,s,z\right) ;x)$ and the strict
single crossing property in $(z;x)$ at each $(s,t)$. (ii) $g(t,s,z,x)$ is
increasing in $x.$
\end{description}

\begin{theorem}[Milgrom-Shannon Monotone Selection Theorem]
\label{thm: Milgrom-Shannon}
Let $f:A\times \Theta \rightarrow 
\mathbb{R}
,$ where $A$ is a lattice and $\Theta $ is a partially ordered set. If $f$
is quasisupermodular\footnote{%
Given a lattice $A,$ a function $f:A\rightarrow 
\mathbb{R}
$ is \emph{quasisupermodular} if (i) $f(a)\geq f(a\wedge b)$ implies $%
f(a\vee b)\geq f(b)$ and (ii) $f(a)>f(a\wedge b)$ implies $f(a\vee b)>f(b).$
If $f$ is supermodular, then it is quasisupermodular.} in $a$ and satisfies
the strict single crossing property in $(a;\theta ),$ then every selection $%
a^{\ast }(\theta )$ from $\arg \max_{a\in A}f(a,\theta )$ is non-decreasing.
\end{theorem}

\begin{theorem}[Stronger Monotone CSE Theorem]
\label{theorem_stronger_monotone_eq} 
Suppose that Assumptions A and B are satisfied. Then, an equilibrium $%
\left\{ \sigma ,\mu ,\tau ,m\right\} $ is stronger monotone if and only if
it passes Criterion D1.
\end{theorem}

\begin{proof}
Given Assumption A, the stronger monotonicity of $\sigma ,\mu ,$ and $\tau $
comes from Lemma \ref{lemma_monotone_eq_A'} and Corollary \ref%
{corollary_monotone_belief}. Given the stronger monotonicity of $\sigma ,\mu
,$ and $\tau $, consider a receiver's matching problem that is $\max_{s\in
S^{\ast }}V(s,x),$ where $V(s,x):=\mathbb{E}_{\mu (s)}\left[ g(\tau
(s),s,z,x)\right] .$ For any $s,s^{\prime }\in S^{\ast }$ such that $%
s>s^{\prime },$ we have that $\tau (s)>\tau (s^{\prime })$ and $z\geq
z^{\prime }$ for any $z\in $ supp $\mu (s)$ and $z^{\prime }\in \mu
(s^{\prime }).$ Therefore, the first three arguments in $g$ are linearly
ordered with respect to $s.$ Given Assumption B(i), this implies that $%
V(s,x) $ satisfies the strict single crossing property. Choose an arbitrary
selection $\xi _{\circ }(x)\in \arg \max_{s\in S^{\ast }}V(s,x).$ Then, by
Milgrom and Shannon's Monotone Selection Theorem, $\xi _{\circ }(x)$ is
non-decreasing in $x$. Note that $\max_{s\in S^{\ast }}V(s,x)$ is a
maximization problem with no individual rationality. For all $x\in X,$ let 
\begin{equation*}
\xi (x)=\left\{ 
\begin{array}{cc}
\xi _{\circ }(x) & \text{if }V(\xi _{\circ }(x),x)\geq 0, \\ 
\eta & \text{otherwise.}%
\end{array}%
\right.
\end{equation*}%
$V(s,x)$ is increasing in $x$ because of Assumption B(ii) and hence we have
that $x<x^{\prime }$ for any $x$ with $\xi (x)=\eta $ and any $x^{\prime }$
with $\xi (x^{\prime })\neq \eta $. This property and the non-decreasing
property of $\xi _{\circ }(x)$ make $\xi (x)$ non-decreasing in $x.$

For any $s\in S^{\ast },$ the set of receiver types who are matched with
senders with $s$ can be expressed as $m(s)=\xi ^{-1}(s):=\{x|\xi (x)=s\}$.
Because $\xi (x)$ is non-decreasing in $x,$ $m$ is non-decreasing with
respect to the stronger set order.
\end{proof}

\bigskip

Without loss of generality, we can focus on stronger monotone equilibria to
derive all D1 equilibria given Assumptions A and B.

A receiver's optimal choice of the action chosen by the sender whom he is
matched with $\xi (x)$ is non-decreasing in $x$ as proved above. It implies
that $m(s)$ for each $s\in S^{\ast }$ is an interval if $m(s)$ is not a
singleton. Therefore, in a stronger monotone CSE, a market matching function
admits separting, pooling, or semi-pooling.

In general, monotone comparative statics (e.g., Milgrom and Shannon (1994),
Milgrom and Roberts (1990), Quah and Strulovici (2009), and Shannon (1995),
etc.) are designed for single-person decision problems, and there is no
assumption of the existence of equilibrium. Of course, such decision
problems can arise in a competitive market or a game environment. Our
approach in Theorem \ref{theorem_stronger_monotone_eq} follows this standard
practice in the literature.

Section \ref{Sec_Eq_w_lower_bound} establishes the existence and uniqueness
of a stronger monotone CSE with quasilinear utility functions given any
interval of feasible reactions. A key for the existence of a stronger
monotone CSE is the existence of (i) the solution for threshold sender types
and their action choices and (ii) the solution for the differential equation
for the equilibrium belief function for the separating part of action
choices by senders. The quasilinearity of utility function allows us to
isolate the differential equation for the equilibrium belief function. If we
do not impose the quasilinearity on utility functions, the equilibrium
belief function and the market reaction function must be derived jointly by
simultaneously solving a coupled non-linear system of two differential
equations.

We can analogously establish the stronger monotone signaling equilibrium
theorem for a \emph{pure-strategy} perfect Bayesian equilibrium (PBE) in the
standard one-sender-one-receiver game.\footnote{%
The proof is analgous to the proof of Theorem \ref%
{theorem_stronger_monotone_eq}. It is available upon request.} In such a
standard game, there is no matching function we introduce for a CSE, a
sender's private type $z$ follows a probability distribution $G,$ and the
receiver has no private type with $g(t,s,z)$ denoting his utility. Denoting
the receiver's reaction-choice function by $\tau $, $\{\sigma ,\tau ,\mu \}$
is a pure-strategy PBE if (i) for all $z\in Z$, $\sigma (z)\in $ $\arg
\max_{s\in S}u(\tau (s),s,z),$ (ii) for all $s\in S$, $\tau (s)\in $ $\arg
\max_{t\in T}\mathbb{E}_{\mu (s)}\left[ g(t,s,z)\right] ,$ (iii) $\mu $ is 
\emph{consistent }(Definition \ref{definition1}.(ii)). Assumption A (the
monotone-supermodular condition for the sender's utility) alone ensures the
stronger monotonicity of a pure-strategy PBE. As we consider a CSE in the
whole market with two-sided heterogeneity, Assumption B (the weakly
monotone-supermodular condition for the receiver's utility) kicks in to
ensure the stronger monotonicity of a CSE.

In fact, Assumption A is the same as the one imposed for a game with one
sender and one receiver in Liu and Pei (2020). Focusing on mixed-strategy
equilibria (i.e., both players use mixed strategies), their interest is to
show only the monotonicity of the sender's equilibrium mixed strategy in the
strong set order in a game with finite $S,$ $Z,$ and $T$. They show that
when the receiver has two feasible reactions (i.e., $\left\vert T\right\vert
=2)$, the monotone-supermodular condition for the sender's utility alone is
sufficient for the monotonicity of the sender's equilibrium mixed strategy. 
It works because every pair of distributions over the receiver's reaction
can be ranked according to the first-order stochastic dominance (FOSD).
Since choosing a higher action is more costly for the sender, she only has
an incentive to do so when it induces a more favorable response from the
receiver. Therefore, the ranking over the sender's equilibrium actions must
coincide with the ranking over the receiver's (mixed) actions that they
induce. Because a high type sender has a stronger preference towards higher
action profiles, she will never choose a strictly lower action than a low
type chooses.

However, when the cardinality of the receiver's reactions is greater than
two, distributions over the receiver's reactions are not totally ordered
according to FOSD. Therefore, the monotone-supermodular condition for the
sender's utility alone is not sufficient for the monotonicity of the
sender's equilibrium mixed action strategy: In addition, we need either the
quasi-preserving property for the receiver's utility (Definition 5 in Liu
and Pei (2020)) or the property of increasing absolute difference over
distributions for the sender's utility (Definition 6 in Liu and Pei (2020)).
If the receiver's reaction choice is deterministic, such a problem does not
arise. Therefore, the cardinality of the receiver's reaction set is not an
issue in establishing the stronger monotone CSE theorem. It also implies
that the monotone-supermodular condition for the sender's utility alone is
sufficient to establish the stronger monotone \emph{pure-strategy} PBE in a
game with one sender and one receiver.

\section{Unique Stronger Monotone CSE\label{Sec_Eq_w_lower_bound}}

In this section, we study the existence of a unique stronger monotone CSE in
a model with quasilinear utility functions. A receiver's utility is $%
g(t,s,z,x)=v(x,s,z)-t$ and a sender's utility is $u(t,s,z)=t-c(s,z)$.
Therefore, utilities are one-to-one transferrable in a match between a
sender and a receiver through a receiver's reaction $t$. $v$ can be
interpreted as gross match surplus and $c$ is the cost of taking an action
for senders. One may think of $t$ as surplus transfer from the receiver to
the sender in a match.

In this section, we also assume (i) $Z=[\underline{z},\overline{z}]\subset 
\mathbb{R}
$ with $\underline{z}<\bar{z}$, (ii) $X=[\underline{x},\overline{x}]\subset 
\mathbb{R}
$ with $\underline{x}<\overline{x}$, (iii) $S=%
\mathbb{R}
_{+},$ (iv) $T=[t_{\ell },t_{h}]\subset 
\mathbb{R}
_{+}\cup \{\infty \}$. Let $0\in S$ be the null action. The set of feasible
reactions $T$ can be a singleton if $t_{\ell }=t_{h}<\infty $ or a compact
interval if $t_{\ell }<t_{h}<\infty $ or unbounded if $t_{\ell
}<t_{h}=\infty $.

\begin{description}
\item[Assumption 1.] \label{Ass1}(i) $c(s,z)$ is increasing in $s$ but
decreasing in $z$ and (ii) $-c(s,z)$ is strictly supermodular in $(s,z)$.
\end{description}

It is easy to see that Assumption 1 implies that the sender's utility is
monotone-supermodular given the form of the utility function, $%
u(t,s,z)=t-c(s,z)$.\footnote{%
If the domain $A$ of a real-valued function $f$ is a subset of $%
\mathbb{R}
^{N},$ then the (strict) supermodularity of $f$ is equivalent to
non-decreasing (increasing) differences (Theorem 2.6.1 and Corollary 2.6.1
in Topkis (1998)), which in turn guarantees the (strict) single crossing
property.}

\begin{description}
\item[Assumption 2.] \label{Ass2}(i)\textbf{\ }$v(x,s,z)$ is supermodular in 
$(x,s,z)$ and strictly supermodular in $(z,x),$ (ii) $v$ is increasing in $%
x. $
\end{description}

\begin{lemma}
\label{lemma_ass_2_b}If Assumption 2 holds, then the receiver's utility is
weakly monotone-supermodular.
\end{lemma}

Because the monotone-supermodular conditions for sender's and receivers'
utilities are implied by Assumptions 1 and 2, Theorem \ref%
{theorem_stronger_monotone_eq} goes through in this section. 

We impose Assumptions 3, 4, 5, and 6 below for the differentiability of the
separating part of a stronger monotone CSE and the existence of a stronger
monotone CSE.

\begin{description}
\item[Assumption 3] \label{ass3}(i) $v$ is non-negative, increasing in $z$,
and non-decreasing in action $s$. (ii) $v$ is differentiable and $v_{s}$ and 
$v_{z}$ are continuous.

\item[Assumption 4] \label{ass4}$c$ is differentiable with $c(0,z)=0$, $%
\lim_{s\rightarrow \infty }c(s,z)=\infty $ for all $z\in \left[ \underline{z}%
,\overline{z}\right] $, and $c_{s}$ is continuous.

\item[Assumption 5] \label{ass5}$v(x,s,z)\geq 0$ for all $(x,s,z).$ If $%
v(x,s,z)$ is increasing in $s$, it is concave in $s$ with $%
\lim_{s\rightarrow 0}v_{s}(x,s,z)=\infty $ and $\lim_{s\rightarrow \infty
}v_{s}(x,s,z)=0$ and $c(s,z)$ is strictly convex in $s$ with $%
\lim_{s\rightarrow 0}c_{s}(s,z)=0$ and $\lim_{s\rightarrow \infty }$ $%
c_{s}(s,z)=\infty $.

\item[Assumption 6] \label{ass6}$0<G^{\prime }(z)<\infty $ for all $z\in %
\left[ \underline{z},\overline{z}\right] $ and $0<H^{\prime }(x)<\infty $
for all $x\in \lbrack \underline{x},\overline{x}]$.
\end{description}

We define the function $n$\ as $n\equiv H^{-1}\circ G$\ so that $%
H(n(z))=G(z) $ for all $z\in \lbrack \underline{z},\overline{z}]$. A
bilaterally efficient action $\zeta (x,z)$ for type $z$ given $x$ maximizes $%
v(x,s,z)-c(s,z)$. 
\begin{equation}
v(\underline{x},\zeta (\underline{x},\underline{z}),z)-c(\zeta (\underline{x}%
,\underline{z}),\underline{z})\geq 0.  \label{constrained_eff_b}
\end{equation}%
We normalize $\zeta (\underline{x},\underline{z})$ and $v(\underline{x}%
,\zeta (\underline{x},\underline{z}),\underline{z})-c(\zeta (\underline{x},%
\underline{z}),\underline{z})$ to $0$, respectively. Given $c(\zeta (%
\underline{x},\underline{z}),\underline{z})=0$ due to Assumption \ref{ass4}, 
$v(\underline{x},\zeta (\underline{x},\underline{z}),\underline{z})-c(\zeta (%
\underline{x},\underline{z}),\underline{z})=0$ implies that $v(\underline{x}%
,\zeta (\underline{x},\underline{z}),\underline{z})=0$ as well. The
reservation utility for each agent is zero. We assume that every agent
enters the market if she can get at least her reservation utility by
entering the market in equilibrium.

We introduce a \emph{well-behaved stronger monotone CSE}, a type of stronger
monotone CSE, that encompasses a separating CSE and a pooling CSE as well. A
stronger monotone CSE is called well-behaved if it is characterized by two
threshold sender types, $z_{\ell }$ and $z_{h}$ such that every sender of
type below $z_{\ell }$ stays out of the market, every sender in $[z_{\ell
},z_{h})$ differentiates themselves with their unique action choice, and
every sender in $[z_{h},\overline{z}]$ pools themselves with the same
action. If $z_{\ell }<z_{h}=\bar{z},$ then a well-behaved CSE is separating.
If $z_{\ell }=z_{h},$ then a well-behaved CSE is pooling. If $z_{\ell
}<z_{h}<\bar{z},$ then it is strictly well-behaved with both separating and
pooling parts in the equilibrium. We shall show that any stronger monotone
CSE (i.e., any D1 CSE) is unique as well as well-behaved. 

\subsection{Separating CSE}

We first start with a stronger monotone separating CSE. Once we characterize
it, the characterization of any stronger monotone well-behaved CSE comes
naturally. For now, let us assume that the lower bound $t_{\ell }$ of the
interval $T$ is less than the maximal value of $v-c$ that can be created by
the highest types $\overline{z}$ and $\overline{x}$. Otherwise all agents
would stay out of the market. 
Let $z_{\ell }$ be the lowest sender type who is matched in equilibrium and $%
s_{\ell }$ her action. The following two inequalities must be satisfied at $%
(s_{\ell },z_{\ell })$: 
\begin{align}
v\left( n\left( z\right) ,s,z\right) -t_{\ell }& \geq 0,  \label{lem1} \\
t_{\ell }-c\left( s,z\right) & \geq 0.  \label{lem2}
\end{align}%
The two cases must be distinguished. If $z_{\ell }=\underline{z},$ then all
types are matched in equilibrium. This is the \emph{first case} and it
happens when $t_{\ell }=0$. In this case, if we have a separating part in
equilibrium, there is no information rent in the lowest match between type $%
\underline{z}$ and type $\underline{x}$. Therefore, the equilibrium action $%
s_{\ell }$ in the lowest match is bilaterally efficient (i.e., $s_{\ell
}=\zeta (\underline{x},\underline{z})=0$) and (\ref{lem1}) and (\ref{lem2})
hold with equality.

If $t_{\ell }>0$, then type $\underline{x}$ cannot achieve a non-negative
value of $v-t_{\ell }$ in a match with type $\underline{z}$ who takes an
action that costs her $t_{\ell }$, then the lowest match must be between
types $z_{\ell }$ and $x_{\ell }:=n(z_{\ell })$ in the interior of both type
distributions. (\ref{lem1}) and (\ref{lem2}) must be also satisfied with
equality at $(s_{\ell },z_{\ell })$. This is the \emph{second case}. If
either one of them, e.g., (\ref{lem1}), is positive, then a receiver whose
type is below but arbitrarily close to $x_{\ell }$ finds it profitable to be
matched with type $z_{\ell }$ instead of staying out of the market. Define $%
\overline{s}>0$ and $\overline{t}>0$ such that $v(\overline{x},s,\overline{z}%
)-t=0$ and $t-c(s,\overline{z})=0.$ We only focus on the lower bound of
feasible reactions $t_{\ell }$ less than $\overline{t}$ because otherwise no
positive measure of senders enters the market.

\begin{lemma}
\label{lemmaA}If $t_{\ell }=0,$ $(s_{\ell },z_{\ell })=(\zeta (\underline{x},%
\underline{z}),\underline{z})$ is a unique solution that satisfies (\ref%
{lem1}) and (\ref{lem2}) with equality. Given $0<t_{\ell }<\overline{t}$,
there exists a unique solution $\left( s_{\ell },z_{\ell }\right) \in 
\mathbb{R}
_{++}\times (\underline{z},\overline{z})$ that solves (\ref{lem1}) and (\ref%
{lem2}) with equality.
\end{lemma}

Now we establish that when $0\leq t_{\ell }<\overline{t}$ and $t_{h}=\infty $%
, a well-behaved stronger monotone CSE is differentiable in Theorem \ref%
{thm_differentiable_sep_eq}.

\begin{theorem}[Differentiability of separating CSE]
\label{thm_differentiable_sep_eq}Fix $0\leq t_{\ell }<\overline{t}$ and $%
t_{h}=\infty $. In any well-behaved stronger monotone CSE with $t_{h}=\infty 
$, (i) $S^{\ast }$ is a compact real interval, $[\sigma (\underline{z}%
),\sigma \left( \bar{z}\right) ]$, (ii) $\tau :S^{\ast }\rightarrow T$ is
increasing and continuous on $S^{\ast }$ and has continuous derivative $%
\tilde{\tau}^{\prime }$ on Int $S^{\ast },$ and (iii) $\mu :S\rightarrow
\Delta (Z)$ is increasing and continuous on $S^{\ast }$ and has continuous
derivative $\mu ^{\prime }$ on Int $S^{\ast }.$
\end{theorem}

In any stronger monotone CSE with $t_{h}=\infty $, the first-order necessary
condition for the sender's equilibrium action choice that solves her problem
in \eqref{WP3} would satisfy that for all $z\in (z_{\ell },\bar{z})$%
\begin{equation}
\tau ^{\prime }(\sigma \left( z\right) )-c_{s}(\sigma \left( z\right) ,z)=0.
\label{FOCS}
\end{equation}%
On the other hand, the equilibrium reaction choice $\tau (s)$ by the
receiver who is matched with a sender with action $s$ solves his problem in %
\eqref{FP} and its first-order necessary condition must satisfy that for all 
$s\in $ Int$S^{\ast }$ 
\begin{equation}
\tau ^{\prime }(s)=v_{s}(m(s),s,\mu \left( s\right) )+v_{z}(m(s),s,\mu
\left( s\right) )\mu ^{\prime }\left( s\right) ,  \label{FOCR}
\end{equation}%
where $m(s)=n(\mu \left( s\right) )$ is the type of the receiver who is
matched with a sender with action $s.$ Note that the equilibrium matching
function $m$ is stronger monotone due to Theorem \ref%
{theorem_stronger_monotone_eq}. Because all senders on the market
differentiate themselves with unique action choices in a stronger monotone
separating CSE, $m$ is strictly increasing over $S^{\ast }$ and the matching
is assortative in terms of the receiver's type and the sender's action (and
the receiver's type and the sender' type as well).

In our two-sided matching model with a continuum of senders and receivers,
it is crucial to incorporate the market clearing condition in the belief
function $\mu $ to prove that it has continuous derivative $\mu ^{\prime }$
on Int $S^{\ast }$ (See the proof of Lemma \ref{lem_differentiable_belief}).
The differentiability of $\tau $ comes from senders' optimal action choices
and the continuity of $\mu $ comes from receivers' optimal choice of a
sender (See the proof of Lemma \ref{lemma3_in_thm1}). Theorem \ref%
{thm_differentiable_sep_eq} is the consequence of Assumptions 1 - 5. As it
will be clear at this end of this section, Theorem \ref%
{thm_differentiable_sep_eq} implies that if a stronger monotone CSE has a
separating part, that part of the CSE is differentiable. Because $\mu $ is
the inverse of $\sigma ,$ the differentiability of $\sigma $ is immediate
from Theorem \ref{thm_differentiable_sep_eq}.(iii).

The full characterization of a stronger monotone separating CSE is
established by Theorem \ref{proposition1} in the appendix. For
the existence of a unique stronger monotone separating CSE, note that
combining (\ref{FOCS}) and (\ref{FOCR}) yields a function $\phi (s,z)$
defined below: 
\begin{equation}
\phi (s,z):=\frac{-\left[ v_{s}\left( n(z),s,z\right) -c_{s}\left(
s,z\right) \right] }{v_{z}\left( n(z),s,z\right) }.
\label{differential_characteristic}
\end{equation}%
This is the first-order ordinary differential equation, $\mu ^{\prime }=\phi
(s,\mu \left( s\right) )$ with the initial condition $(s_{\ell },z_{\ell }).$
The existence of a unique stronger monotone separating CSE comes down to the
existence of a unique solution for $\mu ^{\prime }=\phi (s,\mu \left(
s\right) )$.

\begin{theorem}[Existence of separating CSE]
\label{thm_unique_separating_eq}Fix $0\leq t_{\ell }<\overline{t}$ and $%
t_{h}=\infty $. If $v$\ and $c$\ are such that $\phi $\ defined in (\ref%
{differential_characteristic}) is uniformly Lipshitz continuous, then a
unique stronger monotone separating CSE $\{\tilde{\sigma},\tilde{\mu},\tilde{%
\tau},\tilde{m}\}$ exists.
\end{theorem}

\begin{proof}
Given the full characterization of a stronger monotone separating CSE $\{%
\tilde{\sigma},\tilde{\mu},\tilde{\tau},\tilde{m}\}$ in the appendix, 
$\tilde{\mu}$ is a solution to the first-order
differential equation $\mu ^{\prime }=\phi (s,\mu \left( s\right) ),$ where $%
\phi $\ is defined in (\ref{differential_characteristic}). Given $0\leq
t_{\ell }<\overline{t}$ and $t_{h}=\infty ,$ the initial condition $(s_{\ell
},z_{\ell })$ is unique because of Lemma \ref{lemmaA}. By applying the
Picard-Lindelof Theorem (See Teschl (2012)), we can assure that if $v$\ and $%
c$\ are such that $\phi $\ defined in (\ref{differential_characteristic}) is
uniformly Lipshitz continuous, we have a unique solution $\tilde{\mu}.$
Because of Theorem \ref{thm_differentiable_sep_eq}.(iii), it establishes the
existence of a unique stronger monotone separating CSE.
\end{proof}

\textbf{\bigskip }

Once we drive $\tilde{\mu}$, we can construct the functions $\tilde{\sigma},$
$\tilde{\tau},$ and $\tilde{m}$ implied by $\tilde{\mu}$. $\tilde{\sigma}(z)$
for all $z\in \lbrack z_{\ell },\overline{z}]$ is determined by $\tilde{%
\sigma}(z)=\tilde{\mu}^{-1}(z)$ for all $z\in \lbrack z_{\ell },\overline{z}%
],$ where $\tilde{\mu}^{-1}(z)$ is the type of a sender that satisfies $z=%
\tilde{\mu}\left( \tilde{\mu}^{-1}(z)\right) $ for all $z\in \lbrack z_{\ell
},\overline{z}].$ For $s\in \lbrack s_{\ell },\tilde{\sigma}(\overline{z})],$
we can derive the matching function $\tilde{m}$ according to $\tilde{m}%
(s)=n\left( \tilde{\mu}(s)\right) $. Because $\tilde{\mu}$ is continuous
everywhere and differentiable at all $s\in $ Int $S^{\ast },$ integrating
the right-hand-side of (\ref{FOCR}) with the initial condition with $\tilde{%
\tau}(s_{\ell })=t_{\ell }$ induces%
\begin{equation}
\tilde{\tau}(s)=\int_{s_{\ell }}^{s}\left[ v_{s}(\tilde{m}(y),y,\tilde{\mu}%
\left( y\right) )+v_{z}(\tilde{m}(y),y,\tilde{\mu}\left( y\right) )\tilde{\mu%
}^{\prime }\left( y\right) \right] dy+t_{\ell }.  \label{equilibrium_wage}
\end{equation}

It is worthwhile to mention the difference between Theorems \ref%
{thm_differentiable_sep_eq} and \ref{thm_unique_separating_eq}. Theorem \ref%
{thm_differentiable_sep_eq} establishes the differentiability of a stronger
monotone separating CSE without the Lipshitz continuity and it has its own
contribution in the literature. Focusing on a separating equilibrium,
Hopkins (2012) applied the differentiability results in Mailath (1987) to a
two-sided matching model by imposing the restriction that there is no
complementary between receiver type $x$ and sender action $s$ in the
receiver's utility (equivalently the match surplus when the receiver's
utility is quasilinear with respect to the receiver's reaction). This
restriction gets rid of a matching effect on the marginal productivity of a
sender's action. Such a restriction is not needed for establishing our
differentiability result.

\paragraph{Example 1}

The utility function for the receiver of type $x$ follows $v(x,s,z)-t=xzs-t.$
The sender's utility function is $t-c(s,z)=t-\frac{s^{2}}{z}$ for the sender
of type $z$. Sender type $z$ uniformly distributed over $[0,1],$ whereas
receiver type $x$ is uniformly distributed over $[0,2].$ Then, $n(z)=2z$ is
the type of the receiver who is matched with the sender of type $z$ in a
stronger monotone separating CSE. The differential equation for the belief
function $\mu $ becomes $\mu ^{3}+\mu ^{2}\cdot \mu ^{\prime }\cdot s-s=0.$
Solving this with the initial condition $(z_{\ell },s_{\ell })$ yields 
\begin{equation*}
\tilde{\mu}(s)=\left[ \frac{3s}{4}+\left( \frac{s_{\ell }}{s}\right) \left(
z_{\ell }^{3}-\frac{3s_{\ell }}{4}\right) \right] ^{1/3}
\end{equation*}

If the lower bound of reactions is $t_{\ell }=0.1,$ we have a unique
solution for the initial condition; $z_{\ell }=0.47818$ and $s_{\ell
}=0.21868$ that solve (\ref{lem1}) and (\ref{lem2}), each with equality.
Therefore, almost 48\% of agents stays out of the market. Given $(z_{\ell
},s_{\ell })=(0.47818,0.21868),$ the solution for first-order differential
equation becomes 
\begin{equation*}
\tilde{\mu}(s)=\left[ \frac{3s}{4}-\frac{0.00057172}{s^{3}}\right] ^{1/3}.
\end{equation*}%
$\tilde{\mu}(0.21868)=0.47818$, $\tilde{\mu}(1.3337)=1$, and $\tilde{\mu}$
is increasing in between. Because it is increasing, $\tilde{\sigma}(z)=%
\tilde{\mu}^{-1}(z)$ is the optimal action choice by sender type $z$ for all 
$z\in \lbrack z_{\ell },\overline{z}]$. The market matching function is then 
$\tilde{m}(s)=n\left( \tilde{\mu}(s)\right) =2\left[ \frac{3s}{4}-\frac{%
0.00057172}{s^{3}}\right] ^{1/3}$. The market reaction function $\tilde{\tau}
$ can be derived by (\ref{equilibrium_wage}).

\bigskip

However, if $t_{h}<\tilde{\tau}(\tilde{\sigma}(\bar{z}))$, then we have no
separating CSE. In this case, let $Z(s)$ denote the set of the types of
senders who choose the same action $s$.

\begin{lemma}
\label{lemma_no_bottom_bunching}If $Z(s)$ has a positive measure in a
stronger monotone CSE, then it is an interval with $\max Z(s)=\overline{z}.$
\end{lemma}

Lemma \ref{lemma_binding_upper_bound} below shows that if there is pooling
on the top of the sender side, the reaction to those senders pooled at the
top must be the upper bound of feasible reactions $t_{h}.$

\begin{lemma}
\label{lemma_binding_upper_bound}If $Z(s)$ has a positive measure in a
stronger monotone CSE, then $t_{h}$ is the reaction to the senders of types
in $Z(s)$.
\end{lemma}

We can establish Lemmas \ref{lemma_no_bottom_bunching} and \ref%
{lemma_binding_upper_bound} using only Cho and Sobel monotonicity of $\mu $
without relying on the stronger monotonicity of $\mu $. However, we cannot
derive a D1 equilibrium with Cho and Sobel monotonicity as explained after
Theorem \ref{theorem1}.

Using Lemmas \ref{lemma_no_bottom_bunching} and \ref%
{lemma_binding_upper_bound}, we can establish that the stronger monotone
separating CSE $\{\tilde{\sigma},\tilde{\mu},\tilde{\tau},\tilde{m}\}$ is a
unique stronger monotone CSE if $\tilde{\tau}(\tilde{\sigma}(\bar{z}))\leq
t_{h}.$

\begin{theorem}
\label{thm_uniqueSME}Suppose that $T=[t_{\ell },t_{h}]$ satisfies $0\leq
t_{\ell }<\tilde{\tau}(\tilde{\sigma}(\bar{z}))\leq t_{h}$. Then, a unique
stronger monotone CSE is the well-behaved stronger monotone CSE and it is
separating.
\end{theorem}

Theorem \ref{thm_uniqueSME} extends the uniqueness result in Cho and Sobel
(1990) and Ramey (1996) to a two-sided matching model with a continuum of
senders and receivers.

\begin{remark}
Lemma \ref{lemmaA} and Theorems \ref{thm_unique_separating_eq} and \ref%
{thm_uniqueSME} imply that if $T=[t_{\ell },t_{h}]$ satisfies $0\leq t_{\ell
}<\tilde{\tau}(\tilde{\sigma}(\bar{z}))\leq t_{h},$ a stronger monotone CSE
is unique and separating and it exists.
\end{remark}

If $t_{h}<\tilde{\tau}(\tilde{\sigma}(\bar{z}))$, there are only two types
of non-separating stronger monotone CSEs as shown in Lemma \ref%
{theorem_all_eq_w/o_separating}. The reason is that pooling can happen only
among senders in an interval with $\bar{z}$ being its maximum due to Lemma %
\ref{lemma_no_bottom_bunching}.

\begin{lemma}
\label{theorem_all_eq_w/o_separating}If $t_{h}<\tilde{\tau}(\tilde{\sigma}(%
\bar{z}))$, then, there are two possible stronger monotone CSEs: (i) a
strictly well-behaved stronger monotone CSE and (ii) a stronger monotone
pooling CSE.
\end{lemma}

Lemma \ref{theorem_all_eq_w/o_separating} follows Theorem \ref%
{theorem_stronger_monotone_eq} (Stronger Monotone Equilibrium Theorem II)
and Lemmas \ref{lemma_no_bottom_bunching} and \ref{lemma_binding_upper_bound}%
.

For the uniqueness of a stronger monotone CSE when $t_{h}<\tilde{\tau}(%
\tilde{\sigma}(\bar{z}))$, we impose an additional assumption as follows.

\begin{description}
\item[Assumption 7] \label{ass7 copy(1)}$\lim_{z\rightarrow \underline{z}%
}c(s,z)=\infty $ for all $s>0$, and either (i) or (ii) below is satisfied

(i) $v(x,s,z)=v(x,s^{\prime },z)$ for all $s,s^{\prime }\in 
\mathbb{R}
_{+}$, $v(\underline{x},s,z)=0$ for all $s,z$, and $v(x,s,z)>0$ for all $x>%
\underline{x},$ all $z>\underline{z},$ and all $s\in 
\mathbb{R}
_{+}$.

(ii) $v(x,0,z)=0$ and $v(x,s,z)$ is increasing in $s$ for all $x$ and $z$,
and $v(x,0,z)-c(0,z)\geq 0$ for all $x>\underline{x},$ all $z>\underline{z}$.
\end{description}

\subsection{Pooling CSE}

We first consider a stronger monotone pooling CSE. This is a type of
stronger monotone CSE when $t_{\ell }=t_{h}=t^{\ast }.$ Every seller of type
above $z_{\ell }=z_{h}=z^{\ast }$ enters the market with the pooled action $%
s^{\ast }$.%
\begin{gather}
t^{\ast }-c(s^{\ast },z^{\ast })\geq 0,  \label{pooling_sender1} \\
\mathbb{E}\left[ v\left( n\left( z^{\ast }\right) ,s^{\ast },z^{\prime
}\right) |z^{\prime }\geq z^{\ast }\right] -t^{\ast }\geq 0,\text{ }
\label{pooling_receiver1}
\end{gather}%
where each condition holds with equality if $z^{\ast }>\underline{z}.$

\begin{theorem}
\label{lemma_unique_pooling}Given only a single feasible reaction $0\leq
t^{\ast }<\overline{t},$ a stronger monotone CSE is pooling. Furthermore,
for only a single feasible reaction $0<t^{\ast }<\overline{t},$ the only
possible stronger monotone pooling CSE is a stronger monotone pooling CSE
with $z^{\ast }>\underline{z}$ and $s^{\ast }>0$ that satisfy (\ref%
{pooling_sender1}) and (\ref{pooling_receiver1}), each with equality. For
only a single feasible reaction, $t^{\ast }=0,$ the only possible stronger
monotone pooling CSE is a stronger monotone pooling CSE with $z^{\ast }=%
\underline{z}$ and $s^{\ast }=0$.
\end{theorem}

When there is only one feasible reaction $t^{\ast }\in 
\mathbb{R}
_{+}$, a stronger monotone CSE cannot include a separating part, so all
senders who enter the market choose a pooled action $s^{\ast }$. Whether or
not all senders and receivers enter the market and whether or not all
senders on the market chooses zero action depend on whether the single
feasible reaction is zero or strictly positive.

Assumption 7 plays an important role to establish Theorem \ref%
{lemma_unique_pooling} (See Appendix \ref{App_lemma_unique_pooling} for the
proof). Consider the case with $0<t^{\ast }<\overline{t}.$ It is easy to see
that the lowest sender type $z^{\ast }$ who enters the market must be higher
than the lowest sender type $\underline{z}$. Suppose not, i.e., $z^{\ast }=%
\underline{z}.$ Then, the equilibrium pooled action $s^{\ast }$ must be
zero. Otherwise, the lowest sender type's utility is negative because $%
\lim_{z\rightarrow \underline{z}}c(s,z)=\infty $ for all $s>0$ according to
Assumption 7. However, if $s^{\ast }=0,$ the utility for the lowest receiver
type $n\left( z^{\ast }\right) $ who enter the market is negative because $%
\mathbb{E}\left[ v\left( n\left( z^{\ast }\right) ,s^{\ast },z^{\prime
}\right) |z^{\prime }\geq z^{\ast }\right] =0$ given any one of Assumption
7.(i) and Assumption 7.(ii). This means that the market clearing condition
is not satisfied. Therefore, $z^{\ast }>\underline{z}$, which implies that (%
\ref{pooling_sender1}) and (\ref{pooling_receiver1}) both hold with
equality. This in turn implies $s^{\ast }>0$.

Consider the case with $t^{\ast }=0.$ Then, it is clear that $s^{\ast }=0.$
Otherwise, a sender with $s^{\ast }$ will have utility less than her
reservation utility. If Assumption 7.(i) is satisfied, every receiver type
above $\underline{x}$ gets positive expected utility by matching a sender
with $s^{\ast }$, so every receiver enters the market and it will make every
sender enter the market as well (a sender's utility is exactly her
reservation utility). If Assumption 7.(ii) is satisfied, any receiver who is
matched with a sender with $s^{\ast }=0$ has the utility the same as his
reservation utility. Therefore, all receivers and senders enter the market
and they all get their reservation utility (This is observationally
equivalent to the outcome where no one enters the market).

The existence of a stronger monotone pooling CSE can be established similar
to Lemma \ref{lemmaA}.

\begin{lemma}
\label{lemma_pooling}Given $0<t^{\ast }<\overline{t}$, there exists a unique
solution $\left( s^{\ast },z^{\ast }\right) \in 
\mathbb{R}
_{++}\times (\underline{z},\overline{z})$ that solves (\ref{pooling_sender1}%
) and (\ref{pooling_receiver1}) with equality.
\end{lemma}

\begin{remark}
\label{remark_pooling}For any given $0\leq t^{\ast }<\overline{t}$, Theorem %
\ref{lemma_unique_pooling} and Lemma \ref{lemma_pooling} show that a
stronger monotone CSE is pooling and it is unique and exists.
\end{remark}

\paragraph{Example 2}

We keep the same utility functions in Example 1: $v(x,s,z)-t=xzs-t$ for the
receiver and $t-c(s,z)=t-\frac{s^{2}}{z}$ for the sender Sender type $z$
uniformly distributed over $[0,1],$ whereas receiver type $x$ is uniformly
distributed over $[0,2].$ Suppose that $t^{\ast }=0.1$ is the only possible
reaction receivers can take. Then, a unique stronger monotone CSE is
pooling. We have a unique solution $(z^{\ast },s^{\ast })$ that satisfies (%
\ref{pooling_sender1}) and (\ref{pooling_receiver1}), each with equality and
they are $z^{\ast }=0.36811$ and $s^{\ast }=0.369.$ Any sender type no less
than $z^{\ast }$ chooses a pooled action $s^{\ast }$ and any receiver type
no less than $2z^{\ast }$ is matched with a sender with $s^{\ast }$,
choosing the only reaction $t^{\ast }.$ The threshold types on both sides
are just indifferent between staying out of the market and forming a match
between them with $s^{\ast }$ and $t^{\ast }.$ Anyone above is strictly
better off.

\subsection{Strictly well-behaved CSE}

Now consider a (strictly) well-behaved CSE with both separating and pooling
parts when $t_{\ell }<t_{h}<\tilde{\tau}(\tilde{\sigma}(\bar{z})).$ The
system of equations represented in (\ref{jumping_sellers}) and (\ref%
{jumping_buyers}) is the \emph{key} to understand jumping and pooling in the
upper tail of the match distribution with $t_{h}<\tilde{\tau}(\tilde{\sigma}(%
\bar{z}))$: 
\begin{gather}
t_{h}-c\left( s,z\right) =\tilde{\tau}\left( \tilde{\sigma}\left( z\right)
\right) -c\left( \tilde{\sigma}\left( z\right) ,z\right) ,
\label{jumping_sellers} \\
\mathbb{E}[v(n\left( z\right) ,s,z^{\prime })|z^{\prime }\geq
z]-t_{h}=v\left( n\left( z\right) ,\tilde{\sigma}\left( z\right) ,z\right) -%
\tilde{\tau}\left( \tilde{\sigma}\left( z\right) \right) .
\label{jumping_buyers}
\end{gather}

Let $(s_{h},z_{h})$ denote a solution of (\ref{jumping_sellers}) and (\ref%
{jumping_buyers}). Note that (\ref{jumping_sellers}) makes the type $z_{h}$
sender indifferent between choosing $s_{h}$ for $t_{h}$ and $\tilde{\sigma}%
\left( z_{h}\right) $ for $\tilde{\tau}\left( \tilde{\sigma}\left(
z_{h}\right) \right) .$ The expression on the left hand side of (\ref%
{jumping_sellers}) is the equilibrium utility for the type $z_{h}$ receiver.
The expression on the left hand side of (\ref{jumping_buyers}) is the
utility for the type $n\left( z_{h}\right) $ receiver who chooses a sender
with action $s_{h}$ as his partner by choosing $t_{h}$ for her. This is the
equilibrium utility for type $n(z_{h})$. The expression on the right-hand
side is his utility if he chooses a sender of type $z_{h}$ with action $%
\tilde{\sigma}\left( z_{h}\right) $ as his partner by choosing the reaction $%
\tilde{\tau}\left( \tilde{\sigma}\left( z_{h}\right) \right) $.

Given $t_{\ell }<t_{h}<\tilde{\tau}(\tilde{\sigma}(\bar{z})),$ the pooled
action $s_{h}$ chosen by all senders above type $z_{h}$ is greater than the
action $\tilde{\sigma}\left( z_{h}\right) $ that would have been chosen by
sender type $z_{h}$ in the stronger monotone separating CSE (See Lemma \ref%
{lemma2} in Appendix \ref{Appendix_first_lemma2}) and we have a unique $%
\left( s_{h},z_{h}\right) $ as established in Lemma \ref{lemma_well_behaved}
below.

\begin{lemma}
\label{lemma_well_behaved}Given $t_{\ell }<t_{h}<\tilde{\tau}(\tilde{\sigma}(%
\bar{z}))$, there exists a unique $\left( s_{h},z_{h}\right) \in (\tilde{%
\sigma}\left( z_{h}\right) ,\tilde{\sigma}(\bar{z}))\times \left( z_{\ell },%
\overline{z}\right) $ that satisfies (\ref{jumping_sellers}) and (\ref%
{jumping_buyers}).
\end{lemma}

\bigskip

Theorem \ref{theorem1} below provides the full characterization of a unique
well-behaved CSE. Note that Theorem \ref{theorem1} allows for the
possibility of separating or pooling (as $t_{h}\rightarrow \tilde{\tau}(%
\tilde{\sigma}(\bar{z}))$ or $t_{h}\rightarrow t_{\ell }$) as well as
strictly well-behaved. Let $x_{h}:=n(z_{h})$.

\begin{theorem}
\label{theorem1}Fix a set of feasible reactions $T=[t_{\ell },t_{h}]$ with $%
0\leq t_{\ell }<\tilde{\tau}(\tilde{\sigma}(\bar{z}))<t_{h}$ under which $%
(z_{\ell },s_{\ell })$ denotes the pair of the lower threshold sender type
and her equilibrium action and $(z_{h},s_{h})$\ is the pair of the upper
threshold sender type and her equilibrium action. Then, there exists a
unique well-behaved stronger monotone CSE $\left\{ \hat{\sigma},\hat{\mu},%
\hat{\tau},\hat{m}\right\} $. It is characterized as follows.

\begin{enumerate}
\item $\hat{\sigma}$ follows (i) $\hat{\sigma}(z)=0$ if $z\in \left[ 
\underline{z},z_{\ell }\right) $; (ii) $\hat{\sigma}(z)=s_{\ell }$ if $%
z=z_{\ell }$; (iii) $\hat{\sigma}(z)$ satisfies that $\hat{\tau}^{\prime }(%
\hat{\sigma}\left( z\right) )-c_{s}(\hat{\sigma}\left( z\right) ,z)=0$ if $%
z\in (z_{\ell },z_{h})$; (iv) $\hat{\sigma}(z)=s_{h}$ if $z\in \left[ z_{h},%
\overline{z}\right] $. Further, $\hat{\sigma}\left( z_{h}\right) <s_{h}$.

\item $\hat{\mu}$ follows (i) $\hat{\mu}(s)=G(z|\underline{z}\leq z<z_{\ell
})$ if $s=0$; (ii) $\hat{\mu}(s)=z_{\ell }$ if $s\in (0,\hat{\sigma}(z_{\ell
}))$; (iii) $\hat{\mu}(s)=\hat{\sigma}^{-1}(s)$ if $s\in \left[ \hat{\sigma}%
(z_{\ell }),\hat{\sigma}(z_{h})\right) $; (iv) $\hat{\mu}(s)=z_{h}$ if $s\in
\lbrack \lim_{z\nearrow z_{h}}\hat{\sigma}(z),s_{h})$; (v) $\hat{\mu}%
(s)=G(z|z_{h}\leq z\leq \overline{z})$ if $s=s_{h}$; (vi) $\hat{\mu}(s)=%
\overline{z}$ if $s>s_{h}$.

\item $\hat{\tau}(s)$ with $\hat{\tau}(s_{\ell })=t_{\ell }$ satisfies (i) $%
v_{s}\left( x,s,\hat{\mu}(s)\right) +v_{z}\left( x,s,\hat{\mu}(s)\right) 
\hat{\mu}^{\prime }(s)-\hat{\tau}^{\prime }(s)=0$ at $s=\xi (x)$ for all $%
x\in $ $(x_{\ell },x_{h})$ and (ii) $\hat{\tau}(s)=t_{h}$ if $s\geq s_{h}$.
Further, $\hat{\tau}\left( \hat{\sigma}\left( z_{h}\right) \right) <t_{h}$.

\item $\hat{m}$ follows that (i) $\hat{m}(s)=n(\hat{\mu}(s))$ if $s\in \left[
\hat{\sigma}(z_{\ell }),\hat{\sigma}(z_{h})\right) $, (ii) $\hat{m}(s)=\left[
x_{h},\overline{x}\right] $ if $s=s_{h}.$
\end{enumerate}
\end{theorem}

\begin{figure}[tbp]
\centering\includegraphics[scale=0.4]{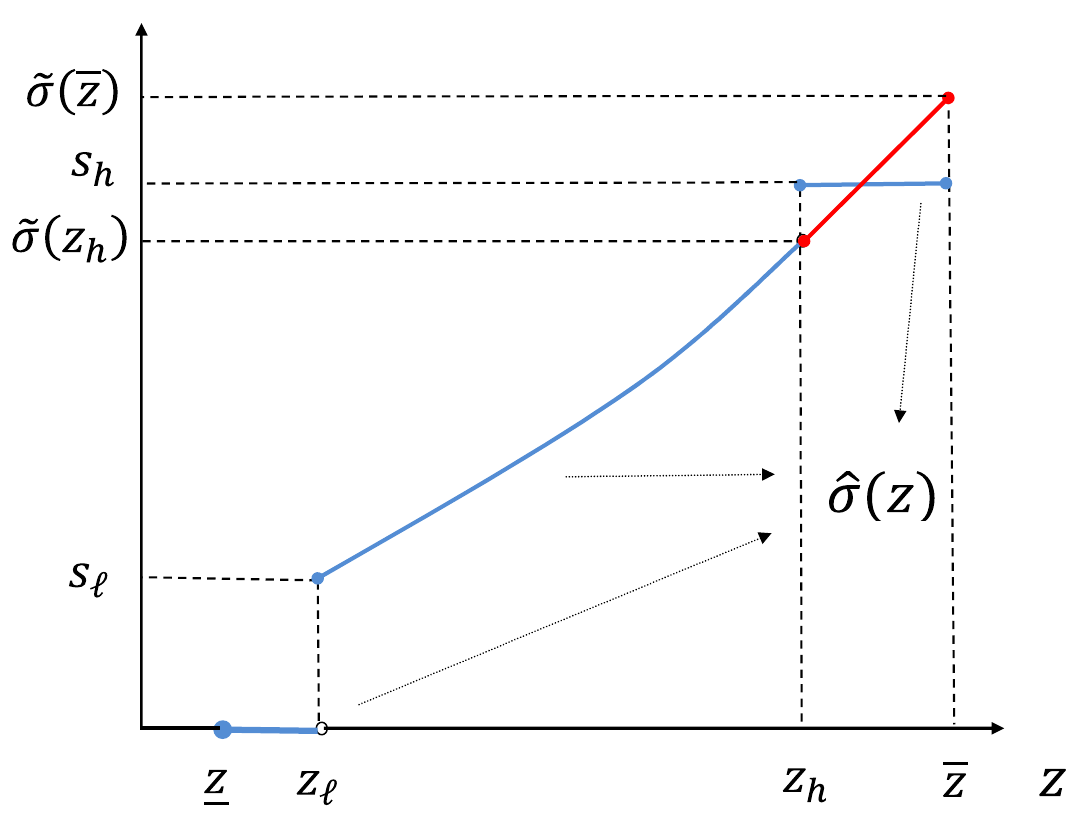}
\caption{Senders' equilibrium actions}
\label{fig:figure1}
\end{figure}

If a well-behaved stronger monotone CSE has both separating and pooling, it
follows the separating CSE with the same $z_{\ell }$ before $z$ hits $z_{h}$
according to Conditions 
1(i)--(iii), 2(i)--(iii), 3(i), and 4(i) in Theorem \ref{theorem1} above. As
Condition 1 in Theorem \ref{theorem1} and Lemma \ref{lemma2} (Appendix \ref%
{Appendix_first_lemma2}) show, in a (strictly) well-behaved stronger
monotone CSE, we have jumping in equilibrium sender actions at the threshold
sender type $z_{h}$, followed by pooling.\footnote{%
There is also jumping in equilibrium reactions at $t_{h}$. See Lemma \ref%
{lemma2} in Appendix \ref{Appendix_first_lemma2}.} In Figure \ref%
{fig:figure1}, the equilibrium sender actions consist of the three different
blue parts.\footnote{%
Note that $\lim_{z\nearrow z_{h}}\hat{\sigma}(z)=\tilde{\sigma}(z_{h})$ in
Figure \ref{fig:figure1}.} Note that equilibrium matching is assortative in
terms of sender action and receiver type (and therefore in terms of sender
type and receiver type) in the separating part of the CSE but it is random
in the pooling part of the CSE. Therefore, there is matching inefficiency in
the pooling part but there may be potential savings in the signaling cost
associated with the pooled action choice by senders above $z_{h}$. 

Because $z_{h}$ is in the interior of the sender's type interval in a
strictly well-behaved stronger monotone CSE, Cho and Sobel monotonicity does
not pin down the belief $\hat{\mu}(s)$ conditional on an off-path action $%
s\in \lbrack \lim_{z\nearrow z_{h}}\hat{\sigma}(z),s_{h})$, whereas the
stronger monotonicity of the belief (see Corollary \ref%
{corollary_monotone_belief} in Section \ref{section_montone_equilibrium})
uniquely pins it down as one that puts all the probability weights on $z_{h}$
as specified in Condition 2(iv). Further, because the stronger monotonicity
of the belief is equivalent to Criterion D1, we only need to show that the
sender type $z_{h}$ has no incentive to deviate to an off-path action in $%
[\lim_{z\nearrow z_{h}}\hat{\sigma}(z),s_{h})$ in order to show that no
sender has an incentive to deviate to such an off-path action.

Theorem \ref{thm_unique_well_behaved} below shows that if $0\leq t_{\ell
}<t_{h}<\tilde{\tau}(\tilde{\sigma}\left( \overline{z}\right) )$, then $%
\left\{ \hat{\sigma},\hat{\mu},\hat{\tau},\hat{m}\right\} $ characterized in
Theorem \ref{theorem1} is a unique stronger monotone CSE.

\begin{theorem}
\label{thm_unique_well_behaved}Fix a set of feasible receiver actions to $%
T=[t_{\ell },t_{h}]$ with $0\leq t_{\ell }<t_{h}<\tilde{\tau}(\tilde{\sigma}%
\left( \overline{z}\right) )$. A unique stronger monotone CSE is $\left\{ 
\hat{\sigma},\hat{\mu},\hat{\tau},\hat{m}\right\} $ and it exists.
\end{theorem}

Theorem \ref{thm_unique_separating_eq} and Lemma \ref{lemmaA} and \ref%
{lemma_well_behaved} establish the existence of a unique (strictly)
well-behaved stronger monotone CSE $\left\{ \hat{\sigma},\hat{\mu},\hat{\tau}%
,\hat{m}\right\} $ characterized in Theorem \ref{theorem1}. Because of Lemma %
\ref{theorem_all_eq_w/o_separating}, we can establish Theorem \ref%
{thm_unique_well_behaved} by showing that there is no stronger monotone
pooling CSE if $0\leq t_{\ell }<t_{h}<\tilde{\tau}(\tilde{\sigma}(\bar{z}))$
(See the proof of Theorem \ref{thm_unique_well_behaved} in Appendix \ref%
{App_thm_unique_well_behaved}).

\paragraph{Example 3}

In Example 1, $t_{\ell }=0.1$ is the lower bound of feasible reactions with
no upper bound. Suppose that $t_{h}=2.5$ is the upper bound of reactions
with the same lower bound. Because $t_{h}<\tilde{\tau}(\tilde{\sigma}(\bar{z}%
)),$ the unique stronger monotone CSE is strictly well-behaved. In this
strictly well-behaved CSE, any sender type above $z_{h}$ chooses the pooled
action $s_{h},$ each sender type in $[z_{\ell },z_{h})$ separates herself by
choosing the same action she would have chosen in the separating CSE in
Example 1, any sender type below From (\ref{jumping_sellers}) and (\ref%
{jumping_buyers}), $s_{h}$ and $z_{h}$ are jointly determined and they are $%
s_{h}=1.12301$ and $z_{h}=0.90275$.

\bigskip

Theorems \ref{thm_uniqueSME}, \ref{lemma_unique_pooling}, and \ref%
{thm_unique_well_behaved} establish the unique stronger monotone CSE given
each type of the feasible reaction sets: (i) If $0\leq t_{\ell }<\tilde{\tau}%
(\tilde{\sigma}(\bar{z}))\leq t_{h}$, then a unique stronger monotone CSE is
well-behaved and separating, (ii) If there is only a single feasible
reaction, then it is pooling, (iii) If $0\leq t_{\ell }<t_{h}<\tilde{\tau}(%
\tilde{\sigma}\left( \overline{z}\right) ),$ then it is strictly
well-behaved.

Furthermore, given our characterization of a unique stronger monotone CSE
with any interval of feasible reactions, Theorem \ref%
{thm_differentiable_sep_eq} implies that if a stronger monotone CSE has a
separating part, that part of the CSE is differentiable.

\section{Concluding remarks\label{sec_discussion}}

This paper studies a monotone CSE in the stronger set order, i.e.,
equilibrium outcomes and beliefs are all monotone in the stronger set order.
We show that if the sender utility is monotone-supermodular and the
receiver's utility is weakly monotone-supermodular, a CSE is stronger
monotone if and only if it passes Criterion D1 (Cho and Kreps (1987), Banks
and Sobel (1987)). We fully characterize a unique stronger monotone CSE and
establishes its existence with quasilinear utility functions. A stronger
monotone CSE is well-behaved and hence characterized by the two threshold
sender types. Furthermore, a separating part of a stronger monotone CSE is
differentiable even if a stronger monotone separating CSE does not exist.

Because the characterization of a stronger monotone CSE is provided given
any interval of feasible reactions, it opens a way to study the planner's
mechanism design problem to regulate receivers' reaction choices in
two-sided matching markets with a continuum of senders and receivers. A
companion paper of ours (Han, Sam, and Shin (2023)) studies such a mechanism
design problem for the planner who announces a mechanism that specifies a
receiver's reaction as a function of his message. In that paper, we
establish the Interval Delegation Principle that ensures no loss of
generality to delegate a receiver's reaction choice from an interval of
reactions. Therefore, the planner's mechanism design problem comes down to
choosing an interval of reactions. This makes it possible for us to use our
results in Section \ref{Sec_Eq_w_lower_bound} for the planner's optimal
delegation problem.

\addcontentsline{toc}{section}{Appendices}

\renewcommand{\thesection}{\Alph{section}} \setcounter{section}{0} %
\setcounter{equation}{0} \renewcommand{\theequation}{A\arabic{equation}}

\begin{center}
{\LARGE \textbf{Appendix}}
\end{center}

\section{Proof of Lemma \protect\ref{lemma_monotone_eq_A'}}

\begin{proof}
For any $s$ and $s^{\prime }$ in $S^{\ast }$ such that $s>s^{\prime },$
consider a sender who chooses $s$ in equilibrium. Then, her utility must
satisfy 
\begin{equation}
u(\tau (s),s,z)\geq u(\tau (s^{\prime }),s^{\prime },z)  \label{monotone0}
\end{equation}%
Because $s>s^{\prime }$ and $u$ is decreasing in $s$ and increasing in $t,$ (%
\ref{monotone0}) implies $\tau (s)>\tau (s^{\prime })$ and hence $\tau $ is
monotone increasing.

Now we prove the monotonicity of $\sigma $ by contradiction. Suppose that
for $z>z^{\prime }$, type $z$ chooses $s^{\prime }\in S^{\ast }$ and type $%
z^{\prime }$ chooses $s\in S^{\ast }$ such that $s>s^{\prime }$ in
equilibrium. This implies $u(\tau (s),s,z^{\prime })\geq u(\tau (s^{\prime
}),s^{\prime },z^{\prime })$. Because $(\tau (s),s)>(\tau (s^{\prime
}),s^{\prime }),$ the strict single crossing property of $u$ in $((t,s);z)$
implies that $u(\tau (s),s,z)>u(\tau (s^{\prime }),s^{\prime },z)$ for any $%
z>z^{\prime }$. This contradicts that $z$ chooses $s^{\prime }$. Therefore, $%
\sigma $ is non-decreasing over types that choose actions in $S^{\ast }$.

For the non-decreasing property of $\sigma $, we only need to show that any $%
z^{\prime }$ with $\eta $ is no higher than $z$ with $s\in S^{\ast }$ in
equilibrium. By contradiction, suppose that there exist $z^{\prime }$ with $%
\eta $ and $z$ with $s\in S^{\ast }$ such that $z^{\prime }>z.$ Then, we
have that $0\leq u(\tau (s),s,z)<u(\tau (s),s,z^{\prime }),$where the weak
inequality holds because $s$ is the optimal choice for type $z$ and the
strict inequality holds due to the monotonicity of $u$ in type. This
contradicts that taking no action (i.e., null action $\eta $) is optimal for
type $z^{\prime }.$ This completes the proof of the non-decreasing property
of $\sigma $.

We prove the monotonicity of supp $\mu (s)$ in the subset of domain, $\sigma
(Z),$ in the stronger set order by contradiction. Suppose that for $%
s>s^{\prime }$, there exist $z\in $ supp $\mu (s)$ and $z^{\prime }\in $
supp $\mu (s^{\prime })$ such that $z^{\prime }>z.$ Because $z\in $\ cl $%
\left\{ \tilde{z}|\sigma \left( \tilde{z}\right) =s\right\} $\ and $%
z^{\prime }\in $\ cl $\left\{ \tilde{z}|\sigma \left( \tilde{z}\right)
=s^{\prime }\right\} $, it contradicts that $\sigma $\ is non-decreasing.
Therefore, $\mu $ is non-decreasing in the subset of domain, $\sigma (Z),$
with respect to the stronger set order.
\end{proof}

\section{Proof of Proposition \protect\ref{theorem_monotone_belief}}

According to the proof of Lemma \ref{lemma_monotone_eq_A'}, $\sigma $ is
non-decreasing if Assumption A is satisfied. If there are types who choose $%
\eta $ and stay out of the market, those types are lower than the types who
choose actions in $S$ because of the non-decreasing property of $\sigma $.
If this happens, let $z_{\eta }:=\min \left\{ z\in Z|\sigma (z)\in S^{\ast
}\right\} $ (or $z_{\eta }:=\inf \left\{ z\in Z|\sigma (z)\in S^{\ast
}\right\} $ if $\min \left\{ z\in Z|\sigma (z)\in S^{\ast }\right\} $ does
not exist).

At any discontinuity point $z$, let $\sigma (z_{+}):=\lim_{k\searrow
z}\sigma (k)$ and $\sigma (z_{-}):=\lim_{k\nearrow z}\sigma (k).$ For the
proof of item 1, we first consider the case where a discontinuity occurs at $%
z>z_{\eta }$ and $\sigma $ is only right continuous at $z$, then, $\sigma
(z_{+})=\sigma (z)$. In this case, $[\sigma (z_{-}),\sigma (z))$ is the
interval of off-path sender actions due to the discontinuity at $z$. We show
that Criterion D1 places zero posterior weight on $z^{\prime }\neq z$.

\textit{Case 1}: We show that $z^{\prime }$ cannot be in the support of $\mu
(s)$ for any $s\in \lbrack \sigma (z_{-}),\sigma (z))$ if $z^{\prime }>z.$
On the contrary, suppose that $z^{\prime }\in $ supp $\mu (s)$ for some $%
s\in \lbrack \sigma (z_{-}),\sigma (z))$ when $z<z^{\prime }$. If $%
z<z^{\prime },$ then we have $z^{\prime \prime }$ such that $z<z^{\prime
\prime }<z^{\prime }$. For the proof, it is sufficient that if type $%
z^{\prime \prime }$ is weakly worse off by deviating to $s\in \lbrack \sigma
(z_{-}),\sigma (z)),$ then type $z^{\prime }$ is strictly worse off with the
same deviation. For a reaction $t$ chosen by the receiver after observing
such $s$, let 
\begin{equation}
u(t,s,z^{\prime \prime })\leq u(\tau (\sigma (z^{\prime \prime })),\sigma
(z^{\prime \prime }),z^{\prime \prime }).  \label{D1_1}
\end{equation}%
Because $s\in \lbrack \sigma (z_{-}),\sigma (z)),$ we have that $s<\sigma
(z).$ Because $\sigma $ is non-decreasing, we have that $\sigma (z)\leq
\sigma (z^{\prime \prime }).$ These two inequality relations yield $s<\sigma
(z^{\prime \prime }).$ Because the first part of Assumption A says that $u$
is decreasing in $s$ and increasing in $t,$ we must have that $t<\tau
(\sigma (z^{\prime \prime }))$ in order to satisfy (\ref{D1_1}). Because $%
s<\sigma (z^{\prime \prime })$ and $t<\tau (\sigma (z^{\prime \prime }))$,
we can use the strict single crossing property of $u$ in Assumption A to
show that (\ref{D1_1}) implies that for $z^{\prime }>z^{\prime \prime }$%
\begin{equation}
u(t,s,z^{\prime })<u(\tau (\sigma (z^{\prime \prime })),\sigma (z^{\prime
\prime }),z^{\prime }).  \label{D1_2}
\end{equation}%
On the other hand, we have that 
\begin{equation}
u(\tau (\sigma (z^{\prime \prime })),\sigma (z^{\prime \prime }),z^{\prime
})\leq u(\tau (\sigma (z^{\prime })),\sigma (z^{\prime }),z^{\prime })
\label{D1_3}
\end{equation}%
in equilibrium. Combining (\ref{D1_2}) and (\ref{D1_3}) yields that for $%
z^{\prime }>z^{\prime \prime },$ 
\begin{equation*}
u(t,s,z^{\prime })<u(\tau (\sigma (z^{\prime })),\sigma (z^{\prime
}),z^{\prime }),
\end{equation*}%
which shows that type $z^{\prime }$ is strictly worse off with the same
deviation. (\ref{Criterion_D12}), the contrapositive of (\ref{Criterion_D1})
in the definition of Criterion D1 implies that any $z^{\prime }>z$ cannot be
in the support of $\mu (s)$ for any $s\in \lbrack \sigma (z_{-}),\sigma (z))$%
.

\textit{Case 2}: We now show that $z^{\prime }$ cannot be in the support of $%
\mu (s)$ for any $s\in \lbrack \sigma (z_{-}),\sigma (z))$ if $z^{\prime
}<z. $ On the contrary, suppose that $z^{\prime }\in $ supp $\mu (s)$ for
some $s\in \lbrack \sigma (z_{-}),\sigma (z))$ when $z^{\prime }<z$. We work
with the original condition (\ref{Criterion_D1}). If $z^{\prime }<z,$ then
we have $z^{\prime \prime }$ such that $z^{\prime }<z^{\prime \prime }<z$.
For a reaction $t$ chosen by the receiver after observing such $s$, let 
\begin{equation}
u(t,s,z^{\prime })\geq u(\tau (\sigma (z^{\prime })),\sigma (z^{\prime
}),z^{\prime }),  \label{D1_4}
\end{equation}%
that is, type $z^{\prime }$ is weakly better off by deviating to some $s\in
\lbrack \sigma (z_{-}),\sigma (z))$. On the other hand, we have that 
\begin{equation}
u(\tau (\sigma (z^{\prime })),\sigma (z^{\prime }),z^{\prime })\geq u(\tau
(\sigma (z^{\prime \prime })),\sigma (z^{\prime \prime }),z^{\prime })
\label{D1_5}
\end{equation}%
in equilibrium. Combining (\ref{D1_4}) and (\ref{D1_5}) yields 
\begin{equation}
u(t,s,z^{\prime })\geq u(\tau (\sigma (z^{\prime \prime })),\sigma
(z^{\prime \prime }),z^{\prime })  \label{D1_6}
\end{equation}%
Because $s>\sigma (z^{\prime \prime }),$ the monotonicity of $u$ in
Assumption A implies that $t>\tau (\sigma (z^{\prime \prime }))$ in order to
satisfy (\ref{D1_6}). Then, applying the strict single crossing property of $%
u$ in Assumption A to (\ref{D1_6}), we have that for $z^{\prime \prime
}>z^{\prime },$ 
\begin{equation*}
u(t,s,z^{\prime \prime })>u(\tau (\sigma (z^{\prime \prime })),\sigma
(z^{\prime \prime }),z^{\prime \prime }),
\end{equation*}%
which shows that the sender of type $z^{\prime \prime }$ is strictly better
off with the same deviation. Criterion D1 implies that any $z^{\prime }<z$
cannot be in the support of $\mu (s)$ for any $s\in \lbrack \sigma
(z_{-}),\sigma (z))$. Therefore, the only $\mu (s)$ conditional on $s\in
\lbrack \sigma (z_{-}),\sigma (z))$ that passes Criterion D1 puts all the
posterior weights on $z$ and hence supp $\mu (s)=\{z\}.$

Item 1 can be proved similarly in the cases where $\sigma $ is only left
continuous at $z$ or $\sigma (z_{-})<\sigma (z)<\sigma (z_{+})$, or in the
case where $\sigma $ is discontinuous at $z_{\eta }$. The only thing we need
to be careful about is the case where $\sigma $ is discontinuous at $z_{\eta
}$. If $\sigma $ is only right continuous at $z_{h},$ the discontinuity at $%
z_{h}$ creates off-path actions $s\in S$ such that $s<\min S^{\ast }$ (or $%
s\leq \inf S^{\ast }$ if $\min S^{\ast }$ does not exist). We can show that
any $z^{\prime }>z_{h}$ cannot be in the support of $\mu (s)$ for any $%
s<\min S^{\ast }$ following the logic of Case 1 above. Consider $z^{\prime
}<z_{\eta }.$ If $z^{\prime }<z,$ then we have $z^{\prime \prime }$ such
that $z^{\prime }<z^{\prime \prime }<z_{h}.$ Note that $\sigma (z^{\prime
})=\sigma (z^{\prime \prime })=\eta $ and hence, equilibrium utilities for
both types are zero. For a reaction $t$ chosen by the receiver after
observing $s\in S\backslash S^{\ast },$ let $u(t,s,z^{\prime })\geq 0.$ By
the increasing property of $u$ in type in Assumption A.(i), $u(t,s,z^{\prime
})\geq 0$ implies that $u(t,s,z^{\prime \prime })>0.$ This shows that any $%
z^{\prime }<z_{\eta }$ cannot be in the support of $\mu (s)$ given Criterion
D1. Therefore, the only $\mu (s)$ conditional on any $s\in S\backslash
S^{\ast }$ that passes Criterion D1 puts all the posterior weights on $z$
and hence supp $\mu (s)=\{z_{h}\}.$ Item 1 can be proved similarly if only
left continuous at $z_{h}$ or $\sigma _{-}(z_{h})<\sigma (z_{h})<\sigma
_{+}(z_{h}).$

For the proof of item 2, we can following the proof of Case 2 to show that
the only $\mu (s)$ conditional on $s>\sigma (\overline{z})$ that passes
Criterion D1 puts all the posterior weights on $\overline{z}$ and hence supp 
$\mu (s)=\{\overline{z}\}$ for $s>\sigma (\overline{z})$. Similarly, for the
proof of item 3, we can follow the proof of Case 1 above to show that only $%
\mu (s)$ conditional on $s<\sigma (\underline{z})$ that passes Criterion D1
puts all the posterior weights on $\underline{z}$ and hence supp $\mu (s)=\{%
\underline{z}\}$ for $s<\sigma (\underline{z}).$

\section{Proof of Corollary \protect\ref{corollary_monotone_belief}}

According to the proof of Lemma \ref{lemma_monotone_eq_A'}, $\mu $ is
non-decreasing in the subset of domain, $\sigma (Z),$ with respect to the
stronger set order if Assumption A is satisfied.

We first show that a non-decreasing $\mu $ in the stronger set order passes
Criterion D1. Consider $s\notin $ $\sigma (Z)$. If $s>\sigma (\overline{z}),$
a non-decreasing $\mu $ in the stronger set order must have $\{\overline{z}%
\} $ as supp $\mu (s)$. On the contrary, suppose that $z\in $ supp $\mu (s)$
for $s>\sigma (\overline{z})$ and $z<\bar{z}.$ This implies that $z<%
\overline{z}$ for $z\in $ supp $\mu (s)$ and $\overline{z}\in $ supp $\mu
(\sigma (\overline{z}))$ but $s>\sigma (\overline{z})$: supp $\mu (s)\ngeq
_{c}$ supp $\mu (\sigma (\overline{z})).$ This contradicts the monotonicity
of $\mu $ in the stronger set order and hence supp $\mu (s)=\{\overline{z}\}$
for $s>\sigma (\overline{z}).$ This passes Criterion D1, which requires it
as in item 2 in Lemma \ref{theorem_monotone_belief}. We can analogously show
that if $s<\sigma (\underline{z}),$ a monotone non-decreasing $\mu $ in the
stronger set order must have $\{\underline{z}\}$ as supp $\mu (s)$ and that
it passes Criterion D1, which requires it as in item 3 in Lemma \ref%
{theorem_monotone_belief}.

Consider the case where $s$ belongs to the interval of off-path sender
actions induced by the discontinuity of $\sigma $ at some $z.$ Consider the
case where $\sigma $ is only right-continuous at $z.$ Given the monotonicity
of $\sigma ,$ we have that 
\begin{equation}
\lim_{k\nearrow z}\sup \text{supp }\mu (\sigma (k))=\min \text{supp }\mu
(\sigma (z))=z.  \label{common_element}
\end{equation}

If $\mu $ is monotone non-decreasing in the stronger set order, supp $\mu
(s^{\prime })$ and supp $\mu (s^{\prime \prime })$ for two different $%
s^{\prime }$ and $s^{\prime \prime }$ have at most one element in common.
Therefore, (\ref{common_element}) implies that a non-decreasing $\mu $ in
the stronger set order must have $\{z\}$ as supp $\mu (s)$ for any $s\in
\lbrack \sigma (z_{-}),\sigma (z)).$ This passes Criterion D1, which
requires it as in item 1 in Lemma \ref{theorem_monotone_belief}. We can
analogously prove that a non-decreasing $\mu $ in the stronger set order
satisfies item 1 in Lemma \ref{theorem_monotone_belief} in the cases where $%
\sigma $ is only left continuous at $z$ or $\sigma (z_{-})<\sigma (z)<\sigma
(z_{+})$. It is straightforward to show that an equilibrium $\mu $ that
passes Criterion D1 is non-decreasing in the stronger set order.

\section{Proof of Lemma \protect\ref{lemma_ass_2_b}}

If Assumption 2 is satisfied, then $g(t,s,z,x)=v(x,s,z)-t$ is supermodular
in all arguments $(t,s,z,x).$ Because each of $T,S,Z,$ and $X$ is a lattice
in $%
\mathbb{R}
,$ we can invoke Theorem 2.6.1 in Topkis (1998) to show that the
supermodularity of $g(t,s,z,x)$ in all arguments implies non-decreasing
differences in $((t,s,z),x),$ which in turn implies the single crossing
property in $((t,s,z);x)$. Because Assumptions 2.(i) implies that $%
g(t,s,z,x) $ has increasing differences in $(z,x)$. Therefore, (ii) in the
weak monotone-supermodular condition for the receiver's utility is
satisfied. (i) in the weak monotone-supermodular condition for the
receiver's utility is satisfied by Assumption 2.(ii).

\section{Proof of Lemma \protect\ref{lemmaA} \label{Appendix_Lemma1}}

Consider the case with $t_{\ell }=0.$ Since there is no restrictions on the
receiver's reactions, every sender enters the market. Furthermore, there is
no information rent in the lowest match between type $\underline{z}$ and
type $\underline{x}$. Therefore, the equilibrium action $s_{\ell }$ in the
lowest match is bilaterally efficient (i.e., $s_{\ell }=\zeta (\underline{x},%
\underline{z})=0$). (\ref{lem1}) and (\ref{lem2}) hold with equality. The
reason is that we normalize $\zeta (\underline{x},\underline{z})$ and $v(%
\underline{x},\zeta (\underline{x},\underline{z}),\underline{z})-c(\zeta (%
\underline{x},\underline{z}),\underline{z})$ to $0$ respectively. Given $%
c(\zeta (\underline{x},\underline{z}),\underline{z})=0$ due to Assumption %
\ref{ass4}, $v(\underline{x},\zeta (\underline{x},\underline{z}),\underline{z%
})-c(\zeta (\underline{x},\underline{z}),\underline{z})=0$ implies that $v(%
\underline{x},\zeta (\underline{x},\underline{z}),\underline{z})=0$ as well.

Now we consider the case with $0<t_{\ell }<\overline{t}$. First, note that
if $t_{\ell }>0$, then (\ref{lem1}) and (\ref{lem2}) must be satisfied with
equality at $(s_{\ell },z_{\ell })$ as pointed out above. When $t_{\ell }>0,$
we cannot have $s_{\ell }=0$ in equilibrium. Given $s_{\ell }=0$, suppose
that $z_{\ell }>\underline{z}$ . Then, those sender types below $z_{\ell }$
will enter the market to enjoy positive utility. Given $s_{\ell }=0,$
suppose that $z_{\ell }=\underline{z}.$ Then, we have $v\left( n\left(
z_{\ell }\right) ,s_{\ell },z_{\ell }\right) -t_{\ell }<0$ because $v\left(
n\left( z_{\ell }\right) ,s_{\ell },z_{\ell }\right) =v(\underline{x},\zeta (%
\underline{x},\underline{z}),\underline{z})=0$. Then, receiver type $n\left(
z_{\ell }\right) $ will leave the market. Therefore, we must have%
\begin{equation*}
s_{\ell }>0.
\end{equation*}

There are two possibilities for $s_{\ell }$: case (i) $0<s_{\ell }<\zeta
(n\left( z_{\ell }\right) ,z_{\ell })$ and case (ii) $s_{\ell }\geq \zeta
(n\left( z_{\ell }\right) ,z_{\ell }).$ First of all, we show that case (i)
is not satisfied in equilibrium. Suppose that case (i) is satisfied. Then,
there is a profitable sender deviation for type $z_{\ell }$. To see this,
suppose that the sender of type $z_{\ell }$ chooses $\zeta (n\left( z_{\ell
}\right) ,z_{\ell })$ instead of $s_{\ell }$ and that the receiver of type $%
n\left( z_{\ell }\right) $ is matched with her by choosing reaction $%
t^{\prime }.$ If $t^{\prime }$ satisfies that 
\begin{eqnarray}
t^{\prime }-c(\zeta (n\left( z_{\ell }\right) ,z_{\ell }),z_{\ell })
&>&t-c(s_{\ell },z_{\ell }),  \label{lemmaA-00} \\
\mathbb{E}_{\mu \left( \zeta (n\left( z_{\ell }\right) ,z_{\ell })\right) }%
\left[ v(n\left( z_{\ell }\right) ,\zeta (n\left( z_{\ell }\right) ,z_{\ell
}),z)\right] -t^{\prime } &>&v(n\left( z_{\ell }\right) ,s_{\ell },z_{\ell
})-t  \label{lemmaA-0}
\end{eqnarray}%
then, there is a profitable sender deviation for type $z_{\ell }$. Define $%
t^{\prime \prime }$ as%
\begin{equation*}
t^{\prime \prime }:=\mathbb{E}_{\mu }\left[ v(n\left( z_{\ell }\right)
,\zeta (n\left( z_{\ell }\right) ,z_{\ell }),z)\right] -\left( v(n\left(
z_{\ell }\right) ,s_{\ell },z_{\ell })-t\right) .
\end{equation*}%
If the receiver chooses $t^{\prime \prime }$, the utility for the sender of
type $z_{\ell }$ is 
\begin{equation*}
\mathbb{E}_{\mu }\left[ v(n\left( z_{\ell }\right) ,\zeta (n\left( z_{\ell
}\right) ,z_{\ell }),z)\right] -\left( v(n\left( z_{\ell }\right) ,s_{\ell
},z_{\ell })-t\right) -c(\zeta (n\left( z_{\ell }\right) ,z_{\ell }),z_{\ell
})
\end{equation*}%
This utility is strictly higher than $t-c(s_{\ell },z_{\ell })$ because 
\begin{multline*}
\mathbb{E}_{\mu \left( \zeta (n\left( z_{\ell }\right) ,z_{\ell })\right) }%
\left[ v(n\left( z_{\ell }\right) ,\zeta (n\left( z_{\ell }\right) ,z_{\ell
}),z)\right] -c(\zeta (n\left( z_{\ell }\right) ,z_{\ell }),z_{\ell }) \\
\geq v(n\left( z_{\ell }\right) ,\zeta (n\left( z_{\ell }\right) ,z_{\ell
}),z_{\ell })-c(\zeta (n\left( z_{\ell }\right) ,z_{\ell }),z_{\ell
})>v\left( n\left( z_{\ell }\right) ,s_{\ell },z_{\ell }\right) -c(s_{\ell
},z_{\ell }),
\end{multline*}%
where the first inequality is satisfied because the infimum of the support
of $\mu \left( \zeta (n\left( z_{\ell }\right) ,z_{\ell })\right) $ is $%
z_{\ell }$ and the second inequality is satisfied because $0<s_{\ell }<\zeta
(n\left( z_{\ell }\right) ,z_{\ell })$ and $v-c$ is strictly concave in $s$
(Assumption 5). Therefore, there exists $\epsilon >0$ such that $t^{\prime
}=t^{\prime \prime }-\epsilon $ that makes both (\ref{lemmaA-0}) and (\ref%
{lemmaA-00}) holds, that is, there is a profitable sender deviation for type 
$z_{\ell }$. This means that case (i) is not satisfied in equilibrium,
leaving case (ii) as the only possibility.

Suppose that case (ii) holds with equality, i.e., $s_{\ell }=\zeta (n\left(
z_{\ell }\right) ,z_{\ell }).$ Note that (\ref{lem1}) and (\ref{lem2}), each
with equality lead to 
\begin{equation}
v\left( n\left( z\right) ,s,z\right) -c\left( s,z\right) =0  \label{lemmaA-1}
\end{equation}%
at $s=s_{\ell }$ and $z=z_{\ell }.$ If $s_{\ell }=\zeta (n\left( z_{\ell
}\right) ,z_{\ell }),$ then (\ref{lemmaA-1}) holds only when $z_{\ell }=0$
and therefore $s_{\ell }=\zeta (n\left( z_{\ell }\right) ,z_{\ell })=0.$
This contradicts $s_{\ell }>0.$ Therefore, case (ii) must hold with strictly
inequality: 
\begin{equation}
s>\zeta (n\left( z\right) ,z)\text{ at }(s,z)=(s_{\ell },z_{\ell }).
\label{lemmaA-11}
\end{equation}

For all $z\in (\underline{z},\overline{z}),$ we have that $v\left( n\left(
z\right) ,\zeta (n\left( z\right) ,z),z\right) -c\left( s,z\right) >0$ and $%
v\left( n\left( z\right) ,\zeta (n\left( z\right) ,z),z\right) -c\left(
s,z\right) <0$ as $s\rightarrow \infty $ (Assumption 5). Furthermore, Given
Assumptions 3 and 4, the left hand side of (\ref{lemmaA-1}) is continuously
differentiable in its domain. By employing Assumption 5, because $s>\zeta
(n(z),z),$ the partial derivative with respect to $s$ of the left hand side
of (\ref{lemmaA-1}) is strictly negative. i.e., for $s>\zeta (n(z),z),$ $%
v_{s}\left( n\left( z\right) ,s,z\right) -c_{s}\left( s,z\right) <0.$ By the
intermediate value theorem, there exists a unique $s(z)$ for all $z\in (%
\underline{z},\overline{z})$ that satisfies (\ref{lemmaA-1}).

Then, it follows from the implicit function theorem that there exists a
unique continuously differentiable function $s(z)>0$ that satisfies (\ref%
{lemmaA-1}) for all $z\in (\underline{z},\overline{z}),$ and 
\begin{equation}
s^{\prime }(z)=\dfrac{c_{z}\left( s(z),z\right) -v_{z}\left( n\left(
z\right) ,s,z\right) -v_{x}\left( n\left( z\right) ,s,z\right) n^{\prime }(z)%
}{v_{s}\left( n\left( z\right) ,s,z\right) -c_{s}\left( s,z\right) }.
\label{lemmaA-2}
\end{equation}%
Given Assumptions 1 (i), 2 (ii), 3 (ii) and 6, the left hand side of (\ref%
{lemmaA-1}) is increasing in $z.$ This implies that the numerator of the
right hand side of (\ref{lemmaA-2}) is negative. Therefore, because $%
s(z)>\zeta (n(z),z),$ it is the case that $s(z)$ is strictly increasing in $%
z $.

Let $\Lambda (z):=v\left( n\left( z\right) ,s(z),z\right) $. $\Lambda (z)$
is continuous and increasing in $s$ because $v$ is continuous and $n(z)$ and 
$s(z)$ are continuous. Because $\Lambda (\underline{z})-t_{\ell }<0$ and $%
\Lambda (\overline{z})-t_{\ell }>0,$ we have a unique $z_{\ell }\in (%
\underline{z},\overline{z})$ such that $\Lambda (z_{\ell })-t_{\ell }=0$ by
the intermediate value theorem. Because of (\ref{lemmaA-1}), $\Lambda
(z_{\ell })-t_{\ell }=0$ also implies that $t_{\ell }-c(s(z_{\ell }),z_{\ell
})=0,$ so that both (\ref{lem1}) and (\ref{lem2}) holds with equality at $%
s_{\ell }=s(z_{\ell })$ and $z_{\ell }.$

\section{Proof of Theorem \protect\ref{thm_differentiable_sep_eq}}

We first show that $\sigma $ is continuous on $[z_{\ell },\bar{z}]$. To
prove that, we start by showing that $\sigma (z)\geq \zeta (x,z)$, the
bilaterally efficient level of type $z$'s action, where $x$ is the type of
the receiver who is matched with the sender of type $z$ in equilibrium.

\begin{lemma}
\label{lemma1_in_thm1}For all $z\in \lbrack z_{\ell },\bar{z}]$, $\sigma
(z)\geq \zeta (x,z)$ in any stronger monotone separating CSE, where $x$ is
the type of the receiver who matches with type $z.$
\end{lemma}

\begin{proof}
We prove by contradiction. Suppose that there exists $z\in \lbrack z_{\ell },%
\bar{z}]$ such that $\sigma (z)<\zeta (x,z)$. There are two possible cases.
The first case is when $\sigma (z)<\zeta (x,z)$ and $\zeta (x,z)\notin
S^{\ast }.$ Then, it is a profitable sender deviation by type $z$ to an
off-path action $\zeta (x,z)$ if 
\begin{equation}
\mathbb{E}_{\mu \left( \zeta (x,z)\right) }\left[ v\left( x,\zeta
(x,z),z^{\prime }\right) \right] -c\left( \zeta (x,z),z\right) >v\left(
x,\sigma (z),z\right) -c\left( \sigma (z),z\right) .  \label{thm1_1}
\end{equation}%
Because of the constrained efficiency of $\zeta (x,z)$ given the strict
concavity of $v-c$ in $s$ (Assumption 3), we have that 
\begin{equation}
v\left( x,\zeta (x,z),z\right) -c\left( \zeta (x,z),z\right) >v\left(
x,\sigma (z),z\right) -c\left( \sigma (z),z\right) .  \label{thm1_2}
\end{equation}%
Further, because $\sigma (z)<\zeta (x,z),$ we have $z^{\prime }\geq z$ for
all $z^{\prime }\in $ supp $\mu \left( \zeta (x,z)\right) $ due to the
stronger monotonicity of $\mu $. Therefore, we have 
\begin{equation}
\mathbb{E}_{\mu \left( \zeta (x,z)\right) }\left[ v\left( x,\zeta
(x,z),z^{\prime }\right) \right] \geq v\left( x,\zeta (x,z),z\right) .
\label{thm1_3}
\end{equation}%
Because of (\ref{thm1_2}) and (\ref{thm1_3}), (\ref{thm1_1}) holds.

Therefore, if there exists $z\in \lbrack z_{\ell },\bar{z}]$ such that $%
\sigma (z)<\zeta (x,z)$, then it must be the second case where $\zeta
(x,z)\in S^{\ast }.$ This implies that there exists $z^{\prime }>z$ such
that $\sigma (z^{\prime })=\zeta (x,z)<\zeta (x^{\prime },z^{\prime }),$
where $x^{\prime }$ is the type of the receiver who matches with type $%
z^{\prime }$, given the increasing property of $\sigma $ in a stronger
monotone separating CSE (implication of Lemma \ref{lemma_monotone_eq_A'}%
.(i)). Because we have $\sigma (z)<\zeta (x,z)$ and $\sigma (z^{\prime
})<\zeta (x^{\prime },z^{\prime }),$ there exists an interval $%
(z_{1},z_{2})\subset (z,z^{\prime })$ such that for all $z^{\prime \prime
}\in (z_{1},z_{2}),$ $\sigma (z^{\prime \prime })<\zeta (x^{\prime \prime
},z^{\prime \prime }),$ where $x^{\prime \prime }$ is the type of the
receiver who matches with type $z^{\prime \prime }$ in equilibrium.

$\sigma $ is increasing on $(z_{1},z_{2})$ in any stronger monotone
separating CSE. Further, $\sigma $ is finite on $(z_{1},z_{2})$ because $%
\sigma (z^{\prime \prime })\leq \zeta (x_{2},z_{2}),$ where $x_{2}$ is the
type of the receiver who matches with type $z_{2}$ in equilibrium. One can
invoke Theorem 7.21 in Wheeden and Zygmund (1977) to show that $\sigma $ is
differentiable with non-negative derivative $\sigma ^{\prime }$ almost
everywhere on $\left( z_{1},z_{2}\right) $. Because $\sigma $ is strictly
increasing in $s\in S^{\ast }$ in a stronger monotone separating CSE, $%
\sigma ^{\prime }$ must be in fact positive almost everywhere on $\left(
z_{1},z_{2}\right) .$

Because $\sigma $ is differentiable with positive $\sigma ^{\prime }$ almost
everywhere on $\left( z_{1},z_{2}\right) $, we can find an interval $%
(z_{1}^{\prime },z_{2}^{\prime })\subset \left( z_{1},z_{2}\right) $ such
that $\sigma $ is differentiable with positive $\sigma ^{\prime }$
everywhere on $\left( z_{1}^{\prime },z_{2}^{\prime }\right) .$ Because $%
\sigma $ is differentiable with positive $\sigma ^{\prime }$ everywhere on $%
\left( z_{1}^{\prime },z_{2}^{\prime }\right) ,$ $\left\{ \sigma (z^{\prime
\prime }):z^{\prime \prime }\in \left( z_{1}^{\prime },z_{2}^{\prime
}\right) \right\} $ is an interval $\left( \sigma \left( z_{1}^{\prime
}\right) ,\sigma \left( z_{2}^{\prime }\right) \right) $ and $\mu $ is
differentiable with positive $\mu ^{\prime }=1/\sigma ^{\prime }$ everywhere
on $\left( \sigma \left( z_{1}^{\prime }\right) ,\sigma \left( z_{2}^{\prime
}\right) \right) $. On the other hand, $\tau :S^{\ast }\rightarrow \lbrack
t_{\ell },t_{h}]$ is increasing according to Lemma \ref{lemma_monotone_eq_A'}%
.(iii). Invoking Theorem 7.21 in Wheeden and Zygmund (1977), we can show
that $\tau $ is differentiable with non-negative $\tau ^{\prime }$ almost
everywhere on $\left( \sigma \left( z_{1}^{\prime }\right) ,\sigma \left(
z_{2}^{\prime }\right) \right) $.

Finally, we can pick an interval $\left( z_{1}^{\circ },z_{2}^{\circ
}\right) \subset \left( z_{1}^{\prime },z_{2}^{\prime }\right) $ such that
(i) $\sigma $ is differentiable with positive $\sigma ^{\prime }$ everywhere
on $\left( z_{1}^{\circ },z_{2}^{\circ }\right) $, (ii) $\mu $ is
differentiable with positive $\mu ^{\prime }=1/\sigma ^{\prime }$ everywhere
on $\left( \sigma \left( z_{1}^{\prime }\right) ,\sigma \left( z_{2}^{\prime
}\right) \right) $ and (iii) $\tau $ is differentiable with non-negative
derivative everywhere on $\left( \sigma \left( z_{1}^{\circ }\right) ,\sigma
\left( z_{2}^{\circ }\right) \right) .$ It implies that for all $z^{\prime
\prime }\in \left( z_{1}^{\circ },z_{2}^{\circ }\right) $, the following
first-order condition must be satisfied: 
\begin{equation}
\tau ^{\prime }(\sigma (z^{\prime \prime }))-c_{s}\left( \sigma (z^{\prime
\prime }),z^{\prime \prime }\right) =0\text{.}  \label{thm1-C}
\end{equation}%
Because (1) $\mu (s)$ is differentiable everywhere in $\left( \sigma \left(
z_{1}^{\circ }\right) ,\sigma \left( z_{2}^{\circ }\right) \right) $ and (2) 
$\tau $ is differentiable everywhere on $\left( \sigma \left( z_{1}^{\circ
}\right) ,\sigma \left( z_{2}^{\circ }\right) \right) $, $\pi $ is
differentiable everywhere on $\left( \sigma \left( z_{1}^{\circ }\right)
,\sigma \left( z_{2}^{\circ }\right) \right) $. If type $x^{\prime \prime }$
chooses a sender with $s\in \left( \sigma \left( z_{1}^{\circ }\right)
,\sigma \left( z_{2}^{\circ }\right) \right) $, the following first-order
condition must be satisfied: 
\begin{equation}
\pi _{s}(s,x^{\prime \prime })=v_{s}\left( x^{\prime \prime },s,\mu
(s)\right) +v_{z}\left( x,s,\mu (s)\right) \mu ^{\prime }(s^{\prime \prime
})-\tau ^{\prime }\left( s^{\prime \prime }\right) =0  \label{thm1-D}
\end{equation}%
Let type $z^{\prime \prime }$ choose $s=\sigma (z^{\prime \prime })\in
\left( \sigma \left( z_{1}^{\circ }\right) ,\sigma \left( z_{2}^{\circ
}\right) \right) ,$ which type $x^{\prime \prime }$ chooses. Combining (\ref%
{thm1-C}) and (\ref{thm1-D}) yields that $v_{s}\left( x,s,\mu (s)\right)
-c_{s}\left( \sigma (z^{\prime \prime }),z^{\prime \prime }\right)
+v_{z}\left( x^{\prime \prime },s,\mu (s)\right) \mu ^{\prime }(s)=0,$ which
cannot hold. The reason is that (a) $v_{s}-c_{s}>0$ because $v-c$ is
strictly concave (Assumption 3) and $\sigma (z^{\prime \prime })<\zeta
(x^{\prime \prime },z^{\prime \prime }),$ (b) $\mu ^{\prime }\geq 0$ and (c) 
$v_{z}>0$ (Assumption 5). Therefore, we cannot have the case where $\sigma
(z)<\zeta (x,z)$, and $\zeta (x,z)\in S^{\ast }.$ This concludes the proof.
\end{proof}

\begin{lemma}\label{lm:conti_sigma}
\label{lemma2_in_thm1}$\sigma $ is continuous on $[z_{\ell },\bar{z}]$ and
hence $S^{\ast }=[s_{\ell },\sigma \left( \bar{z}\right) ]$ in any stronger
monotone separating CSE
\end{lemma}

\begin{proof}
We prove by contradiction. Suppose that $\sigma $ is discontinuous at some $%
z\in \lbrack z_{\ell },\bar{z}]$. We consider the case where $\sigma $ is
only right continuous at $z,$ so that $\sigma \left( z_{+}\right) =\sigma
\left( z\right) >\sigma \left( z_{-}\right) .$ Because $\sigma (z)\geq \zeta
(x,z)$ for all $z\in \lbrack z_{\ell },\bar{z}]$ by Lemma \ref%
{lemma1_in_thm1}, it implies that $\sigma \left( z\right) >\zeta (x,z),$
where $x$ is the type of a receiver who matches with type $z.$ This
discontinuity creates an off-path action interval $[\sigma \left(
z_{-}\right) ,\sigma \left( z\right) )$. The stronger monotone belief $\mu $
implies that $\mu (s)$ puts all the weights on $z$ conditional on $s\in
\lbrack \sigma \left( z_{-}\right) ,\sigma \left( z\right) )$ because of
Lemma \ref{theorem_monotone_belief}.1. Because $\sigma \left( z\right) $ is
inefficiently high (i.e., $\sigma \left( z\right) >\zeta (x,z)$), there
exists $s\in \lbrack \sigma \left( z_{-}\right) ,\sigma \left( z\right) )$
such that 
\begin{equation}
v(x,s,z)-c(s,z)>v(x,\sigma \left( z\right) ,z)-c(\sigma \left( z\right) ,z),
\label{thm1_4}
\end{equation}%
due to the strict concavity of $v-c$ in $s$ (Assumption 3). (\ref{thm1_4})
shows the existence of a profitable sender deviation by type $z$ to an off
path action $s\in \lbrack \sigma _{-}\left( z\right) ,\sigma \left( z\right)
)$. One can analogously show the existence of a profitable sender deviation
by type $z$ in the case where $\sigma $ is only left continuous at $z$ or $%
\sigma \left( z_{-}\right) <\sigma \left( z\right) <\sigma \left(
z_{+}\right) .$ Therefore, $\sigma $ is continuous at all $z\in \lbrack
z_{\ell },\bar{z}].$

Because $\sigma $ is increasing over $[z_{\ell },\bar{z}]$ in any stronger
monotone separating CSE, the continuity of $\sigma $ at all $z\in \lbrack
z_{\ell },\bar{z}]$ implies a compact real interval $S^{\ast }=[s_{\ell
},\sigma \left( \bar{z}\right) ]$
\end{proof}

\begin{lemma}
\label{lemma3_in_thm1}$\tau :S^{\ast }\rightarrow T$ is increasing and
continuous on $S^{\ast }$ and has continuous derivative $\tau ^{\prime }$ on
Int $S^{\ast }$ in any stronger monotone separating CSE.
\end{lemma}

\begin{proof}
The increasing property of $\tau $ is from Lemma \ref{lemma_monotone_eq_A'}%
.(iii). We prove the continuity of $\tau $ by contradiction. Suppose that $%
\tau $ is discontinuous at $s\in S^{\ast }.$ Consider the case where $\tau $
is only right continuous at $s$. Let $z$ be the type of a sender who chooses 
$s$ in equilibrium, i.e., $\sigma (z)=s$. Because $\tau (\sigma \left(
z_{-}\right) )<\tau (\sigma \left( z\right) )$ and $c$ and $\sigma $ are
continuous (Assumption 5.(i) and Lemma \ref{lemma2_in_thm1}), there exists $%
z^{\prime }<z$ such that $\tau (\sigma \left( z^{\prime }\right) )-c\left(
\sigma \left( z^{\prime }\right) ,z^{\prime }\right) <\tau (\sigma \left(
z\right) )-c\left( \sigma \left( z\right) ,z^{\prime }\right) ,$ which
contradicts the optimality of $\sigma \left( z^{\prime }\right) $ for type $%
z^{\prime }.$ We can analogously prove that the discontinuity of $\tau $
contradicts the optimality of the sender's action choice in the case where $%
\tau $ is left right continuous at $s$ or $\tau (s_{-})<\tau (s)<\tau
(s_{+}) $. Therefore, $\tau :S^{\ast }\rightarrow T$ is continuous
everywhere on $S^{\ast }$.

We prove the differentiability by contradiction as well. Suppose that $\tau $
is not differentiable at some $\hat{s}\in $ Int $S^{\ast }=\left( s_{\ell
},\sigma \left( \bar{z}\right) \right) $.

$\tau :S^{\ast }\rightarrow T$ is increasing and hence it is not
differentiable only at finitely many points in Int $S^{\ast }=\left( s_{\ell
},\sigma \left( \bar{z}\right) \right) $ due to Theorem 7.21 in Wheeden and
Zygmund (1977). This implies that if $\tau $ is not differentiable at $\hat{s%
}$, there exists two intervals $(s_{1},\hat{s}),$ $(\hat{s},s_{2})\subset $
Int $S^{\ast }$ where $\tau $ is differentiable. Because $\tau $ is
differentiable at any point in $(s_{1},\hat{s})\cup (\hat{s},s_{2})$ and $c$
is differentiable everywhere (Assumption 4), the optimality of $s=\sigma (z)$
implies that that the first-order condition $\tau ^{\prime }(s)=c_{s}\left(
s,\mu (s)\right) $ for all $s=\sigma (z)\in (s_{1},\hat{s})\cup (\hat{s}%
,s_{2})$. Because $c_{s}$ is continuous (Assumption 4) and $\mu =\sigma
^{-1} $ is continuous on $S^{\ast }$, this implies that 
\begin{equation}
\tau ^{\prime }(\hat{s}_{-})=c_{s}\left( \hat{s},\mu (\hat{s})\right) =\tau
^{\prime }(\hat{s}_{+}).  \label{thm1_6}
\end{equation}%
Because $\tau $ is continuous, (\ref{thm1_6}) implies that $\tau $ is
differentiable at $\hat{s},$ which contradicts the non-differentiability of $%
\tau $ at $\hat{s}.$ Therefore, $\tau $ must be differentiable everywhere on
Int $S^{\ast }.$

Because $\tau $ is differentiable everywhere on Int $S^{\ast },$ the
first-order condition $\tau ^{\prime }\left( s\right) =c_{s}\left( s,\mu
(s)\right) $ must be satisfied for all $s\in $ Int $S^{\ast }$ in
equilibrium. $\mu $ is continuous on $S^{\ast }$ because it is the inverse
of $\sigma $ over $S^{\ast }$ and $\sigma $ is continuous (Lemma \ref%
{lemma2_in_thm1}). Further $c_{s}$ is continuous (Assumption 5.(i)).
Therefore, $\tau ^{\prime }\left( s\right) =c_{s}\left( s,\mu (s)\right) $
is continuous on Int $S^{\ast }.$
\end{proof}

\begin{lemma}
\label{lem_differentiable_belief}$\mu :S\rightarrow \Delta (Z)$ is
increasing and continuous on $S^{\ast }$ and has continuous derivative $\mu
^{\prime }$ on Int $S^{\ast }.$
\end{lemma}

\begin{proof}
$\sigma $ is continuous on $Z$ and $S^{\ast }$ is a compact real interval $%
[\sigma (z_{\ell }),\sigma \left( \bar{z}\right) ]$ (Lemma \ref%
{lemma2_in_thm1}). Given Lemma \ref{lemma_monotone_eq_A'}.(i), $\sigma $ is
increasing over $Z$ in a stronger monotone \emph{separating} CSE. Therefore,
Lemma \ref{lemma2_in_thm1} implies that $\mu \left( s\right) $ (the support
of $\mu \left( s\right) $ to be precise) for all $s\in $ $S^{\ast }$ is the
inverse of $\sigma (z)$ so that $\mu $ is increasing and continuous on $%
S^{\ast }$.

Because $\mu $ is increasing on $S^{\ast }$and $\mu (s)\in Z$ for $s\in
S^{\ast },$ we can apply Theorem 7.21 in Wheeden and Zygmund (1977) to show
that $\mu $ is differentiable almost everywhere on Int $S^{\ast }$. Let us
prove that $\mu $ is differentiable everywhere on Int $S^{\ast }.$ Suppose
that $\mu $ is not differentiable at $\check{s}\in $ Int $S^{\ast }$.
Because $\mu $ is not differentiable at only finitely many points, there
exists $s_{1},s_{2}\in $ Int $S^{\ast }$ such that $\mu $ is differentiable
everywhere on $(s_{1},\check{s})$ and $(\check{s},s_{2})$.

Because (1) $v$ is differentiable with respect to $s$ and $z$ (Assumptions
3.(ii)), (2) $\mu $ is differentiable everywhere on $(s_{1},\check{s})$ and $%
(\check{s},s_{2})$, and (3) $\tau $ is differentiable everywhere on Int $%
S^{\ast }$ (Lemma \ref{lemma3_in_thm1}), $\pi (s,x):=v\left( x,s,\mu
(s)\right) -\tau \left( s\right) ,$ $x$ is differentiable everywhere on $%
(s_{1},\check{s})\cup (\check{s},s_{2})$.

Let $x$ be the type of a receiver who matches with a sender with $s=\xi (x).$
For the receiver's matching problem, the following first-order condition is
satisfied: for all $s=\xi (x)\in (s_{1},\check{s})\cup (\check{s},s_{2})$:%
\begin{equation}
\pi _{s}(s,x)=v_{s}\left( x,s,\mu (s)\right) +v_{z}\left( x,s,\mu (s)\right)
\mu ^{\prime }(s)-\tau ^{\prime }(s)=0  \label{thm1_8}
\end{equation}

Suppose that $\xi (x)=\xi (x^{\prime })=s$ for some $s\in S^{\ast }$ with $%
x>x^{\prime }$. It implies that $\xi (x^{\prime \prime })=s$ for all $%
x^{\prime \prime }\in \lbrack x,x^{\prime }]$ because $\xi $ is
non-decreasing in a stronger monotone CSE (Theorem \ref%
{theorem_stronger_monotone_eq}) given Assumptions 1 and 2. Then, the market
clearing condition is not satisfied because $H\left( [x,x^{\prime }]\right)
>G(\{s\})=0$. Therefore, $\xi $ is increasing on $X$.

Then, the \emph{market-clearing condition} implies that $\xi =\sigma \circ
n^{-1}.$ Because $\sigma $ is continuous on $Z$ (Lemma \ref{lemma2_in_thm1})
and $n^{-1}$ is continuous on $X$ (implication of Assumption 6), $\xi $ is
continuous on $[z_{\ell },\bar{z}].$ Because $\xi $ is increasing and
continuous on $Z$, $\xi ^{-1}=n\circ \mu $ is increasing and continuous on $%
S^{\ast }.$

Given $v_{z}>0$ (Assumption 3.(i)), replacing $x$ with $\xi ^{-1}(s)$ in (%
\ref{thm1_8}) yields that 
\begin{equation}
\mu ^{\prime }(s)=\frac{-\left[ v_{s}\left( \xi ^{-1}(s),s,\mu (s)\right)
-\tau ^{\prime }(s)\right] }{v_{z}\left( \xi ^{-1}(s),s,\mu (s)\right) },%
\text{ }\forall s\in (s_{1},\check{s})\cup (\check{s},s_{2}).  \label{thm1_9}
\end{equation}%
In addition to the continuity of $\xi ^{-1}$ on $S^{\ast },$ $v_{s}$, $v_{z}$%
, $\tau ^{\prime }$, and $\mu $ are continuous (Assumption 3.(ii) and Lemmas %
\ref{lemma2_in_thm1} and \ref{lemma3_in_thm1}). Therefore, from (\ref{thm1_9}%
), we have that 
\begin{equation}
\mu ^{\prime }(\check{s}_{-})=\frac{-\left[ v_{s}\left( \xi ^{-1}(\check{s}),%
\check{s},\mu (\check{s})\right) -\tau ^{\prime }(\check{s})\right] }{%
v_{z}\left( \xi ^{-1}(\check{s}),\check{s},\mu (\check{s})\right) }=\mu
^{\prime }(\check{s}_{+})  \label{thm1_10}
\end{equation}%
Because $\mu $ is continuous on $S^{\ast }$, (\ref{thm1_10}) implies that $%
\mu $ is differentiable at $\check{s},$ which contradicts the
non-differentiability of $\mu $ at $\check{s}$. Therefore, $\mu $ must be
differentiable everywhere on Int $S^{\ast }$ in any stronger monotone
separating CSE. Further, the continuity of $v_{s}$, $v_{z}$, $\tau ^{\prime
} $, $\xi ^{-1},$ and $\mu $ implies that $\mu ^{\prime }$ is continuous on
Int $S^{\ast }$.
\end{proof}

\section{Proof of Lemma \protect\ref{lemma_no_bottom_bunching}}

Let $Z(s)$ be the set of the types of senders who choose the same action $s$
and it has a positive measure. We start with the case where there exists $%
\max Z(s)$. Let $z^{\circ }:=\max Z(s).$ We first show that $z^{\circ }=%
\overline{z}$. Let $x^{\circ }:=\max X(s)$, where $X(s)$ be the set of types
of receivers who are matched with a sender with $s$ in equilibrium. We prove
by contradiction. Suppose that bunching does not happen on the top, i.e., $%
z^{\circ }<\overline{z}$. Then we have that 
\begin{equation}
s=\sigma (z^{\circ })\leq \lim_{z\searrow z^{\circ }}\sigma (z)
\label{no_bottom_bunching1}
\end{equation}%
This is due to the monotonicity of $\sigma $ in Lemma \ref%
{lemma_monotone_eq_A'}.(i). We like to show that (\ref{no_bottom_bunching1})
holds with strict inequality, i.e., $s<\lim_{z\searrow z^{\circ }}\sigma (z)$%
. In equilibrium, we have that for any $z>z^{\circ },$%
\begin{gather}
\tau (s)-c(s,z^{\circ })\geq \lim_{z\searrow z^{\circ }}\left[ \tau (\sigma
(z))-c(\sigma (z),z^{\circ })\right]  \label{no_bottom_bunching1_0} \\
\mathbb{E[}v(x^{\circ },s,z^{\prime }|z^{\prime }\in Z(s)]-\tau (s)\geq
\lim_{z\searrow z^{\circ }}\left( \mathbb{E[}v(x^{\circ },\sigma
(z),z^{\prime \prime }|z^{\prime \prime }\in Z(\sigma (z))]-\tau (\sigma
(z))\right)  \label{no_bottom_bunching1_1'}
\end{gather}%
For any $\sigma (z)\geq s$, we have that $z^{\prime \prime }\geq z^{\circ
}=\max Z(s)$ for any $z^{\prime \prime }\in Z(\sigma (z))$ because of the
monotonicity of $\sigma $ (Lemma \ref{lemma_monotone_eq_A'}.(i)). Further $%
Z(s)$ has a positive measure. Therefore, the monotonicity of $v$ in
Assumption 3.(i) implies that, for any $z>z^{\circ }$ 
\begin{equation}
\mathbb{E[}v(x^{\circ },s,z^{\prime }|z^{\prime }\in Z(s)]<\mathbb{E[}%
v(x^{\circ },\sigma (z),z^{\prime \prime }|z^{\prime \prime }\in Z(\sigma
(z))].  \label{no_bottom_bunching1_2}
\end{equation}%
(\ref{no_bottom_bunching1_1'}) and (\ref{no_bottom_bunching1_2}) imply that 
\begin{equation}
\tau (s)<\lim_{z\searrow z^{\circ }}\tau (\sigma (z))
\label{no_bottom_bunching1_3}
\end{equation}%
Because $c$ is decreasing in $s$ (Assumption 1.(i)), (\ref%
{no_bottom_bunching1_0}) and (\ref{no_bottom_bunching1_3}) induces that%
\begin{equation}
s=\sigma (z^{\circ })<\lim_{z\searrow z^{\circ }}\sigma (z).
\label{no_bottom_bunching1_4}
\end{equation}%
Therefore, any $s^{\prime }\in (s,\lim_{z\searrow z^{\circ }}\sigma (z))$ is
not chosen in equilibrium given that the monotonicity of $\sigma $.

The support of $\mu (s)$ is $Z(s)$. On the other hand, we have that $%
\lim_{z\searrow z^{\circ }}\inf $ supp$(\mu (\sigma (z)))=z^{\circ }$. This
implies that there is the unique stronger monotone belief on the sender's
type conditional on any $s^{\prime }\in (s,\lim_{z\searrow z^{\circ }}\sigma
(z))$ and it is equal to $\mu (s^{\prime })=z^{\circ }$.

Suppose that the sender of type $z^{\circ }$ deviates to action $s+\epsilon
\in (s,\lim_{z\searrow z^{\circ }}\sigma (z))$. A receiver of type $x$ who
is currently matched with a sender with $s$ receives the matching utility of 
$\mathbb{E[}v(x,s,z|z\in Z(s)]-\tau (s)$. Note that $\tau (s)<t_{h}$ given (%
\ref{no_bottom_bunching1_3}). Therefore, there is a profitable deviation for
the sender of type $z^{\circ }$ if 
\begin{equation}
v(x,s+\epsilon ,z^{\circ })-c(s+\epsilon ,z^{\circ })>\mathbb{E[}%
v(x,s,z|z\in Z(s)]-c(s,z^{\circ }).  \label{no_bottom_bunching2}
\end{equation}%
Because $v$ and $c$ are continuous in the sender's action, $v$ is increasing
in $z$, and $Z(s)$ has a positive measure, we have that 
\begin{equation}
\lim_{\epsilon \searrow 0}\left( v(x,s+\epsilon ,z^{\circ })-c(s+\epsilon
,z^{\circ })\right) >\mathbb{E[}v(x,s,z|z\in Z(s)]-c(s,z^{\circ })
\label{no_bottom_bunching3}
\end{equation}%
Because $v$ and $c$ are continuous in the sender's action, (\ref%
{no_bottom_bunching3}) implies that there exists $\epsilon $ such that (\ref%
{no_bottom_bunching2}) is satisfied. This contradicts that $s$ is an
equilibrium signal chosen by all senders whose types are in $Z(s)$.

We can analogously prove that there exists a profitable sender deviation if $%
z^{\circ }<\overline{z}$ when $z^{\circ }$ is defined as $\sup Z(s)$ rather
than $\max Z(s)$.

Assumption 6 implies that there is no atom in the sender type distribution.
Therefore, $Z(s)$ is an interval with $\max Z(s)=\overline{z}$ due to the
monotonicity of $\sigma $ (Lemma \ref{lemma_monotone_eq_A'}.(i)).

\section{Proof of Lemma \protect\ref{lemma_binding_upper_bound}}

Let $z^{\ast }$ be the minimum of $Z(s)$ (We can analogously prove the lemma
for the case where $z^{\ast }$ is infimum of $Z(s)$). If $Z(s)$ has a
positive measure, we have that $z^{\ast }<\overline{z}$ given Assumption 6
on $G.$ Let $t^{\ast }$ be the reaction to action $s$ chosen by the positive
measure of senders. We prove by contradiction.

On the contrary, suppose that $t^{\ast }<t_{h}$ in a stronger monotone CSE.
Because type $\overline{z}$ is one of senders who choose $s$ and $\overline{z%
}$ is the maximum of sender types, the stronger monotonicity of $\mu $
implies that $\mu (s^{\prime })=\overline{z}$ for any $s^{\prime }>s.$
Suppose that the sender of type $\overline{z}$ deviates to $s+\epsilon $ for
small $\epsilon >0.$ The type of this sender is believed to be $\overline{z}$%
. Suppose that the receiver of type $\overline{x}$ is matched with the
sender with $s+\epsilon $. A profitable upward deviation for a sender is
equivalent to the existence of $t\in \lbrack t_{\ell },t_{h}]$ and $\epsilon
>0$ such that 
\begin{eqnarray}
v\left( \overline{x},s+\epsilon ,\overline{z}\right) -t &>&\mathbb{E}\left[
v\left( \overline{x},s,z^{\prime }\right) |z^{\ast }\leq z^{\prime }<%
\overline{z}\right] -t^{\ast },  \label{PSD_simple0_1} \\
t-c(s+\epsilon ,\overline{z}) &>&t^{\ast }-c(s,\overline{z}),
\label{PSD_simple0_2}
\end{eqnarray}%
which yield 
\begin{equation}
v\left( \overline{x},s+\epsilon ,\overline{z}\right) -c(s+\epsilon ,%
\overline{z})>\mathbb{E}\left[ v\left( \overline{x},s,z^{\prime }\right)
|z^{\ast }\leq z^{\prime }<\overline{z}\right] -c(s,\overline{z}).
\label{PSD_simple}
\end{equation}

Given $G^{\prime }(z)>0$ for all $z\in Z$ (Assumption 6), the monotonicity
of $v$ in $z$ (Assumption 3.(i)) implies that 
\begin{equation}
v\left( \overline{x},s,\overline{z}\right) -c(s,\overline{z})>\mathbb{E}%
\left[ v\left( \overline{x},s,z^{\prime }\right) |z^{\ast }\leq z^{\prime }<%
\overline{z}\right] -c(s,\overline{z}).  \label{PSD_simple1}
\end{equation}%
Because $v$ and $c$ are continuous in the sender action$,$ (\ref{PSD_simple1}%
) ensures the existence of $\epsilon >0$ that satisfies (\ref{PSD_simple}).
Because $t^{\ast }<t_{h}$, (\ref{PSD_simple}) implies that there exists $t$
such that $t^{\ast }<t<t_{h}$ and it satisfies (\ref{PSD_simple0_1}) and (%
\ref{PSD_simple0_2}). Therefore, the only way to prevent such an upward
deviation by the sender is $t^{\ast }=t_{h}$.

\section{Proof of Theorem \protect\ref{thm_uniqueSME}}

Lemma \ref{thm_unique_separating_eq} leads to the existence of the unique
stronger monotone separating CSE. The remaining question is whether there
are other stronger monotone CSEs. Lemma \ref{lemma_no_bottom_bunching} is
still valid. Therefore, if there is bunching in sender action $s$, it must
be among senders in a type interval $Z(s)$ with $\overline{z}$ as its
maximum. However, such bunching is not sustained because if we follow the
logic in the proof of Lemma \ref{lemma_binding_upper_bound}, we can show
that there is a profitable small upward deviation from $s$ for the sender of
type-$\overline{z}$. Therefore, there is no additional stronger monotone CSE.

\section{Proof of Theorem \protect\ref{lemma_unique_pooling}\label%
{App_lemma_unique_pooling}}

Fix any single feasible reaction $0\leq t^{\ast }<\overline{t}.$ Then any
stronger monotone CSE is pooling because any action choice is rewarded by a
single reaction $t^{\ast }.$

Now consider $0<t^{\ast }<\overline{t}.$ We first show that $z^{\ast }>%
\underline{z}$. On the contrary, suppose that $z^{\ast }=\underline{z}.$
Then, $s^{\ast }=0.$ Otherwise (i.e., $s^{\ast }>0$), (\ref{pooling_sender1}%
) is not satisfied because $\lim_{z\rightarrow \underline{z}}c(s,z)=\infty $
for all $s>0$ in Assumption 7. If Assumption 7.(i) is satisfied, $\mathbb{E}%
\left[ v\left( n\left( z^{\ast }\right) ,s^{\ast },z^{\prime }\right)
|z^{\prime }\geq z^{\ast }\right] =0$ because $n\left( z^{\ast }\right) =%
\underline{x}.$ Therefore, (\ref{pooling_receiver1}) is not satisfied. If
Assumption 7.(ii) is satisfied, $s^{\ast }=0$ implies that $\mathbb{E}\left[
v\left( n\left( z^{\ast }\right) ,s^{\ast },z^{\prime }\right) |z^{\prime
}\geq z^{\ast }\right] =0.$ Therefore, (\ref{pooling_receiver1}) is not
satisfied. It means that if $t^{\ast }>0,$ then $z^{\ast }>\underline{z}$.
Given $t^{\ast }>0,$ let $z^{\ast }>\underline{z}$ be the threshold sender
type in a pooling CSE and hence (\ref{pooling_sender1}) and (\ref%
{pooling_receiver1}), each with equality. It means that $s^{\ast }>0.$ If a
sender reduces her action below $s^{\ast },$ the stronger monotone belief
implies that her type is believed to be $z^{\ast }$ and no receiver is
willing to match with her at $t^{\ast }.$ Furthermore, no sender wants to
choose her action above $s^{\ast }$ because the reaction is fixed to $%
t^{\ast }.$ Given binding (\ref{pooling_sender1}) and (\ref%
{pooling_receiver1}), no agent who stays out of the market wants to enter it
and vice versa no agent who enters the market stays out of the market.

If $t^{\ast }=0,$ then we must have $s^{\ast }=0.$ Otherwise the sender with 
$s^{\ast }$ will have utility less than her reservation utility. Suppose
that Assumption 7.(i) is satisfied. Then, every receiver of type above $%
\underline{x}$ gets positive (expected) utility by matching with a sender.
Therefore, every receiver wants to enter the market. Then, every sender must
enter the market as well. Therefore, $z^{\ast }=\underline{z}.$ Suppose that
Assumption 7.(ii) is satisfied. Because $s^{\ast }=0,$ any receiver who is
matched with a sender with $s^{\ast }=0$ gets the same utility as his
reservation utility. All receivers and sender receive zero utility
regardless of their decisions on market entry (So, the aggregate net surplus
is always zero). Because agents enter the market whenever they are
indifferent between entering the market and staying out, agents enter the
market, $z^{\ast }=\underline{z}.$ It is clear to see that no agent want to
leave the market.

\section{Proof of Lemma \protect\ref{lemma_pooling}}

By Theorem \ref{lemma_unique_pooling}, we must have $s^{\ast }>0.$ Let $%
V(n(z),s,z):=\mathbb{E}\left[ v\left( n\left( z\right) ,s,z^{\prime }\right)
|z^{\prime }\geq z\right] .$ If $t^{\ast }>0,$ then $(s^{\ast },z^{\ast })$
must satisfy%
\begin{align}
t^{\ast }-c(s,z)& =0,  \label{lem100} \\
V(n(z),s,z)-t^{\ast }& =0.\text{ }  \label{lem200}
\end{align}%
First, we show that if there is a solution $(s^{\ast },z^{\ast })$ that
solves (\ref{lem100}) and (\ref{lem200}), it is unique. Consider the
receiver's indifference curve $\left\{ (s,z)\in 
\mathbb{R}
_{++}\times \left[ \underline{z},\overline{z}\right] :V\left( n\left(
z\right) ,s,z\right) -t^{\ast }=0\right\} $. The slope of this indifference
curve is 
\begin{equation}
\frac{dz}{ds}=-\frac{V_{s}}{V_{z}}=-\frac{\mathbb{E}v_{s}}{\mathbb{E}%
v_{x}n^{\prime }+\frac{\partial \mathbb{E}\left[ v|z^{\prime }>z\right] }{%
\partial z}}\leq 0.  \label{MRS_buyer}
\end{equation}%
Given $v_{s}\geq 0$, $v_{x}>0,$ $v_{z}>0$ (Assumptions 2.(ii), 3.(i), and
5), the sign above holds because of $n^{\prime }>0$. Note that $n^{\prime
}>0 $ holds because of Assumption 6 ($G^{\prime }(z)>0$ for all $z$ and $%
H^{\prime }(x)>0$ for all $x$). On the other hand, $\left\{ (s,z)\in 
\mathbb{R}
_{+}\times \left[ \underline{z},\overline{z}\right] :t^{\ast
}-c(s,z)=0\right\} $ is the sender's indifference curve based on (\ref%
{lem200}) and its slope is 
\begin{equation}
\frac{dz}{ds}=-\frac{c_{s}}{c_{z}}>0  \label{MRS_seller}
\end{equation}%
where the sign holds due to Assumption 1.(i). (\ref{MRS_buyer}) and (\ref%
{MRS_seller}) imply that the two indifference curves intersect at most once
and the intersection becomes a unique solution for the system of equations, (%
\ref{lem100}) and (\ref{lem200}).

Now, we show the existence of a solution satisfying (\ref{lem100}) and (\ref%
{lem200}). Putting (\ref{lem100}) and (\ref{lem200}) together gives 
\begin{equation}
V\left( n\left( z\right) ,s,z\right) -c\left( s,z\right) =0  \label{lemmaA-3}
\end{equation}%
for all $\left( z,s\right) \in (\underline{z},\overline{z})\times 
\mathbb{R}
_{++}.$ Assumption 3 and 4 implies that the left hand side of (\ref{lemmaA-3}%
) is continuously differentiable in the rectangle $(\underline{z},\overline{z%
})\times 
\mathbb{R}
_{++}.$ The left hand side of (\ref{lemmaA-3}) is nonnegative at $s=0$. It
is also strictly negative as $s\rightarrow \infty $ because of Assumption 5.
Furthermore, we have that $\max_{s}\left[ V\left( n\left( z\right)
,s,z\right) -c\left( s,z\right) \right] >\max_{s}\left[ v(n\left( \underline{%
z}\right) ,s,\underline{z})-c\left( s,\underline{z}\right) \right] =0.$
Therefore, by the intermediate value theorem there exists $s(z)\in \mathbb{R}%
_{++}$ that satisfies (\ref{lemmaA-3}). Furthermore, $s(z)$ is continuous.
To show that $s(z)$ is continuous, suppose towards a contradiction that $%
s(z) $ is not continuous. Then there is some $\tilde{z}\in (\underline{z},%
\overline{z})$ such that $s(\tilde{z})\neq \lim_{z\rightarrow \tilde{z}%
}s(z):=\tilde{s}.$ This leads to a contradiction since $0=\lim_{z\rightarrow 
\tilde{z}}\left[ V\left( n\left( z\right) ,s(z),z\right) -c\left(
s(z),z\right) \right] =V\left( n\left( \tilde{z}\right) ,\tilde{s},\tilde{z}%
\right) -c\left( \tilde{s},\tilde{z}\right) \neq 0.$ The first equality
holds because $s(z)$ is a solution to (\ref{lemmaA-3}). The second equality
is holds because $V\left( n\left( z\right) ,s,z\right) -c\left( s,z\right) $
is continuous by Assumption 3 and 4. Let $\Lambda (z):=V\left( n\left(
z\right) ,s(z),z\right) $. Since $v$ is continuous and $n(z)$ and $s(z)$ are
continuous, $\Lambda (z)$ is continuous. Because $\Lambda (\underline{z}%
)-t^{\ast }<0$ and $\Lambda (\overline{z})-t^{\ast }>0,$ there exists $%
z^{\ast }\in (\underline{z},\overline{z})$ such that $\Lambda (z^{\ast
})-t^{\ast }=0$ by the intermediate value theorem. Because of (\ref{lemmaA-3}%
), $\Lambda (z^{\ast })-t^{\ast }=0$ also implies that $t^{\ast
}-c(s(z^{\ast }),z^{\ast })=0,$ so that both (\ref{lem100}) and (\ref{lem200}%
) holds with equality at $s^{\ast }=s(z^{\ast })$ and $z^{\ast }.$

\section{Lemma \protect\ref{lemma2}\label{Appendix_first_lemma2}}

\begin{lemma}
\label{lemma2}If there exists a solution $(s_{h},z_{h})$ of (\ref%
{jumping_sellers}) and (\ref{jumping_buyers}) with $z_{\ell }<z_{h}<\bar{z}$%
, then $\tilde{\tau}\left( \tilde{\sigma}\left( z_{h}\right) \right) <t_{h}<%
\tilde{\tau}\left( \tilde{\sigma}\left( \bar{z}\right) \right) $ and $\tilde{%
\sigma}\left( z_{h}\right) <s_{h}<\tilde{\sigma}\left( \bar{z}\right) $%
.\bigskip
\end{lemma}

\begin{proof}
When all senders of type $z_{h}$ or above choose the same action $s_{h}$ in
equilibrium, we have that for all $z>z_{h}$%
\begin{equation}
t_{h}-c\left( s_{h},z\right) \geq \tilde{\tau}\left( \tilde{\sigma}\left(
z_{h}\right) \right) -c\left( \tilde{\sigma}\left( z_{h}\right) ,z\right) .
\label{jumping_sellers2}
\end{equation}%
Since (\ref{jumping_sellers}) holds at $(s_{h},z_{h})$, (\ref%
{jumping_sellers}) and (\ref{jumping_sellers2}) imply that for all $z>z_{h},$%
\begin{equation*}
-c\left( s_{h},z\right) +c\left( \tilde{\sigma}\left( z_{h}\right) ,z\right)
\geq -c\left( s_{h},z_{h}\right) +c\left( \tilde{\sigma}\left( z_{h}\right)
,z_{h}\right) ,
\end{equation*}%
which implies that $s_{h}\geq \tilde{\sigma}\left( z_{h}\right) $ by the
strict supermodularity of $-c$ (Assumption 1.(ii)). Because $s_{h}\geq 
\tilde{\sigma}\left( z_{h}\right) ,$ both (\ref{jumping_sellers}) and (\ref%
{jumping_sellers2}) imply that $t_{h}\geq \tilde{\tau}\left( \tilde{\sigma}%
\left( z_{h}\right) \right) .$

Because $s_{h}\geq \tilde{\sigma}\left( z_{h}\right) ,$ we have that $%
\mathbb{E}[v(n\left( z_{h}\right) ,s_{h},z^{\prime })|z^{\prime }\geq
z_{h}]>v(n\left( z_{h}\right) ,\tilde{\sigma}\left( z_{h}\right) ,z_{h}).$ (%
\ref{jumping_buyers}) at $(s,z)=(s_{h},z_{h})$ is written as $\mathbb{E}%
[v(n\left( z_{h}\right) ,s_{h},z^{\prime })|z^{\prime }\geq
z_{h}]-t_{h}=v(n\left( z_{h}\right) ,\tilde{\sigma}\left( z_{h}\right)
,z_{h})-\tilde{\tau}\left( \tilde{\sigma}\left( z_{h}\right) \right) $,
which implies $t_{h}>\tilde{\tau}\left( \tilde{\sigma}\left( z_{h}\right)
\right) $ given $\mathbb{E}[v(n\left( z_{h}\right) ,s_{h},z^{\prime
})|z^{\prime }\geq z_{h}]>v(n\left( z_{h}\right) ,\tilde{\sigma}\left(
z_{h}\right) ,z_{h})$. If $t_{h}>\tilde{\tau}\left( \tilde{\sigma}\left(
z_{h}\right) \right) ,$ (\ref{jumping_sellers}) at $(s_{h},z_{h})$ implies
that $s_{h}>\tilde{\sigma}\left( z_{h}\right) .$

For all $z,$ let $s_{h}^{s}(z)$ be the value of $s$ that satisfies $%
t_{h}-c\left( s_{h}^{s}(z),z\right) =\tilde{\tau}\left( \tilde{\sigma}\left(
z\right) \right) -c\left( \tilde{\sigma}\left( z\right) ,z\right) .$Because $%
t_{h}<\tilde{\tau}\left( \tilde{\sigma}\left( \bar{z}\right) \right) ,$ $%
s_{h}^{s}(\bar{z})<\tilde{\sigma}\left( \bar{z}\right) .$ On the other hand, 
$s_{h}=s_{h}^{s}(z_{h})<s_{h}^{s}(\bar{z})$ because $z_{h}<\bar{z}$ and $%
s_{h}^{s}$ is increasing in $z.$ Therefore, we have that $s_{h}<\tilde{\sigma%
}\left( \bar{z}\right) .$
\end{proof}

\section{Proof of Lemma \protect\ref{lemma_well_behaved}}

First, we show that if there is a solution $(s_{h},z_{h})$ of (\ref%
{jumping_sellers}) and (\ref{jumping_buyers}), it is unique. Consider the
following set for senders: 
\begin{equation}
\left\{ (s,z)\in 
\mathbb{R}
_{++}\times Z:t_{h}-c\left( s,z\right) =\tilde{\tau}\left( \tilde{\sigma}%
\left( z\right) \right) -c\left( \tilde{\sigma}\left( z\right) ,z\right) 
\text{, }s>\tilde{\sigma}\left( z\right) \right\}  \label{set1}
\end{equation}%
Because $(s_{h},z_{h})$ must satisfy (\ref{jumping_sellers}) and $s_{h}>%
\tilde{\sigma}\left( z_{h}\right) $ (Lemma \ref{lemma2}), $(s_{h},z_{h})$
must belong to the set in (\ref{set1}). Applying the envelope theorem for $%
\tilde{\sigma}$, to the total differential of $t_{h}-c\left( s,z\right) =%
\tilde{\tau}\left( \tilde{\sigma}\left( z\right) \right) -c\left( \tilde{%
\sigma}\left( z\right) ,z\right) $ yields the slope of the equation as%
\begin{equation}
\frac{dz}{ds}=-\frac{c_{s}(s,z)}{c_{z}(s,z)-c_{z}\left( \tilde{\sigma}\left(
z\right) ,z\right) }>0\text{ for all }s>\tilde{\sigma}\left( z\right) ,
\label{slope_sellers}
\end{equation}%
where the sign holds because $c_{s}>0$ and $c_{z}(s,z)-c_{z}\left( \tilde{%
\sigma}\left( z\right) ,z\right) <0$ due to Assumption 1.(ii).

Consider the following set for receivers: 
\begin{equation}
\left\{ (s,z)\in 
\mathbb{R}
_{++}\times Z:%
\begin{array}{c}
\mathbb{E}[v(n\left( z\right) ,s,z^{\prime })|z^{\prime }\geq
z]-t_{h}=v\left( n\left( z\right) ,\tilde{\sigma}\left( z\right) ,z\right) -%
\tilde{\tau}\left( \tilde{\sigma}\left( z\right) \right) , \\ 
s>\tilde{\sigma}\left( z\right)%
\end{array}%
\right\} \text{.}  \label{set2}
\end{equation}%
Because $(s_{h},z_{h})$ must satisfy (\ref{jumping_buyers}) and $s_{h}>%
\tilde{\sigma}\left( z_{h}\right) $ (Lemma \ref{lemma2}), $(s_{h},z_{h})$
must belong to the set in (\ref{set2}).

Applying the envelope theorem for $\tilde{\sigma},$ and $\tilde{\tau}$ to
the total differential of $\mathbb{E}[v(n\left( z\right) ,s,z^{\prime
})|z^{\prime }\geq z]-t_{h}=v\left( n\left( z\right) ,\tilde{\sigma}\left(
z\right) ,z\right) -\tilde{\tau}\left( \tilde{\sigma}\left( z\right) \right) 
$, we can express the slope of the equation as 
\begin{equation}
\frac{dz}{ds}=-\frac{\mathbb{E}v_{s}}{\mathbb{E}v_{x}n^{\prime }+\frac{%
\partial \mathbb{E}\left[ v|z^{\prime }>z\right] }{\partial z}-\left(
v_{x}n^{\prime }+v_{z}\right) }\leq 0\text{ for all }s>\tilde{\sigma}\left(
z\right) ,  \label{slope_buyers}
\end{equation}%
where the sign holds because $-\mathbb{E}v_{s}\leq 0$ ($v_{s}\geq 0$
according to Assumption 3.(i)) and the denominator is positive due to
Assumptions 2, 3.(i), and 3.(iv) given $n^{\prime }>0$.

(\ref{slope_sellers}) and (\ref{slope_buyers}) imply that the two sets in (%
\ref{set1}) and (\ref{set2}) have at most one element in common. This
implies that if there is a solution $(s_{h},z_{h})$ that solves (\ref%
{slope_sellers}) and (\ref{slope_buyers}), it must be unique.

Now, we demonstrate the existence of solution. (\ref{jumping_sellers}) and (%
\ref{jumping_buyers}) together become 
\begin{equation}
F(z,s)=0,  \label{lemmaA-4}
\end{equation}%
where $F\left( z,s\right) :=\mathbb{E}[v(n\left( z\right) ,s,z^{\prime
})|z^{\prime }\geq z]-v\left( n\left( z\right) ,\tilde{\sigma}\left(
z\right) ,z\right) -c\left( s,z\right) +c\left( \tilde{\sigma}\left(
z\right) ,z\right) .$ Note that $F\left( z,s\right) $ is strictly positive
at $s=\tilde{\sigma}\left( z\right) $ because $\mathbb{E}[v(n\left( z\right)
,\tilde{\sigma}\left( z\right) ,z^{\prime })|z^{\prime }\geq z]>v\left(
n\left( z\right) ,\tilde{\sigma}\left( z\right) ,z\right) $ and is negative
as $s\rightarrow \infty $ by Assumption 5. Because $F\left( z,s\right) $ is
continuously differentiable on $\left( z_{\ell },\overline{z}\right) \times
(s_{\ell },\tilde{\sigma}(\bar{z}))$ given Assumptions 3 and 4, this implies
that there exists $s(z)\in (\tilde{\sigma}\left( z\right) ,\tilde{\sigma}(%
\bar{z}))$ for each $\left( z_{\ell },\overline{z}\right) $ that satisfies (%
\ref{lemmaA-4}) by the intermediate value theorem.

To show $s(z)$ is continuously differentiable, for an arbitrary small $%
\varepsilon >0,$ pick a point $\left( \overline{z}-\varepsilon ,\tilde{\sigma%
}\left( \overline{z}-\varepsilon \right) \right) $ in the rectangle $\left(
z_{\ell },\overline{z}\right) \times (s_{\ell },\tilde{\sigma}(\bar{z})).$
Then, as $\varepsilon $ gets closer to zero, we have $F\left( \overline{z}%
-\varepsilon ,\tilde{\sigma}\left( \overline{z}-\varepsilon \right) \right) =%
\mathbb{E}[v(n\left( \overline{z}-\varepsilon \right) ,\tilde{\sigma}\left( 
\overline{z}-\varepsilon \right) ,z^{\prime })|z^{\prime }\geq \overline{z}%
-\varepsilon ]-v\left( n\left( \overline{z}-\varepsilon \right) ,\tilde{%
\sigma}\left( \overline{z}-\varepsilon \right) ,\overline{z}-\varepsilon
\right) -c\left( \tilde{\sigma}\left( \overline{z}-\varepsilon \right) ,%
\overline{z}-\varepsilon \right) +c\left( \tilde{\sigma}\left( \overline{z}%
-\varepsilon \right) ,\overline{z}-\varepsilon \right) =0.$ By continuity of 
$v_{s}$, $n$, $\tilde{\sigma}$, and $c_{s}$, we also have $F_{s}\left( 
\overline{z}-\varepsilon ,\tilde{\sigma}\left( \overline{z}-\varepsilon
\right) \right) =\mathbb{E}[v_{s}(n\left( \overline{z}-\varepsilon \right) ,%
\tilde{\sigma}\left( \overline{z}-\varepsilon \right) ,z^{\prime
})|z^{\prime }\geq \overline{z}-\varepsilon ]-c_{s}\left( \tilde{\sigma}%
\left( \overline{z}-\varepsilon \right) ,\overline{z}\right) <0$ according
to Assumption 5. Therefore, according to the implicit function theorem there
exists an open rectangle $I\times J\subset \left( z_{\ell },\overline{z}%
\right) \times (s_{\ell },\tilde{\sigma}(\bar{z}))$ that contains the point $%
\left( \overline{z}-\varepsilon ,\tilde{\sigma}\left( \overline{z}%
-\varepsilon \right) \right) $, and a unique continuously differentiable
function $s(z)\in J$ defined on $I$ that satisfies (\ref{lemmaA-4}).

Because $c$, $\tilde{\tau}$, $\tilde{\sigma}$, and $s(z)$ are continuous, $%
\Lambda \left( z,s(z)\right) :=t_{h}+c\left( s(z),z\right) -\tilde{\tau}%
\left( \tilde{\sigma}\left( z\right) \right) +c\left( \tilde{\sigma}\left(
z\right) ,z\right) $ is also continuous. Because $t_{h}-c\left( s(\overline{z%
}),\overline{z}\right) <\tilde{\tau}\left( \tilde{\sigma}\left( \overline{z}%
\right) \right) -c\left( \tilde{\sigma}\left( \overline{z}\right) ,\overline{%
z}\right) $, and $t_{h}-c\left( s(z_{\ell }),z_{\ell }\right) >\tilde{\tau}%
\left( \tilde{\sigma}\left( z_{\ell }\right) \right) -c\left( \tilde{\sigma}%
\left( z_{\ell }\right) ,z_{\ell }\right) ,$ there exists $z_{h}\in (z_{\ell
},\overline{z})$ such that $\Lambda \left( z_{h},s(z_{h})\right) =0$ by the
intermediate value theorem. Because $F(z_{h},s(z_{h}))=0,$ $\Lambda \left(
z_{h},s(z_{h})\right) =0$ implies that $\mathbb{E}[v(n\left( z_{h}\right)
,s(z_{h}),z^{\prime })|z^{\prime }\geq z_{h}]-t_{h}=v\left( n\left(
z_{h}\right) ,\tilde{\sigma}\left( z_{h}\right) ,z_{h}\right) -\tilde{\tau}%
\left( \tilde{\sigma}\left( z_{h}\right) \right) .$

\section{Proof of Theorem \protect\ref{thm_unique_well_behaved}\label%
{App_thm_unique_well_behaved}}

Theorem \ref{thm_unique_separating_eq} and Lemma \ref{lemmaA} and \ref%
{lemma_well_behaved} establish the existence of a unique (strictly)
well-behaved stronger monotone CSE $\left\{ \hat{\sigma},\hat{\mu},\hat{\tau}%
,\hat{m}\right\} $ characterized in Theorem \ref{theorem1}.

To show that there is no other stronger monotone CSE, note that there is no
separating CSE with $T=[t_{\ell },t_{h}]$ with $t_{\ell }<t_{h}<\tilde{\tau}%
\left( \tilde{\sigma}\left( \overline{z}\right) \right) .$ Therefore, it is
sufficient to show that there is no pooling CSE because of Lemma \ref%
{theorem_all_eq_w/o_separating}. On contrary, suppose that there exists a
pooling CSE. Because of Lemma \ref{lemma_binding_upper_bound}, $t_{h}$ is
the equilibrium reaction for senders with pooled action $s^{\ast }.$
Therefore, $x^{\ast }=n(z^{\ast })$ and the following system of equations is
satisfied in a pooling CSE: 
\begin{gather}
t_{h}-c(s^{\ast },z^{\ast })\geq 0,  \label{bottom_seller1_0} \\
\mathbb{E}\left[ v\left( n\left( z^{\ast }\right) ,s^{\ast },z^{\prime
}\right) |z^{\prime }\geq z^{\ast }\right] -t_{h}\geq 0,\text{ }
\label{bottom_buyer1_0}
\end{gather}%
where both inequalities hold with equality if $z^{\ast }>\underline{z}$.

Suppose that $z^{\ast }>\underline{z}$. Then, (\ref{bottom_seller1_0}) and (%
\ref{bottom_buyer1_0}) hold with equality. Further, because both $t_{h}$ and 
$z^{\ast }$ are positive, $s^{\ast }$ must be positive from (\ref%
{bottom_seller1_0}) with equality. On the other hand, there should be no
profitable downward deviation for senders. Therefore, 
\begin{equation*}
v(n(z^{\ast }),s,z^{\ast })-c(s,z^{\ast })\leq \mathbb{E}\left[ v\left(
n\left( z^{\ast }\right) ,s^{\ast },z^{\prime }\right) |z^{\prime }\geq
z^{\ast }\right] -c(s^{\ast },z^{\ast })\text{ for all }s<s^{\ast }
\end{equation*}%
Because (\ref{bottom_seller1_0}) and (\ref{bottom_buyer1_0}) hold with
equality, this becomes 
\begin{equation}
v(n(z^{\ast }),s,z^{\ast })-c(s,z^{\ast })\leq 0\text{ for all }s<s^{\ast }.
\label{no_pooling}
\end{equation}

If $v(x,s,z)$ satisfies Assumption 7(i), then, $v(n(z^{\ast }),0,z^{\ast
})-c(0,z^{\ast })=v(n(z^{\ast }),0,z^{\ast })>0.$ Therefore, (\ref%
{no_pooling}) is violated. If $v$ and $c$ are satisfies Assumption 7(ii),
then there exists $s<s^{\ast }$ such that $v(n(z^{\ast }),s,z^{\ast
})-c(s,z^{\ast })>0.$ Therefore, (\ref{no_pooling}) is violated.

Therefore, if there is a stronger monotone pooling CSE, it must be the case
where $z^{\ast }=\underline{z}.$ In this case, $s^{\ast }=0$. Otherwise, the
sender type $z$ arbitrarily close to $0$ will get negative utility because $%
t_{h}<\infty $ and $\lim_{z\rightarrow \underline{z}}c(s,z)=\infty $ for all 
$s>0$ (Assumption 7).

Given $z^{\ast }=\underline{z}$ and $s^{\ast }=0,$ every sender will get
positive utility upon being matched. We distinguish the two cases. If
Assumption 7(ii) is satisfied, then $s^{\ast }=0$ makes the surplus equal to
zero upon being matched, so no receiver is willing to pay $t_{h}>0.$
Therefore there is no pooling CSE. If Assumption 7(i) is satisfied, then we
must have $n(z^{\ast })>\underline{x}$ in order to make (\ref%
{bottom_buyer1_0}) hold, because $v(\underline{x},0,z)=0$ for all $z.$ This
implies that there are more senders than receivers, and hence the market
clearing condition is not satisfied. Therefore, there is no pooling CSE.

\section{Stronger monotone separating CSE\label{appendix_separating_CSE}}

Here we present the stronger monotone separating CSE $\{\tilde{\sigma},%
\tilde{\mu},\tilde{\tau},\tilde{m}\}$ given $z_{\ell }$ induced by $t_{\ell
} $ and $t_{\ell }<t_{h}=\infty $. Once we establish the stronger monotone
CSE, it is convenient to establish the (strictly) well-behaved CSE with the
same $z_{\ell }$ but $t_{h}<\tilde{\tau}\left( \tilde{\sigma}\left( 
\overline{z}\right) \right) $.

\begin{theorem}
\label{proposition1}The necessary and sufficient conditions for a stronger
monotone separating CSE $\left\{ \tilde{\sigma},\tilde{\mu},\tilde{\tau},%
\tilde{m}\right\} $ are

\begin{enumerate}
\item $\tilde{\sigma}\left( \underline{z}\right) =\zeta (\underline{x},%
\underline{z})$, and $\tilde{\sigma}\left( z\right) $ satisfies that $\tilde{%
\tau}^{\prime }(\tilde{\sigma}\left( z\right) )-c_{s}(\tilde{\sigma}\left(
z\right) ,z)=0$ for all $z\in $ Int $Z.$

\item For $s\in \lbrack 0,\tilde{\sigma}\left( \underline{z}\right) )$, $%
\tilde{\mu}(s)=\underline{z};$ for all $s\in S^{\ast }$, $\tilde{\mu}(s)=%
\tilde{\sigma}^{-1}(s)$, where $\tilde{\sigma}^{-1}(s)$ satisfies $\tilde{%
\sigma}(\tilde{\sigma}^{-1}(s))=s$ for all $s\in S^{\ast };$ for all $s>%
\tilde{\sigma}\left( \overline{z}\right) ,$ $\tilde{\mu}(s)=\overline{z}$.

\item $\xi (x)$ satisfies 
\begin{equation}
v_{s}\left( x,s,\tilde{\mu}(s)\right) +v_{z}\left( x,s,\tilde{\mu}(s)\right) 
\tilde{\mu}^{\prime }(s)-\tilde{\tau}^{\prime }(s)=0  \label{receiver_FOC}
\end{equation}%
at $s=\xi (x)$ for all $x\in $ Int $X$.

\item $\tilde{\tau}$ with $\tilde{\tau}(\tilde{\sigma}\left( \underline{z}%
\right) )=t_{\ell }$ clears the market given $\tilde{m}$ such that $\tilde{m}%
(s)=\xi ^{-1}(s)=n\left( \tilde{\mu}(s)\right) $ for all $s\in S^{\ast }$.
\end{enumerate}
\end{theorem}

The proof is below.

\paragraph{Proof of Condition 1}

If there are types who stay out of the market, they must be below $z_{\ell }$
given that $c$ is decreasing in $z$ (Assumption 1.(i) in the main text). 
Note that type $z_{\ell }$ is indifferent between staying out of
the market and taking action $s_{\ell }$ because they satisfy \eqref{lem2}. 
Since $c$ is decreasing in $z$ (Assumption 1.(i)), it means that any type in $[\underline{z},z_{\ell
})$ is strictly better off by staying out of the market instead of taking
action $s_{\ell }$.

Given that $\tau $ is continuous on $S^{\ast }$ and differentiable on Int $%
S^{\ast }$ (Theorem \ref{thm_differentiable_sep_eq}.(ii) in the main text) and $c$ is
differentiable (Assumption 5.(i)), it is clear
that $\tilde{\tau}^{\prime }(\tilde{\sigma}\left( z\right) )-c_{s}(\tilde{%
\sigma}\left( z\right) ,z)=0$ for all $z\in (z_{\ell },\bar{z})$ is a
necessary condition for $\tilde{\sigma}\left( z\right) $ to be an optimal
action for type $z\in \left[ z_{\ell },\bar{z}\right] $ among all actions in 
$S^{\ast }=[\tilde{\sigma}\left( z_{\ell }\right) ,\tilde{\sigma}\left( \bar{%
z}\right) ]$. We show that it is also a sufficient condition for $\tilde{%
\sigma}\left( z\right) $ to be an optimal action for type $z\in \left[
z_{\ell },\bar{z}\right] $ among all actions in $S^{\ast }=[\tilde{\sigma}%
\left( z_{\ell }\right) ,\tilde{\sigma}\left( \bar{z}\right) ]$. We need to
be careful about the boundary condition. A part of Condition 1 in Theorem \ref{thm: Milgrom-Shannon}
is that $\tilde{\sigma}\left( z\right) $
satisfies that 
\begin{equation}
\tilde{\tau}^{\prime }(\tilde{\sigma}\left( z\right) )-c_{s}(\tilde{\sigma}%
\left( z\right) ,z)=0\text{ for all }z\in (z_{\ell },\bar{z}).
\label{lemma1-1}
\end{equation}%
Applying the strict supermodularity of $-c$ (Assumption 1.(ii)) to (\ref{lemma1-1}) yields that 
\begin{eqnarray}
\tilde{\tau}^{\prime }(\tilde{\sigma}\left( z^{\prime }\right) )-c_{s}(%
\tilde{\sigma}\left( z^{\prime }\right) ,z) &\gtrless &0\text{ if }z^{\prime
}\lessgtr z\text{, }\forall z,z^{\prime }\in (z_{\ell },\overline{z})
\label{lemma1-3} \\
\tilde{\tau}(\tilde{\sigma}\left( z\right) )-c(\tilde{\sigma}\left( z\right)
,z) &>&\tilde{\tau}(\tilde{\sigma}\left( z^{\prime }\right) )-c(\tilde{\sigma%
}\left( z^{\prime }\right) ,z)\text{ }\forall z,z^{\prime }\in (z_{\ell },%
\bar{z})\text{ s.t. }z\neq z^{\prime }.  \label{thm1-1}
\end{eqnarray}

(\ref{lemma1-3}) implies that $\tilde{\sigma}$ is increasing in $z\in
(z_{\ell },\overline{z})$. Because $\tilde{\sigma}$ is continuous on $%
S^{\ast }$ (Lemma \ref{lm:conti_sigma}), it implies that $\tilde{%
\sigma}$ is increasing over $\left[ z_{\ell },\overline{z}\right] .$ Because 
$\tilde{\sigma}$ is increasing over $\left[ z_{\ell },\overline{z}\right] $
and $\tilde{\tau}$ and $c$ are continuous (Theorem \ref{thm_differentiable_sep_eq}.(ii) and Assumption
5.(i)), (\ref{thm1-1}) implies that 
\begin{equation}
\tilde{\tau}(\tilde{\sigma}\left( z\right) )-c(\tilde{\sigma}\left( z\right)
,z)>\tilde{\tau}(\tilde{\sigma}\left( z^{\prime }\right) )-c(\tilde{\sigma}%
\left( z^{\prime }\right) ,z)\text{ }\forall z,z^{\prime }\in \lbrack
z_{\ell },\bar{z}]\text{ s.t. }z\neq z^{\prime }.  \label{thm1-2}
\end{equation}%
(\ref{thm1-2}) shows that $\tilde{\sigma}\left( z\right) $ is be an optimal
action for type $z\in \left[ z_{\ell },\bar{z}\right] $ among all actions in 
$S^{\ast }=[\tilde{\sigma}\left( z_{\ell }\right) ,\tilde{\sigma}\left( \bar{%
z}\right) ]$, ignoring the individual rationality.

To show the individual rationality, let $\tilde{U}(z):=\tilde{\tau}(\tilde{%
\sigma}\left( z\right) )-c(\tilde{\sigma}\left( z\right) ,z)$. Because of
\eqref{lem2} in the main text, we have that $\tilde{U}(z_{\ell })=\tilde{%
\tau}(\tilde{\sigma}\left( z_{\ell }\right) )-c(\tilde{\sigma}\left( z_{\ell
}\right) ,z_{\ell })=0.$ It is clear that $\tilde{U}(z)>\tilde{\tau}(\tilde{%
\sigma}\left( z_{\ell }\right) )-c(\tilde{\sigma}\left( z_{\ell }\right) ,z)>%
\tilde{U}(z_{\ell })=0$ for all $z\in (z_{\ell },\bar{z}]$, where the first
inequality holds because of (\ref{thm1-2}) and the second inequality holds
because $c$ is decreasing in $z$ (Assumption 1.(i)).

We need to show that type $z$ has no incentive to deviate to $s\notin $ $%
\sigma (Z)$ to complete the proof that $\tilde{\sigma}\left( z\right) $ is
an optimal action for type $z$ among all actions in $S.$ We defer it to the
end. First we start with the stronger monotone belief $\tilde{\mu}$

\paragraph{Proof of Condition 2}

Note that $\sigma (Z)=\{0\}\cup \lbrack s_{\ell },\tilde{\sigma}\left( 
\overline{z}\right) ]$. Therefore, we can derive the belief on the
equilibrium path as follows: (i) for $s=0$, $\tilde{\mu}(s)=G(z|z<z_{\ell })$
and (ii) for all $s\in \lbrack s_{\ell },\tilde{\sigma}\left( \overline{z}%
\right) ]$, $\tilde{\mu}(s)=\tilde{\sigma}^{-1}(s)$, where $\tilde{\sigma}%
^{-1}(s)$ satisfies $\tilde{\sigma}(\tilde{\sigma}^{-1}(s))=s.$ This is part
of Condition 2 in Theorem \ref{proposition1} so that consistency is
satisfied. There are two intervals of off path sender actions: $(0,s_{\ell
}) $ and $(\tilde{\sigma}\left( \overline{z}\right) ,\infty )$. The
monotonicity of the belief in the stronger set order uniquely determines the
belief conditional on $s\notin $ $\sigma (Z)$: (iii) for $s\in (0,s_{\ell }),%
\tilde{\mu}(s)=z_{\ell }$ and (iv) for $s\in (\tilde{\sigma}\left( \overline{%
z}\right) ,\infty ),\tilde{\mu}(s)=\overline{z}.$

The belief function $\tilde{\mu}$ satisfying (i) -(iv), which is Condition 2
in Theorem \ref{proposition1}, is the unique monotone equilibrium belief in
the stronger set order given $\sigma $ in Condition 1

\paragraph{Proof of Conditions 3}

If there are types who stay out of the market, they must be below $x_{\ell }$
given that $v$ is increasing in $x$ (Assumption 2.(ii) in the main text). Note that type $x_{\ell }$ is indifferent between staying out of
the market and matching with the sender of type $z_{\ell }$ because they
satisfy \eqref{lem1}. Since $v$ is increasing in $x$
(Assumption 2.(ii)), it means that any type in $[%
\underline{x},x_{\ell })$ is strictly better off by staying out of the
market.

Consider the matching problem for type $x\in \lbrack x_{\ell },\overline{x}]$%
. Note that $v,$ $c,\gamma $, $\tilde{\tau}$ and $\tilde{\mu}$ are
continuous and differentiable (Assumptions 4.(ii) and 5.(i) and Theorem \ref{thm_differentiable_sep_eq}). Therefore, it is clear that for all $x\in
(x_{\ell },\overline{x}),$%
\begin{equation}
\pi _{s}(\xi (x),x)=v_{s}\left( x,\xi (x),\tilde{\mu}\left( \xi (x)\right)
\right) +v_{z}\left( x,\xi (x),\tilde{\mu}\left( \xi (x)\right) \right) 
\tilde{\mu}^{\prime }\left( \xi (x)\right) -\tilde{\tau}^{\prime }\left( \xi
(x)\right) =0  \label{thm1-3'}
\end{equation}%
is a necessary condition for $\xi (x)$ to be an optimal choice of a matching
partner (in terms of her action) for type $x\in \left[ x_{\ell },\overline{x}%
\right] $ among all actions in $S^{\ast }$.

We show that (\ref{thm1-3'}) is also a sufficient condition for $\xi (x)$ to
be an optimal action of a matching partner for type $x\in \left[ x_{\ell },%
\overline{x}\right] $ among all actions in $S^{\ast }$. Applying the
supermodularity of $v$ (Assumptions 2.(i)) to (%
\ref{thm1-3'}) yields that 
\begin{eqnarray}
\pi _{s}(\xi (x^{\prime }),x) &\gtrless &0\text{ if }x^{\prime }\lessgtr x%
\text{ }\forall x,x^{\prime }\in \left( x_{\ell },\overline{x}\right) ,
\label{thm1-3A} \\
\pi (\xi (x),x) &>&\pi (\xi (x^{\prime }),x)\text{ }\forall x,x^{\prime }\in
(x_{\ell },\bar{x})\text{ s.t. }x\neq x^{\prime }  \label{thm1-3}
\end{eqnarray}%
(\ref{thm1-3A}) implies that $\xi (x)$ is increasing on $\left( x_{\ell },%
\overline{x}\right) $. Given the increasing property of $\xi $ on $\left(
x_{\ell },\overline{x}\right) $, $\xi $ must be continuous on $\left[
x_{\ell },\overline{x}\right] $. Otherwise, senders in the interval created
by a discontinuity of $\xi $ are not matched in equilibrium and it violates
the market clearing condition.

The continuity of $\xi $ on $\left[ x_{\ell },\overline{x}\right] $ makes it
increasing on $\left[ x_{\ell },\overline{x}\right] $ because $\xi $ is
increasing on $x\in \left( x_{\ell },\overline{x}\right) $. Together with
the continuity of $\xi $ over $\left[ x_{\ell },\overline{x}\right] $, the
continuity of $v$ and $\tilde{\tau}$ (Assumptions 3.(ii) and \eqref{lem2}) makes $\pi (\xi (x^{\prime }),x)$ continuous in $%
x^{\prime }\in \left[ x_{\ell },\overline{x}\right] $ and $x\in \left[
x_{\ell },\overline{x}\right] $. Therefore, (\ref{thm1-3}) implies that 
\begin{equation}
\pi (\xi (x),x)>\pi (\xi (x^{\prime }),x)\text{ }\forall x,x^{\prime }\in %
\left[ x_{\ell },\bar{x}\right] \text{ s.t. }x\neq x^{\prime }.
\label{thm1-4}
\end{equation}%
(\ref{thm1-4}) shows that $\xi (x)$ is be an optimal choice of a matching
partner for type $x\in \left[ z_{\ell },\bar{z}\right] $ among all actions
in $S^{\ast }=[\tilde{\sigma}\left( z_{\ell }\right) ,\tilde{\sigma}\left( 
\bar{z}\right) ]$, ignoring the individual rationality.

To show the individual rationality, let $\tilde{\pi}(x):=\pi (\xi (x),x)$.
Because of \eqref{lem1}, we have that $\pi (\xi (x_{\ell
}),x_{\ell })=0.$ It is clear that $\pi (\xi (x),x)>\pi (\xi (x_{\ell
}),x)>\pi (\xi (x_{\ell }),x_{\ell })=0$ for all $x\in (x_{\ell },\bar{x}]$,
where the first inequality holds because of (\ref{thm1-4}) and the second
inequality holds because $v$ is increasing in $x$ (Assumption 2.(ii)).

\paragraph{Proof of Condition 4}

It is straightforward that $\tilde{m}$ in in Condition 4 is a unique
measure-preserving matching function given that $\xi $ and $\tilde{\sigma}$
are both increasing.

\paragraph{Proof of no profitable sender deviation to an off-path action}

Applying Lemma 1 and Corollary \ref{corollary_monotone_belief}, the monotone
belief in the stronger set order is the unique belief that pass Criterion
D1. Because $\tilde{\mu}(s)$ for $s\notin $ $\sigma (Z)$ is a degenerate
probability distribution with a singleton as its support, as suggested in
Corollary \ref{corollary_monotone_belief1}, we only need to check if the type
of the sender in that support has an incentive to deviate in order to check
if any sender has an incentive to deviate to such $s.$

Now let us prove no profitable sender deviation to an off-path action.
Conditional on $s\in (0,s_{\ell }),$ it is believed that the sender who
chose $s$ is $\tilde{\mu}(s)=z_{\ell }.$ According to Corollary 1, if the sender of type $z_{\ell }$ has no profitable
deviation to $s\in (0,s_{\ell })$, then no one else does. Therefore, we only
need to check if the sender of type $z_{\ell }$ has an profitable deviation
to $s\in (0,s_{\ell }).$ If she reduces her action down to $s\in (0,s_{\ell
})$, no receiver wants her because he has to transfer at least $t_{\ell }$
but he can be matched with a sender with $s_{\ell }$ at $t_{\ell }.$
Therefore, there is no sender profitable deviation to $s$.

Now, let us examine if there is a profitable sender deviation to $s\in (%
\tilde{\sigma}\left( \overline{z}\right) ,\infty ).$ First, agents at the
action choice stage expect $\tilde{m}$ to be increasing and the equilibrium
market reaction function $\tilde{\tau}:S^{\ast }\rightarrow T$ satisfies 
\begin{equation}
\tilde{\tau}^{\prime }\left( s\right) =v_{s}(\tilde{m}(s),s,\tilde{\mu}%
\left( s\right) )+v_{z}(\tilde{m}(s),s,\tilde{\mu}\left( s\right) )\tilde{\mu%
}^{\prime }\left( s\right) .  \label{differential_tau}
\end{equation}%
Because the support of $\tilde{\mu}(s)$ for $s>\tilde{\sigma}(\overline{z})$
is a singleton, $\overline{z},$ we only need to check if the sender of type $%
\overline{z}$ has an incentive to choose $s>\tilde{\sigma}(\overline{z})$.
Because there is a continuum of receivers with different reactions and
types, we need to check which receiver is willing to transfer the largest
amount to the sender with $s>\tilde{\sigma}(\overline{z})$. The type $x$
receiver's maximum willingness to transfer is%
\begin{equation*}
t(s,x)=v(x,s,\overline{z})-\left[ v(x,\tilde{\sigma}(n^{-1}(x))),n^{-1}(x))-%
\tilde{\tau}\left( \tilde{\sigma}(n^{-1}(x))\right) \right] .
\end{equation*}%
Because $s>\tilde{\sigma}(\overline{z})$ and $\overline{z}\geq n^{-1}(x),$
we have that $t(s,x)>\tilde{\tau}\left( \tilde{\sigma}(n^{-1}(x))\right) $.

Given $\tilde{\tau}^{\prime }$ in (\ref{differential_tau}), taking the
derivative of $t(s,x)$ with respect to $x$ yields%
\begin{equation}
t_{x}(s,x)=v_{x}(x,s,\overline{z})-v_{x}(x,\tilde{\sigma}%
(n^{-1}(x))),n^{-1}(x))>0.  \label{maximum_pay1}
\end{equation}%
Note that $v_{x}(x,s,z)$ is non-decreasing in $s$ and increasing in $z$,
given Assumption 2.(i) - $v(b,x,s,z)$ is
supermodular in $(b,x,s,z)$ and strictly supermodular in $(z,x)$. Because $s>%
\tilde{\sigma}(n^{-1}(x)))$ and $\overline{z}\geq n^{-1}(x)$ for all $x\geq
x_{\ell }$, this implies that $t_{x}(s,x)$ is positive for any $s>\tilde{%
\sigma}(\overline{z}),$ as in (\ref{maximum_pay1}). It in turn implies that
the maximum amount of transfer that the the receiver of type $\overline{x}$
is willing to make is the largest. Then, given Criterion D1, we only need to
check if the sender of type $\overline{z}$ has a profitable deviation to $s>%
\tilde{\sigma}\left( \overline{z}\right) $ while keeping her current match
partner, the receiver of type $\overline{x}$.

There is a profitable sender deviation to $s>\tilde{\sigma}(\overline{z})$
for $\overline{z}$ if and only if for some $s>\tilde{\sigma}(\overline{z})$ 
\begin{equation}
v(\overline{x},s,\overline{z})-c(s,\overline{z})>v(\overline{x},\tilde{\sigma%
}(\overline{z}))-c(\tilde{\sigma}(\overline{z}),\overline{z})\text{.}
\label{PSD_condition}
\end{equation}%
However, the inequality above is not satisfied for any $s>\tilde{\sigma}(%
\overline{z}).$ The reason is that the information rent, the last term in (%
\ref{differential_tau}), makes $\tilde{\sigma}(\overline{z})$ larger than
the constrained efficient action level for the sender of type $\overline{z}.$
Therefore, if $s>\tilde{\sigma}(\overline{z}),$ then $v-c$ is even smaller
given the strict concavity of $v-c$ in $s$ (Assumption 3).

\section{Proof of Theorem \ref{theorem1}}

Note that when $t_{h}<\tilde{\tau}\left( \tilde{\sigma}\left( \overline{z}%
\right) \right) ,$ $\left\{ \hat{\sigma},\hat{\mu},\hat{\tau},\hat{m}%
\right\} $ follows the separating CSE $\{\tilde{\sigma},\tilde{\mu},\tilde{%
\tau},\tilde{m}\}$ with the same $z_{\ell }$ before $z$ hits $z_{h}$.
Therefore, we will use some of the proof of Theorem \ref{proposition1}.
Because there is a jump to $s_{h}$ and every sender of type $z_{h}$ or
higher chooses the same action $s_{h},$ $\lim_{z\nearrow z_{h}}\hat{\sigma}%
\left( z\right) =\tilde{\sigma}(z_{h}).$ Therefore, we will use $\tilde{%
\sigma}(z_{h})$ instead of $\lim_{z\nearrow z_{h}}\hat{\sigma}\left(
z\right) $ for simplicity of notation.

It is straightforward to show that the beliefs in Condition 2 of Theorem \ref{theorem1}
in the main text satisfies the consistency and the stronger
monotonicity. There are three off-path sender action intervals, $(0,s_{\ell
}),$ $[\tilde{\sigma}(z_{h}),s_{h})$ and $(s_{h},\infty ).$ The \emph{%
stronger monotonicity} of beliefs (Lemma \ref{lemma_monotone_eq_A'} and Corollary \ref{corollary_monotone_belief}) uniquely pins down the singleton support of a belief $\hat{\mu}%
(s)$ conditional on $s$ in each off-path sender action interval. Further, we
only need to check the type-$\hat{\mu}(s)$ sender's incentive to deviate to $%
s$ in any off-path action interval, thanks to Corollary \ref{corollary_monotone_belief1}.

\subsection{Sender's optimal action choice}

In subsection (a) below, we first show that there is no profitable deviation
to an off-path action for every sender if they choose actions according to $%
\hat{\sigma}$.

In the remaining subsections, we show that $\hat{\sigma}(z)$ solves Problem
\eqref{WP3} if Problem \eqref{WP3} admits a solution; $\hat{\sigma}%
(z)=0$ otherwise. Note that $\hat{\sigma}(z)$ solves Problem \eqref{WP3} for $z\in
S^{\ast }[s_{\ell },\tilde{\sigma}(z_{h}))\cup s_{h}$. When it does, the
sender's equilibrium utility $\hat{U}(z)$ is increasing and positive for $%
z>z_{\ell }$ starting from $\hat{U}(z_{\ell })=0$ due to the envelope
theorem. Therefore, the constraint in Problem \eqref{WP3} is satisfied.

\paragraph{(a) No profitable sender deviation to an off-path action}

There are three intervals of actions that are not observed in equilibrium: $%
(0,s_{\ell })$, $[\tilde{\sigma}(z_{h}),s_{h})$, and $(s_{h},\infty )$.
First, consider a deviation to $s>s_{h}.$ Since the belief $\hat{\mu}(s)=%
\overline{z}$ for $s>s_{h}$ passes Criterion D1, we only need to check if
the sender of type $\overline{z}$ has an incentive to deviate to such $s$ in
order to establish that there is no profitable sender deviation to such $s$.
Suppose that the sender of type $\overline{z}$ increases her action above $%
s_{h}.$ The maximum transfer she can receive is $t=t_{h}.$ Because $\hat{%
\sigma}(\overline{z})<s$ and $\hat{\tau}(\hat{\sigma}(\overline{z}))=t_{h},$
we have that $\hat{\tau}(\hat{\sigma}(\overline{z}))-c(\hat{\sigma}(%
\overline{z}),\overline{z})>t_{h}-c(s,\overline{z})$. Therefore, the sender
of type $\overline{z}$ cannot gain by changing her action to $s>s_{h}.$

Second, consider a deviation to $s\in (0,s_{\ell })$. In this case, the
belief is $\hat{\mu}(s)=z_{\ell }.$ Note that $\hat{\mu}(s_{\ell })=z_{\ell
}.$ Since a receiver can be matched with a sender with $z_{\ell }$ whose
type is believed to be $z_{\ell }$, transferring the lower bound of
transfers, $t_{\ell }$ to her, no receiver wants a sender with $s<s_{\ell }$
whose type is believed to be $z_{\ell },$ transferring $t_{\ell }$ to her.
Therefore, there is no profitable deviation to $s\in (0,s_{\ell })$.

Third, consider a deviation to $s\in \lbrack \tilde{\sigma}(z_{h})),s_{h}).$
In this case, the belief is $\hat{\mu}(s)=z_{h}$. Suppose that the sender of
type $z_{h}$ decreases her action from $s_{h}$ to $s\in (\tilde{\sigma}%
(z_{h})),s_{h})$. She can be matched with a receiver of type $x\in \lbrack
x_{\ell },x_{h})$ or a receiver of type $x\in \lbrack x_{h},\overline{x}].$

We first show that there is no profitable sender deviation to $s\in \lbrack 
\tilde{\sigma}(z_{h})),s_{h}),$ being matched with any receiver of type $%
x\in $ $[x_{\ell },x_{h}).$ Let $t(s,x)$ be the maximum amount of transfer
that a receiver of type $x\in \lbrack x_{\ell },x_{h})$ is willing to make
to a sender with $s\in (\tilde{\sigma}(z_{h})),s_{h})$. Following the proof
of no profitable sender deviation in the stronger monotone separating CSE,
we can show that the receiver's maximum willingness increases as $x$
approaches $x_{h}$ from the left. Therefore, the supremum of the amount of
transfers to the sender with $s$ is 
\begin{equation}
t(s,x_{h})=v\left( x_{h},s,z_{h}\right) -\left( v\left( x_{h},\tilde{\sigma}%
(z_{h}),z_{h}\right) -\tilde{\tau}(\tilde{\sigma}(z_{h}))\right) .
\label{PSD3'}
\end{equation}%
Because $s>\tilde{\sigma}(z_{h}),$ we have that $t(s,x_{h})>\tilde{\tau}(%
\tilde{\sigma}(z_{h}))$. Given the upper bound of transfers $t_{h}$, the
receiver of type $x_{h}$ cannot transfer $t(s,x_{h})$ if $t(s,x_{h})>t_{h}$.
Suppose that the receiver can always transfer $t(s,x_{h})$ as if there is no
upper bound of transfers. If a sender of type $z_{h}$ has no incentive to
deviate to $s$ when there is no upper bound of transfers, then she also has
no incentive to deviate to $s$ when there is the upper bound of transfers.

Therefore, the sender of type $z_{h}$ has an incentive to deviate to $s\in (%
\tilde{\sigma}(z_{h})),s_{h})$ if and only if 
\begin{equation}
t(s,x_{h})-c(s,z_{h})>t_{h}-c(s_{h},z_{h})=\tilde{\tau}(\tilde{\sigma}%
(z_{h}))-c(\tilde{\sigma}(z_{h}),z_{h}),  \label{PSD4}
\end{equation}%
where the equality comes from (18). (\ref{PSD3'}%
) and (\ref{PSD4}) together implies that the sender of type $z_{h}$ has an
incentive to deviate to $s\in (\tilde{\sigma}(z_{h})),s_{h})$ if and only if%
\begin{equation}
v\left( x_{h},s,z_{h}\right) -c(s,z_{h})>v\left( x_{h},\tilde{\sigma}%
(z_{h}),z_{h}\right) -c(\tilde{\sigma}(z_{h}),z_{h}).  \label{PSD5}
\end{equation}%
$\tilde{\sigma}(z_{h})$ is the action chosen by type $z_{h}$ in the stronger
monotone separating CSE. The first-order conditions for action choices by
type $z_{h}$ and type $x_{h}$ in Conditions 1 and 4 in Theorem \ref%
{proposition1} imply that%
\begin{equation*}
v_{s}\left( x_{h},\tilde{\sigma}(z_{h}),z_{h}\right) +v_{z}\left( x_{h},%
\tilde{\sigma}(z_{h}),z_{h}\right) \tilde{\mu}^{\prime }\left( \tilde{\sigma}%
(z_{h})\right) -c_{s}(\tilde{\sigma}(z_{h}),z_{h})=0.
\end{equation*}%
Because $\tilde{\mu}^{\prime }>0$ for all $s\in $ Int $S^{\ast }$
(implication of Theorem \ref{thm_differentiable_sep_eq}.(iii), the equality
above means that 
\begin{equation}
v_{s}\left( x_{h},\tilde{\sigma}(z_{h}),z_{h}\right) -c_{s}(\tilde{\sigma}%
(z_{h}),z_{h})<0  \label{PSD6}
\end{equation}%
Because $s>\tilde{\sigma}(z_{h})$ and $v-c$ is strictly concave in $s$
(Assumption 3), (\ref{PSD6}) implies that (\ref%
{PSD5}) is not satisfied. Therefore, there is no profitable sender deviation
to $s\in \lbrack \tilde{\sigma}(z_{h})),s_{h}),$ being matched with any
receiver of type $x\in $ $[x_{\ell },x_{h}).$

Finally, we show that there is no profitable sender deviation to any
off-path action $s$ in $[\tilde{\sigma}(z_{h}),s_{h})$, followed by matching
with a receiver in $[x_{h},\overline{x}]$. $[x_{h},\overline{x}]$ is the
interval of receiver types who are matched with senders on the top. The
maximum amount of the reaction that the receiver of type $x\in \lbrack x_{h},%
\overline{x}]$ is willing to choose is 
\begin{equation}
T(s,x)=v\left( x,s,z_{h}\right) -\left( \mathbb{E}[v(x,s_{h},z^{\prime
})|z^{\prime }\geq z_{h}]-t_{h}\right)  \label{decreasing_T}
\end{equation}%
Because $z_{h}\leq z^{\prime }$ and $s<s_{h},$ we have that $T(s,x)<t_{h}$
and we can also apply Assumption 2.(i) to show
that $T(s,x)$ decreases in $x$. Therefore, if and only if 
\begin{equation}
v\left( x_{h},s,z_{h}\right) -c(s,z_{h})\leq \mathbb{E}[v(x_{h},s_{h},z^{%
\prime })|z^{\prime }\geq z_{h}]-c(s_{h},z_{h})\text{ for all }s\in \lbrack 
\tilde{\sigma}(z_{h}),s_{h})  \label{suff_cond1}
\end{equation}%
is satisfied, the sender of type $z_{h}$ has no profitable deviation to any $%
s\in \lbrack \tilde{\sigma}(z_{h})),s_{h})$, followed by matching with a
receiver in $[x_{h},\overline{x}]$. Consequently, Corollary \ref{corollary_monotone_belief1} implies that no sender has an incentive to deviate to any
off-path action $s$ in $[\tilde{\sigma}(z_{h}),s_{h})$ if and only if (\ref%
{suff_cond1}) is satisfied for all $s\in \lbrack \tilde{\sigma}%
(z_{h}),s_{h}) $. Because (18) and (19) are
satisfied in equilibrium, we have that 
\begin{equation}
\mathbb{E}[v(x_{h},s_{h},z^{\prime })|z^{\prime }\geq
z_{h}]-c(s_{h},z_{h})=v\left( x_{h},\tilde{\sigma}(z_{h}),z_{h}\right) -c(%
\tilde{\sigma}(z_{h}),z_{h}).  \label{suff_cond2'}
\end{equation}%
Applying (\ref{suff_cond2'}) to (\ref{suff_cond1}) yields that for $s\in
\lbrack \tilde{\sigma}(z_{h}),s_{h})$,%
\begin{equation}
v\left( x_{h},s,z_{h}\right) -c(s,z_{h})\leq v\left( x_{h},\tilde{\sigma}%
(z_{h}),z_{h}\right) -c(\tilde{\sigma}(z_{h}),z_{h}),
\end{equation}%
which is always satisfied given Assumption 5 (strict concavity of $v-c$ in $%
s $) because $\tilde{\sigma}(z_{h})$ is greater than the bilaterally
efficient action $\zeta (x_{h},z_{h})$. Therefore, there is no profitable
sender deviation to any off-path action $s$ in $[\tilde{\sigma}%
(z_{h}),s_{h}) $, followed by matching with a receiver in $[x_{h},\overline{x%
}]$.

\paragraph{(a) action choice in $S^{\ast }$ by \textbf{the sender of type }$%
z_{h}$}

\noindent The sender's equilibrium action is $s_{h}$. Note that $S^{\ast
}=[s_{\ell },\tilde{\sigma}(z_{h}))\cup s_{h}.$ Because of (18), her utility is the same as $\tilde{\tau}\left( \tilde{\sigma%
}\left( z_{h}\right) \right) -c\left( \tilde{\sigma}\left( z_{h}\right)
,z_{h}\right) ,$ which is her utility in the stronger monotone separating
CSE. Suppose that she chooses $s\in \lbrack s_{\ell },\tilde{\sigma}%
(z_{h})). $ Because the transfer schedule $\hat{\tau}$ is the same as $%
\tilde{\tau}$ in the stronger monotone separating CSE, we can apply Theorem %
\ref{proposition1} to show that the sender of type $z_{h}$ has no incentive
to decrease her action to $s\in \lbrack s_{\ell },\tilde{\sigma}(z_{h}))$.
Therefore, $s_{h}$ solves Problem \eqref{WP3}.

\paragraph{(b) action choice in $S^{\ast }$ \textbf{by the sender of type }$%
z\in (z_{h},\overline{z}]$}

The sender's equilibrium action is $s_{h}$. From (a) above, we know that
that for all $s\in \lbrack s_{\ell },\tilde{\sigma}(z_{h})),$ 
\begin{equation}
t_{h}-c(s_{h},z_{h})\geq \hat{\tau}\left( s\right) -c(s,z_{h}).
\label{thm2_17}
\end{equation}%
Applying Assumption 1.(ii) to (\ref{thm2_17}) yields that $%
t_{h}-c(s_{h},z_{h})>\hat{\tau}\left( s\right) -c(s,z_{h})$ for all $z>z_{h}$
and all $s\in \lbrack s_{\ell },\tilde{\sigma}(z_{h}))$, which shows that
the sender of type $z>z_{h}$ has no incentive to change her action to any
other action in $S^{\ast }.$ Therefore, $s_{h}$ solves Problem \eqref{WP3}.

\paragraph{(c) action choice in $S^{\ast }$ by \textbf{the sender of type }$%
z\in \lbrack z_{\ell },z_{h})$}

Because $z<z_{h},$ the sender's action is lower than $\tilde{\sigma}\left(
z_{h}\right) .$ Because (i) the reaction schedule $\hat{\tau}$ for the
action in $[s_{\ell },\tilde{\sigma}\left( z_{h}\right) )$ is the same as
the one in the equilibrium with only the lower bound of transfers and (ii)
the sender's action $\hat{\sigma}\left( z\right) $ is the same $\tilde{\sigma%
}\left( z\right) $ that she would have chosen in the stronger monotone
separating CSE, we apply the proof of Condition 1 in Theorem \ref%
{proposition1} to show that for any $s\in \lbrack s_{\ell },\tilde{\sigma}%
\left( z_{h}\right) )$ and $s\neq \hat{\sigma}\left( z\right) $%
\begin{equation}
\hat{\tau}\left( \hat{\sigma}\left( z\right) \right) -c\left( \hat{\sigma}%
\left( z\right) ,z\right) >\hat{\tau}\left( s\right) -c\left( s,z\right) .
\label{thm2-17}
\end{equation}%
Therefore, the sender has no incentive to change her action to another
action in $[s_{\ell },\tilde{\sigma}\left( z_{h}\right) )$.

Now suppose that the sender changes her action to $s_{h}.$ We know that for
the sender of type $z_{h},$%
\begin{equation*}
t_{h}-c(s_{h},z_{h})=\hat{\tau}\left( \tilde{\sigma}\left( z_{h}\right)
\right) -c\left( \tilde{\sigma}\left( z_{h}\right) ,z_{h}\right) .
\end{equation*}%
Because $s_{h}>\tilde{\sigma}\left( z_{h}\right) $ by Lemma \ref{lemma2} and $z_{h}>z,$ applying Assumption 1.(ii) to the equation above yields that%
\begin{equation}
t_{h}-c(s_{h},z)<\hat{\tau}\left( \tilde{\sigma}\left( z_{h}\right) \right)
-c\left( \tilde{\sigma}\left( z_{h}\right) ,z\right) .  \label{thm2_20}
\end{equation}%
Combining (\ref{thm2-17}) at $s=\tilde{\sigma}\left( z_{h}\right) $ and (\ref%
{thm2_20}) yields that $\hat{\tau}\left( s\right) -c\left( s,z\right)
>t_{h}-c(s_{h},z),$ which shows that the sender of type $z\in \lbrack
z_{\ell },z_{h})$ has no incentive to increase her action to $s_{h}$.
Therefore, $s_{h}$ solves Problem \eqref{WP3}.

\paragraph{(d) No action choice by \textbf{the sender of type }$z\in \lbrack 
\protect\underline{z},z_{\ell })$}

The sender chooses no action in equilibrium: $\hat{\sigma}\left( z\right) =0$%
. Consider a change to $s_{\ell }.$ We know that for the sender of type $%
z_{\ell }$, $t_{\ell }-c(s_{\ell },z_{\ell })=0.$ Because $z<z_{\ell },$
applying Assumption 1.(ii) yields that 
\begin{equation}
t_{\ell }-c(s_{\ell },z)<0,  \label{thm2_20'}
\end{equation}%
which implies that the sender's utility is lower than zero, so the sender
cannot gain by increasing her action to $s_{\ell }$.

Consider a change to $s\in S^{\ast }$ with $s>s_{\ell }.$ From the previous
section, we know that for all $s\in S^{\ast }$ with $s>s_{\ell }$ 
\begin{equation*}
t_{\ell }-c(s_{\ell },z_{\ell })>\hat{\tau}\left( s\right) -c(s,z_{\ell }).
\end{equation*}%
Because $z<z_{\ell },$ applying Assumption 1.(ii) to the inequality relation
above yields that for all $s\in S^{\ast }$ with $s>s_{\ell }$%
\begin{equation}
t_{\ell }-c(s_{\ell },z)>\hat{\tau}\left( s\right) -c(s,z).
\label{thm2_20''}
\end{equation}%
Combining (\ref{thm2_20'}) and (\ref{thm2_20''}) yields that $0>\hat{\tau}%
\left( s\right) -c(s,z)$ for $s\in S^{\ast }$ with $s>s_{\ell },$ which
shows that a change to any $s\in S^{\ast }$ with $s>s_{\ell }$ lowers the
sender's utility. Therefore, for the sender of type $z\in \lbrack \underline{%
z},z_{\ell })$, there is no solution for Problem \eqref{WP3}.

\subsection{Receiver's optimal matching choice}

Applying the envelope theorem to the receiver's equilibrium utility $\hat{\Pi%
}(x)$, we can show that $\hat{\Pi}(x)$ is increasing and positive for $%
x>x_{\ell }$ starting from $\hat{\Pi}(x_{\ell })=0$. The receiver's matching
problem can be seen as: which sender with an action in $S^{\ast }$ does he
want to match with as formulated in (1) in the main text?

\paragraph{(a) Optimal \textbf{matching choice by the receiver of type }$%
x_{h}$}

The equilibrium partner is a sender with $s_{h}$ as his partner. Suppose
that the receiver wants to choose a sender with $s\in \lbrack s_{\ell },%
\tilde{\sigma}(z_{h}))$ as his partner. According to \eqref{jumping_buyers}, the receiver's 
equilibrium utility with a sender with $s_{h}$
satisfies 
\begin{equation}
\mathbb{E}\left[ \left. v\left( x_{h},s_{h},z\right) \right\vert z\geq z_{h}%
\right] -t_{h}=v\left( x_{h},\tilde{\sigma}\left( z_{h}\right) ,z_{h}\right)
-\tilde{\tau}\left( \tilde{\sigma}\left( z_{h}\right) \right) .
\label{thm2_22}
\end{equation}%
The proof of Condition 3 of Theorem \ref{proposition1} shows that in the
stronger monotone separating CSE, we have that for any $s\in \lbrack s_{\ell
},\tilde{\sigma}(z_{h}))$,%
\begin{equation}
v\left( x_{h},\tilde{\sigma}\left( z_{h}\right) ,z_{h}\right) -\tilde{\tau}%
\left( \tilde{\sigma}\left( z_{h}\right) \right) >v\left( x_{h},s,\tilde{\mu}%
\left( s\right) \right) -\tilde{\tau}\left( s\right) .  \label{thm2_23}
\end{equation}%
Because $\hat{\mu}(s)=\tilde{\mu}\left( s\right) $ and $\hat{\tau}(s)=\tilde{%
\tau}\left( s\right) $ for any $s\in \lbrack s_{\ell },\tilde{\sigma}%
(z_{h})) $, (\ref{thm2_22}) and (\ref{thm2_23}) together show that for any $%
s\in \lbrack s_{\ell },\tilde{\sigma}(z_{h}))$%
\begin{equation}
\mathbb{E}\left[ \left. v\left( x_{h},s_{h},z\right) \right\vert z\geq z_{h}%
\right] -t_{h}>v\left( x_{h},s,\hat{\mu}\left( s\right) \right) -\hat{\tau}%
\left( s\right) ,  \label{thm2_23A}
\end{equation}%
which shows that the receiver has no incentive to change his reaction to be
matched with a sender with $s\in \lbrack s_{\ell },\tilde{\sigma}(z_{h}))$.

\paragraph{(b) Optimal matching choice by \textbf{the receiver of type }$%
x>x_{h}$}

The equilibrium utility for the receiver of type $x$ with a sender with $%
s_{h}$ as his partner is $\mathbb{E}\left[ \left. v(x,s_{h},z)\right\vert
z\geq z_{h}\right] -t_{h}.$ Suppose that the receiver changes his partner to
a sender with $s\in \lbrack s_{\ell },\tilde{\sigma}\left( z_{h}\right) ).$
Given $s\in \lbrack s_{\ell },\tilde{\sigma}\left( z_{h}\right) ),$ the
belief on the sender's type is $\hat{\mu}\left( s\right) <z_{h}$ and his
utility is $v(x,s,\hat{\mu}\left( s\right) )-\hat{\tau}\left( s\right) $.
Therefore, we need to examine the sign of the utility difference:%
\begin{equation}
\mathbb{E}\left[ \left. v(x,s_{h},z)\right\vert z\geq z_{h}\right] -t_{h}-%
\left[ v(x,s,\hat{\mu}\left( s\right) )-\hat{\tau}\left( s\right) \right]
\label{thm2_26}
\end{equation}%
Applying the envelope theorem for $b_{e}(x,s_{h})$ and $\gamma (x,s,\hat{\mu}%
\left( s\right) ),$ we can express the partial derivative of (\ref{thm2_26})
with respect to $x$%
\begin{equation}
\mathbb{E}\left[ \left. v_{x}(x,s_{h},z)\right\vert z\geq z_{h}\right]
-v_{x}(x,s,\hat{\mu}\left( s\right) )>0.  \label{thm2_27}
\end{equation}%
To show the positive sign in (\ref{thm2_27}), note that $s_{h}>s.$
Therefore, Assumption 2.(i) implies that $v_{x}(x,s_{h},z)>v_{x}(x,s,\hat{\mu%
}\left( s\right) )$ for any $z\geq z_{h}>\hat{\mu}\left( s\right) ,$ which
leads to $\mathbb{E}\left[ \left. v_{x}(x,s_{h},z)\right\vert z\geq z_{h}%
\right] >v_{x}(x,s,\hat{\mu}\left( s\right) ).$

Because the utility difference in (\ref{thm2_26}) is zero at $x=x_{h}$ and $%
s=\tilde{\sigma}\left( z_{h}\right) ,$ (\ref{thm2_27}) implies that (\ref%
{thm2_26}) is positive for $x>x_{h}$, which means that the receiver of type $%
x>x_{h}$ has no incentive to change his partner to a sender with $s\in
\lbrack s_{\ell },\tilde{\sigma}\left( z_{h}\right) ).$

\paragraph{(c) Optimal matching choice by \textbf{the receiver of type }$%
x\in \lbrack x_{\ell },x_{h})$}

The equilibrium outcomes for receivers of type $x\in \lbrack x_{\ell
},x_{h}) $ and senders of types in $[z_{\ell },z_{h})$ are the same as the
outcomes in the stronger monotone separating CSE, including actions,
reactions and matching. Therefore, from the proof of Conditions 3 in Theorem %
\ref{proposition1}, we know that the utility for the receiver will be lower
by changing his partner to any sender with $s\in \lbrack s_{\ell },\tilde{%
\sigma}\left( z_{h}\right) )$.

Suppose that the receiver changes his partner to a sender with $s_{h}$. To
see if the receiver prefers such a change, first note that

\begin{equation}
\mathbb{E}\left[ \left. v(x_{h},s_{h},z)\right\vert z\geq z_{h}\right]
-t_{h}=v(x_{h},\tilde{\sigma}\left( z_{h}\right) ,z_{h})-\tilde{\tau}\left( 
\tilde{\sigma}\left( z_{h}\right) \right)  \label{thm2_33}
\end{equation}%
Given $s_{h}>\tilde{\sigma}\left( z_{h}\right) $ and $z\geq z_{h}$, we can
apply Assumption 2.(i) to show that for $x<x_{h}$ 
\begin{equation}
\mathbb{E}\left[ \left. v(x,s_{h},z)\right\vert z\geq z_{h}\right]
-t_{h}<v(x,\tilde{\sigma}\left( z_{h}\right) ,z_{h})-\tilde{\tau}\left( 
\tilde{\sigma}\left( z_{h}\right) \right) .  \label{thm2_35}
\end{equation}%
From the proof of Conditions 3 in Theorem \ref{proposition1}, we also know
that for $x<x_{h}$ 
\begin{equation}
v(x,\tilde{\sigma}\left( z_{h}\right) ,z_{h})-\tilde{\tau}\left( \tilde{%
\sigma}\left( z_{h}\right) \right) <v\left( x,\tilde{\sigma}\left(
n^{-1}(x)\right) ,n^{-1}(x)\right) -\tilde{\tau}\left( \tilde{\sigma}\left(
n^{-1}(x)\right) \right)  \label{thm2_36}
\end{equation}%
Combining (\ref{thm2_35}) and (\ref{thm2_36}) yields 
\begin{equation}
\mathbb{E}\left[ \left. v(x,s_{h},z)\right\vert z\geq z_{h}\right]
-t_{h}<v\left( n^{-1}(x),x,\tilde{\sigma}\left( n^{-1}(x)\right)
,n^{-1}(x)\right) -\tilde{\tau}\left( \tilde{\sigma}\left( n^{-1}(x)\right)
\right)  \label{thm2_37}
\end{equation}

The expression on the right-hand side of (\ref{thm2_37}) is indeed the same
as the equilibrium utility for the receiver of type $x<x_{h}$ in the
well-behaved CSE. Therefore, a receiver of type $x\in \lbrack x_{\ell
},x_{h})$ strictly prefers a sender with $\hat{\sigma}\left(
n^{-1}(x)\right) =\tilde{\sigma}\left( n^{-1}(x)\right) $ as his partner.

\paragraph{(d) Optimal reaction choice \textbf{by the receiver of type }$%
x\in \lbrack \protect\underline{x},x_{\ell })$}

The receiver of type $x\in \lbrack \underline{x},x_{\ell })$ is unmatched in
equilibrium. Suppose that the receiver decides to choose a sender with $%
s_{\ell }$ as his partner. We know that $v\left( x_{\ell },s_{\ell },\hat{\mu%
}\left( s_{\ell }\right) \right) -t_{\ell }=0$. This implies that for $x\in
\lbrack \underline{x},x_{\ell })$, 
\begin{equation}
v\left( x,s_{\ell },\hat{\mu}\left( s_{\ell }\right) \right) -t_{\ell }<0.
\label{thm2_38''}
\end{equation}%
Therefore, the receiver of type $x$ $\in \lbrack \underline{x},x_{\ell })$
has no incentive to choose a sender with $s_{\ell }$ as his partner.

Suppose that the receiver of type $x\in \lbrack \underline{x},x_{\ell })$
chooses a sender with $s\in \left( s_{\ell },\tilde{\sigma}\left(
z_{h}\right) \right) $ as his partner. According to Subsection (c) above, we
know that for any $s\in \left( s_{\ell },\tilde{\sigma}\left( z_{h}\right)
\right) $, 
\begin{equation}
v\left( x_{\ell },s,\hat{\mu}\left( s\right) \right) -\tau (s)<v\left(
x_{\ell },s_{\ell },\hat{\mu}\left( s_{\ell }\right) \right) -t_{\ell }
\label{thm2_39}
\end{equation}%
Applying Assumption 2.(i) to (\ref{suff_cond1}) yields that for any $s\in
\left( s_{\ell },\tilde{\sigma}\left( z_{h}\right) \right) $ and any $x\in
\lbrack \underline{x},x_{\ell })$%
\begin{equation}
v\left( x,s,\hat{\mu}\left( s\right) \right) -\tau (s)<v\left( x,s_{\ell },%
\hat{\mu}\left( s_{\ell }\right) \right) -t_{\ell }.  \label{thm2_41'}
\end{equation}%
Because the expression on the right hand side of (\ref{suff_cond2'}) is the
same as the expression on the left hand side of (\ref{decreasing_T}), we can
conclude that $v\left( x,s,\hat{\mu}\left( s\right) \right) -\tau (s)<0$ for 
$x<x_{\ell }$ and $s\in \left( s_{\ell },\tilde{\sigma}\left( z_{h}\right)
\right) ,$ which show that the receiver's utility becomes negative to choose
a sender with $s\in \left( s_{\ell },\tilde{\sigma}\left( z_{h}\right)
\right) $ as his partner.

Finally, suppose that the receiver of type $x$ $\in \lbrack \underline{x}%
,x_{\ell })$ chooses a sender with $s_{h}$ as his partner. According to
Subsection (c) above, we know that 
\begin{equation}
\mathbb{E}\left[ \left. v(x_{\ell },s_{h},z)\right\vert z\geq z_{h}\right]
-t_{h}<v\left( x_{\ell },s_{\ell },\hat{\mu}\left( s_{\ell }\right) \right)
-t_{\ell }  \label{thm2_42}
\end{equation}%
Given $s_{h}>s_{\ell },$ $z_{h}>\hat{\mu}\left( s_{\ell }\right) ,$ applying
Assumption 2.(i) implies that for $x<x_{\ell }$ 
\begin{equation}
\mathbb{E}\left[ \left. v(x,s_{h},z)\right\vert z\geq z_{h}\right]
-t_{h}<v\left( x,s_{\ell },\hat{\mu}\left( s_{\ell }\right) \right) -t_{\ell
}  \label{thm2_45}
\end{equation}%
Because the expression on the right hand side of (\ref{PSD3'}) is the same
as the expression on the left hand side of (\ref{decreasing_T}), we can
conclude that $\mathbb{E}\left[ \left. v(x,s_{h},z)\right\vert z\geq z_{h}%
\right] -t_{h}<0$ for $x<x_{\ell }$. This shows that the receiver's utility
is negative with a sender a sender with $s_{h}$ as his partner. This
concludes that no receiver of type $x\in \lbrack \underline{x},x_{\ell })$
wants to choose any sender in the market as his partner.

\end{document}